\pdfoutput=1
\documentclass[a4paper,11pt]{article}
\usepackage{jheppub,multirow}

\newcommand{\be}{\begin{equation}}
\newcommand{\ee}{\end{equation}}
\def\bsp#1\esp{\begin{split}#1\end{split}}
\def\bpm{\begin{pmatrix}}
\def\epm{\end{pmatrix}}
\newcommand {\superiso} {{\tt SuperIso}}
\newcommand {\spheno} {{\tt SPheno}}

\preprint{CERN-PH-TH-2015-208, LAPTH-034/15}
\title{General squark flavour mixing: constraints, phenomenology and
benchmarks}

\author[a]{Karen~De~Causmaecker}
\author[b,c,d]{\!\!,\ Benjamin~Fuks}
\author[e]{\!\!,\ Bj\"orn~Herrmann}
\author[f,g,*]{\!\!,\ Farvah~Mahmoudi~~~~~~,\!\!\!\!\!\!\!\!\!\!\note[*]{Also Institut Universitaire de France, 103 boulevard Saint-Michel, 75005 Paris, France}}
\author[h]{Ben~O'Leary}
\author[h]{\!\!,\ Werner~Porod}
\author[i]{\!\!,\ Sezen~Sekmen}
\author[j,k]{\!\! and Nadja~Strobbe}

\affiliation[a]{Theoretische Natuurkunde, IIHE/ELEM and International Solvay Institutes, Vrije Universiteit Brussel, Pleinlaan 2, B-1050 Brussels, Belgium}
\affiliation[b]{Sorbonne Universit\'es, UPMC Univ.~Paris 06, UMR 7589, LPTHE, F-75005 Paris, France}
\affiliation[c]{CNRS, UMR 7589, LPTHE, F-75005 Paris, France}
\affiliation[d]{Institut Pluridisciplinaire Hubert Curien/D\'epartement Recherches Subatomiques,
    Universit\'e de Strasbourg/CNRS-IN2P3, 23 Rue du Loess, F-67037 Strasbourg, France}
\affiliation[e]{LAPTh, Universit\'e Savoie Mont Blanc, CNRS, 9 Chemin de Bellevue, F-74941 Annecy-le-Vieux, France}
\affiliation[f]{Universit{\' e} de Lyon, Universit{\' e} Lyon 1, F-69622 Villeurbanne Cedex, France;
  Centre de Recherche Astrophysique de Lyon, CNRS, UMR 5574, Saint-Genis Laval Cedex, F-69561, France;
  Ecole Normale Sup{\' e}rieure de Lyon, France}
\affiliation[g]{CERN Theory Division, Physics Department, CH-1211 Geneva 23, Switzerland}
\affiliation[h]{Institut f\"ur Theoretische Physik und Astrophysik, Universit\"at W\"urzburg, D-97074 W\"urzburg, Germany}
\affiliation[i]{Kyungpook National University, Department of Physics, Daegu, 702-701 Korea}
\affiliation[j]{Ghent University, Department of Physics and Astronomy, Proeftuinstraat 86, B-9000 Gent, Belgium}
\affiliation[k]{Fermi National Accelerator Laboratory, Batavia, 60510-5011, USA}

\abstract{
  We present an extensive study of non-minimal flavour violation in the squark
  sector in the framework of the Minimal Supersymmetric Standard Model. We
  investigate the effects of multiple non-vanishing flavour-violating elements
  in the squark mass matrices by means of a Markov Chain Monte Carlo scanning
  technique and identify parameter combinations that are favoured by both current
  data and theoretical constraints. We then detail the resulting distributions
  of the flavour-conserving and flavour-violating model
  parameters.
  Based on this analysis, we propose a set of benchmark scenarios relevant for future
  studies of non-minimal flavour violation in the Minimal Supersymmetric Standard
  Model.
}

\emailAdd{karen.de.causmaecker@vub.ac.be}
\emailAdd{fuks@lpthe.jussieu.fr}
\emailAdd{herrmann@lapth.cnrs.fr}
\emailAdd{nazila@cern.ch}
\emailAdd{ben.oleary@physik.uni-wuerzburg.de}
\emailAdd{porod@physik.uni-wuerzburg.de}
\emailAdd{sezen.sekmen@cern.ch}
\emailAdd{nadja.strobbe@cern.ch}


\begin{document}
\maketitle
\flushbottom



\section{Introduction \label{sec:Intro}}

Among all possible extensions to the Standard Model of particle physics,
weak scale supersymmetry~\cite{Nilles:1983ge,Haber:1984rc} remains one of the
most popular and best studied options. The quest for the superpartners of the
Standard Model degrees of freedom is one of the hot topics of the
current high-energy physics experimental programme and many search channels are
hence investigated at the Large Hadron Collider
(LHC)~\cite{AtlasSusy, CmsSusy}. Since no signal of supersymmetry has been found
so far, the results have been interpreted either in terms of limits on specific
setups like constrained versions of the Minimal Supersymmetric Standard Model
(MSSM) or in terms of simplified model spectra inspired by the MSSM. As a
result, either supersymmetric particles are constrained to reside at scales that
are not reachable in proton-proton
collisions at a centre-of-mass energy of 8~TeV, or the
spectrum must present specific properties that allow the superpartners to evade
detection as in the case of compressed supersymmetric spectra or
non-minimal realizations of supersymmetry. In this work, we follow this
latter guiding principle and explore to which extent deviations from the
minimal flavour-violation paradigm~\cite{Hall:1985dx,D'Ambrosio:2002ex,
Cirigliano:2005ck} are allowed by current data.

In minimally flavour-violating
supersymmetry, all the flavour properties of the model stem from the
diagonalization of the Yukawa matrices that yields different gauge and mass
bases for the (s)quarks and (s)leptons. Flavour violation is thus entirely
encompassed in the CKM and PMNS matrices. There is however no theoretical
motivation for the flavour structure of a supersymmetric model to be the same as
in the Standard Model. For instance, when supersymmetry is embedded in a Grand Unified
framework, new sources of
flavour violation could be allowed~\cite{Gabbiani:1988rb}. The soft supersymmetry-breaking mass
and trilinear coupling matrices of the sfermions could therefore comprise
non-diagonal flavour-violating entries that are not related to the CKM and PMNS
matrices. We adopt such a setup, which is referred to as non-minimally
flavour-violating (NMFV) supersymmetry, and study the impact of these additional
flavour-violating soft terms.

In recent years, the consequences of
non-minimal flavour violation in the squark sector have been
investigated in various areas. NMFV effects on
low-energy observables such as rare decays (see, \textit{e.g.}
Ref.~\cite{Buchalla:2008jp} and references therein) or
oblique parameters~\cite{Heinemeyer:2004by} have been considered,
and the potential signatures at the LHC have been investigated~\cite{Bozzi:2007me,
Dittmaier:2007uw,Fuks:2008ab,Hurth:2009ke,Bartl:2009au,Plehn:2009it,
Bartl:2010du, Bruhnke:2010rh, Bartl:2011wq, Fuks:2011dg, Bartl:2012tx,
Blanke:2013uia,Aebischer:2014lfa,Backovic:2015rwa}.
More recently, the existing constraints on possible non-vanishing flavour-mixing
parameters have been updated~\cite{Arana-Catania:2014ooa,Kowalska:2014opa}.
These results have been derived under the restriction that only few
off-diagonal elements of the squark mass matrices are non-zero, and that
at most two of them are varied at the same time.
One would however generally expect that several of the flavour-violating
entries could be non-vanishing, especially if the flavour structure is generated by
some mechanism at a higher scale. Consequently, a
comprehensive study of the most general NMFV configuration of the MSSM, where
all flavour-violating Lagrangian terms are taken into account and confronted to
current data and theoretical constraints, is in order. A first step in this
direction is achieved with this work.

We consider the most general mixings between second and third generation
squarks. Any non-CKM induced mixing with the first generation is ignored
as a result of constraints imposed by kaon data~\cite{Ciuchini:2007ha}.
Choosing a phenomenological approach, we model the flavour-violating
effects under investigation by a set of 19 free parameters defined at the TeV
scale and identify, by means of a Markov Chain Monte Carlo parameter scanning
technique, the regions of the parameter space that are favoured in
light of current data.
The rest of this paper is organised as follows. We first review in
Section~\ref{sec:Model} the squark sector of the MSSM when NMFV is allowed and
present our parameterisation of the effects under study.
In Section~\ref{sec:Setup}, we describe the machinery
that is employed to explore the 19-dimensional NMFV MSSM parameter space and
present the experimental constraints that are imposed. Our results are discussed
in Section~\ref{sec:Results}, we propose NMFV MSSM benchmark scenarios
for the LHC Run II in Section~\ref{sec:Benchmark} and our conclusions are given in
Section~\ref{sec:Conclusion}.


\section{The squark sector with general flavour mixing \label{sec:Model}}

Starting from the most general MSSM Lagrangian, the super-CKM basis is defined
by rotating the quark and squark superfields in flavour space in a way in which
the quark mass matrices $m_u$ and $m_d$ are diagonal. Squark and quark flavours
are thus aligned, although the squark mass matrices are not necessarily
diagonal. In the
$(\tilde u_L,\tilde c_L,\tilde t_L,\tilde u_R,\tilde c_R,\tilde t_R)$ and
$(\tilde d_L,\tilde s_L,\tilde b_L,\tilde d_R,\tilde s_R,\tilde b_R)$
bases, the up-type and down-type squark mass matrices ${\cal M}^2_{\tilde{u}}$
and ${\cal M}^2_{\tilde{d}}$ are given by
\be\bsp
  {\cal M}_{\tilde{u}}^2 ~&=~ \bpm
     V_{\rm CKM} M^2_{\tilde{Q}} V^{\dag}_{\rm CKM} + m^2_{u} + D_{\tilde{u},L}
     ~~&~~
     \frac{v_u}{\sqrt{2}} T_u^\dag - m_u \frac{\mu}{\tan\beta} \\
     \frac{v_u}{\sqrt{2}} T_u - m_u \frac{\mu^*}{\tan\beta}
     ~~&~~
     M^2_{\tilde{U}} + m^2_{u} + D_{\tilde{u},R}
  \epm\ , \\
  {\cal M}_{\tilde{d}}^2 ~&=~ \bpm
    M^2_{\tilde{Q}} + m^2_{d} + D_{\tilde{d},L}
    ~~&~~
    \frac{v_d}{\sqrt{2}} T_d^\dag - m_d \mu \tan\beta \\
    \frac{v_d}{\sqrt{2}} T_d - m_d \mu^* \tan\beta
    ~~&~~
    M^2_{\tilde{D}} + m^2_{d} + D_{\tilde{d},R}
  \epm \ ,
\esp\label{eq:M2q} \ee
respectively. In these notations, we have introduced the soft
supersymmetry-breaking squark mass matrices $M^2_{\tilde{Q}}$,
$M^2_{\tilde{U}}$ and $M^2_{\tilde{D}}$ for left-handed, up-type right-handed
and down-type right-handed squarks respectively,
as well as the matrices $T_u$ and $T_d$ that embed the
trilinear soft interactions of the up-type and down-type squarks with the Higgs
sector. While these five matrices are defined to be flavour diagonal in usual
constrained versions of the MSSM, our NMFV framework allows them to be general
and possibly flavour-violating. Moreover, $V_{\rm CKM}$ stands for the CKM
matrix, $\mu$ denotes the superpotential
Higgs(ino) mass parameter and $\tan\beta=\frac{v_u}{v_d}$ is the ratio of the
vacuum expectation values of the neutral components of the two Higgs doublets.
Finally, the squark mass matrices also include (flavour-diagonal) $D$-term
contributions
\be
  D_{\tilde{q},L} = m_Z^2 \left( I_q - e_q \sin^2\theta_W \right) \cos 2\beta
  \qquad\text{and}\qquad
  D_{\tilde{q},R} = m_Z^2 e_q \sin^2\theta_W \cos 2\beta\ ,
\ee
where $m_Z$ is the $Z$-boson mass, $\theta_W$ is the weak mixing angle and
$e_q$ and $I_q$ (with $q=u,d$) are the electric charge and the weak isospin
quantum numbers of the (s)quarks.

In order to reduce the number of supersymmetric input
parameters, we assume that the first two generations of squarks are degenerate so
that the (flavour-conserving) soft masses are determined by six free parameters,
\be\bsp
  \big(M_{\tilde Q}\big)_{11}=\big(M_{\tilde Q}\big)_{22}\equiv
     M_{\tilde Q_{1,2}} \ , \qquad
  \big(M_{\tilde Q}\big)_{33}\equiv M_{\tilde Q_{3}} \ ,\\
  \big(M_{\tilde U}\big)_{11}=\big(M_{\tilde U}\big)_{22}\equiv
     M_{\tilde U_{1,2}} \ , \qquad
  \big(M_{\tilde U}\big)_{33}\equiv M_{\tilde U_{3}} \ ,\\
  \big(M_{\tilde D}\big)_{11}=\big(M_{\tilde D}\big)_{22}\equiv
     M_{\tilde D_{1,2}}\ , \qquad
  \big(M_{\tilde D}\big)_{33}\equiv M_{\tilde D_{3}} \ .
\esp\ee 
Moreover, we define the diagonal components of the trilinear couplings $T_q$
relatively to the Yukawa matrices $Y_q$,
\be (T_q)_{ii} = (A_q)_{ii} (Y_q)_{ii}\ .\ee
We then neglect the first and second generation Yukawa couplings so that
only the trilinear coupling parameters related to third generation squarks are
considered as free parameters. We take them equal for simplicity,
so that we have
\be
  (A_u)_{33} \equiv A_t \ , \qquad
  (A_d)_{33}\equiv A_b
  \qquad\text{and}\qquad
  A_t = A_b \equiv A_f \ .
\ee

We now turn to the off-diagonal elements of the squark mass matrices. In order
to be compliant with kaon data, we ignore any mixing involving one of the
first generation squarks~\cite{Ciuchini:2007ha}. Next, following standard
prescriptions~\cite{Gabbiani:1996hi}, we normalise the remaining non-diagonal entries of
the squared squark mass matrices with respect to the diagonal ones and
parameterise all considered NMFV effects by seven dimensionless quantities
\be\bsp
  \delta_{LL}=
     \frac{(M^2_{\tilde{Q}})_{23}}{(M_{\tilde{Q}})_{22}( M_{\tilde{Q}})_{33}}\ ,
  \quad
  \delta^u_{RR}=
     \frac{(M^2_{\tilde{U}})_{23}}{(M_{\tilde{U}})_{22}( M_{\tilde{U}})_{33}}\ ,
  \quad
  \delta^d_{RR}=
     \frac{(M^2_{\tilde{D}})_{23}}{(M_{\tilde{D}})_{22}( M_{\tilde{D}})_{33}}\ , \\
  \delta^u_{RL}= \frac{v_u}{\sqrt{2}}
     \frac{(T_{u})_{23}}{(M_{\tilde{Q}})_{22}( M_{\tilde{U}})_{33}}\ ,
  \qquad
  \delta^u_{LR}= \frac{v_u}{\sqrt{2}}
     \frac{(T_{u})_{32}}{(M_{\tilde{Q}})_{33}( M_{\tilde{U}})_{22}}\ ,
  \qquad\qquad \\
  \delta^d_{RL}= \frac{v_d}{\sqrt{2}}
     \frac{(T_{d})_{23}}{(M_{\tilde{Q}})_{22}( M_{\tilde{D}})_{33}}\ ,
  \qquad
  \delta^d_{LR}= \frac{v_d}{\sqrt{2}}
     \frac{(T_{d})_{32}}{(M_{\tilde{Q}})_{33}( M_{\tilde{D}})_{22}}\ .
  \qquad\qquad
\esp\label{eq:NMFVparams}\ee

The physical squark states $\tilde{u}_i$ and $\tilde{d}_i$ (with $i=1,\ldots,6$)
are obtained by diagonalizing the squared squark mass matrices
${\cal M}_{\tilde{u}}^2$ and ${\cal M}_{\tilde{d}}^2$ according to
\be
  {\rm diag}\big(m^2_{\tilde{q}_1},m^2_{\tilde{q}_2},\dots,m^2_{\tilde{q}_6}\big)
     ~=~ {\cal R}_{\tilde{q}} {\cal M}_{\tilde{q}}^2 {\cal R}^{\dag}_{\tilde{q}}
  \qquad\text{for}\quad  q=u,d \ .
\ee
By convention the mass eigenstates are taken ordered such that
$m^2_{\tilde{q}_1} < \dots < m^2_{\tilde{q}_6}$. The $6\times 6$ rotation
matrices ${\cal R}_{\tilde u}$ and ${\cal R}_{\tilde d}$ carry the information
about the flavour decomposition of the squarks,
\be\bsp
  \bpm \tilde u_1&  \tilde u_2&  \tilde u_3&  \tilde u_4&  \tilde u_5 &  \tilde
    u_6 \epm^t =&\ {\cal R}_{\tilde u} \bpm \tilde u_L&  \tilde c_L&  \tilde t_L&  \tilde u_R & 
    \tilde c_R&  \tilde t_R \epm^t \ ,\\
  \bpm \tilde d_1&  \tilde d_2&  \tilde d_3&  \tilde d_4&  \tilde d_5 &  \tilde
    d_6 \epm^t =&\ {\cal R}_{\tilde d} \bpm \tilde d_L&  \tilde s_L&  \tilde b_L&  \tilde d_R & 
    \tilde s_R&  \tilde b_R \epm^t \ ,
\esp\ee
and their different entries directly appear in couplings of the
squarks to the other particles (see, {\it e.g.} Refs.~\cite{Bozzi:2007me, Bruhnke:2010rh}).

In addition, the gaugino sector is chosen to be determined by a single
parameter, the bino mass $M_1$. The wino and gluino tree-level masses $M_2$ and $M_3$ are then obtained
by making use of a relation inspired by Grand-Unified theories,
\be 
	M_1 = \frac12 M_2 = \frac16 M_3\ . 
\ee
The slepton sector is defined in a flavour-conserving fashion, so that the
soft terms consist of three (diagonal) mass parameters that we set to a
common value
\be
  (M_{\tilde \ell})_{11} =  (M_{\tilde \ell})_{22} = 
    (M_{\tilde \ell})_{33} \equiv M_{\tilde \ell} \ ,
\ee
and the slepton trilinear coupling matrix to the
Higgs sector $T_\ell$ contains a single non-zero entry,
\be
  (T_\ell)_{33} = Y_\tau A_\tau \equiv Y_\tau A_f \ .
\ee
All flavour-conserving trilinear sfermion interactions with the Higgs bosons are
consequently driven by a single input parameter $A_f$. The model description
is completed by the definition of the Higgs sector that is parameterised
in terms of the $\mu$ parameter, $\tan\beta$ and the pole mass of
the pseudoscalar Higgs boson $m_A$.


\section{Setup and constraints \label{sec:Setup}}

\subsection{Numerical setup}

In the previous section, we have defined a simplified parameterisation of general
NMFV MSSM scenarios in terms of 16 soft supersymmetry-breaking parameters to
which we have supplemented three parameters related to the Higgs sector.
Turning to the Standard Model sector, the QCD interaction strength is
computed from the value of the strong coupling constant at the
$Z$-pole $\alpha_s(m_Z)$, while we choose as three independent electroweak inputs the
electromagnetic coupling constant evaluated at the $Z$-pole $\alpha(m_Z)$, the
Fermi constant $G_F$ and the $Z$-boson mass $m_Z$. The fermion sector is defined
by the pole mass of the top quark $m_t^{\rm pole}$, the $\overline{\rm MS}$
mass of the bottom (charm) quark $m_b$ ($m_c$) evaluated at the $m_b$ ($m_c$)
scale and the $\overline{\rm MS}$ masses of the three lightest quarks evaluated
at a scale of 2~GeV. Finally, we include in our parameterisation the masses
of the electron ($m_e$), the muon ($m_\mu$) and the tau ($m_\tau$) and we
calculate the CKM matrix using the Wolfenstein parameters  $\lambda^{\rm CKM}$,
$A^{\rm CKM}$, $\bar \rho^{\rm CKM}$ and $\bar\eta^{\rm CKM}$.

\begin{table}
\renewcommand{\arraystretch}{1.2}
  \begin{center}
    \begin{tabular}{c|c}
        Parameter & Value\\
      \hline
      \hline
        $\alpha^{-1}(m_Z)$ & 127.934 \\
        $m_Z$ & 91.1876~\text{GeV} \\
        $G_F$ & $1.16637\times 10^{-5}~\text{GeV}^{-2}$ \\
      \hline
        $m_c^{\overline{\rm MS}}(m_c)$ & 1.25~\text{GeV} \\
        $m_s^{\overline{\rm MS}}(2~\text{GeV})$ & 120~\text{MeV} \\
        $m_u^{\overline{\rm MS}}(2~\text{GeV})$ & 3~\text{MeV} \\
        $m_d^{\overline{\rm MS}}(2~\text{GeV})$ & 7~\text{MeV}
    \end{tabular}
    \qquad
    \begin{tabular}{c|c}
        Parameter & Value\\
      \hline
      \hline
        $m_e$ & 510.9989~\text{keV} \\
        $m_\mu$ & 105.6583~\text{MeV} \\
        $m_\tau$ & 1.77699~\text{GeV} \\
      \hline
        $\lambda^{\rm CKM}$ & 0.2272 \\
        $A^{\rm CKM}$ & 0.818 \\
        $\bar{\rho}^{\rm CKM}$ & 0.221 \\
        $\bar{\eta}^{\rm CKM}$ & 0.34
    \end{tabular}
  \end{center}
\caption{Standard Model sector of our NMFV MSSM parameter space.
  \label{tab:SMparams}}
\end{table}

For our exploration of the NMFV MSSM, we start
by fixing the Standard Model parameters to the values provided in the
review of the Particle Data Group~\cite{Beringer:1900zz}, as shown in
Table~\ref{tab:SMparams}. We then allow $\alpha_s(m_Z)$,
$m_t^{\rm pole}$ and $m_b^{\overline{\rm MS}}(m_b)$ to vary according to
Gaussian profiles with their measured central values and errors taken as means
and widths,
as shown in Table~\ref{tab:ParamRanges}. We vary randomly all the 19
supersymmetric parameters in the intervals presented in this table
and finally calculate the resulting
mass spectrum of the supersymmetric particles at the one-loop level 
using the publicly available spectrum generator \spheno~\cite{Porod:2003um,Porod:2011nf}.
We can in this way widely cover the regions of the parameter space where the
electroweak symmetry is successfully broken (with, \textit{e.g.}\ sub-TeV values
for the $\mu$ parameter) and whose signatures are in principle observable
at the LHC within the next few years (\textit{i.e.}\
with not dramatically large soft sfermion
masses). Moreover, the trilinear coupling parameter $A_f$ and
the $\delta$-parameters have been chosen to prevent all off-diagonal
elements of the squark mass matrices from being too large, so that tachyonic
mass-eigenstates are avoided.

In order to explore the 22-dimensional parameter space summarized in
Table~\ref{tab:ParamRanges}, we rely on a Markov Chain Monte Carlo (MCMC)
scanning technique~\cite{MCMC1, Metropolis:1953am, Hastings:1970aa} and
impose on each of the studied setups a set of constraints that is described in
the next subsection. In this scanning procedure, a given point is accepted or
rejected based on the comparison of the products of likelihoods of this point
with that of the previous point, where each of the likelihoods is associated with
a specific constraint accounting for measurements and theoretical predictions
in the NMFV MSSM framework.

\begin{table}
\renewcommand{\arraystretch}{1.2}
  \begin{center}
    \begin{tabular}{c|c}
        Parameter & Scanned range \\
      \hline
      \hline
        $\alpha_s(m_Z)$ & ${\cal N}(0.1184,0.0007)$ \\
        $m_t^{\rm pole}$ & ${\cal N}(173.3,1.3928)$~GeV\\
        $m_b(m_b)$ & ${\cal N}(4.19,0.12)$~GeV\\
      \hline
        $M_{\tilde{Q}_{1,2}}$ & [300, 3500]~GeV \\
        $M_{\tilde{Q}_{3}}$ & [100, 3500]~GeV\\
        $M_{\tilde{U}_{1,2}}$ & [300, 3500]~GeV \\
        $M_{\tilde{U}_{3}}$ & [100, 3500]~GeV \\
        $M_{\tilde{D}_{1,2}}$ & [300, 3500]~GeV \\
        $M_{\tilde{D}_{3}}$ & [100, 3500]~GeV \\
      \hline
        \multirow{2}{*}{$A_f$} & [-10000, 10000]~GeV \\
           & \text{or $|A_f| < 4 \max\{M_{\tilde{q}}, M_{\tilde{\ell}}\}$ }
		   \vspace*{5.5mm}
    \end{tabular}
    \qquad\qquad
    \begin{tabular}{c|c}
        Parameter & Scanned range \\
      \hline
      \hline
        $\tan \beta$ & [10, 50] \\
        $\mu$ & [100, 850]~GeV \\
        $m_{A}$ & [100, 1600]~GeV \\
        $M_1$ & [100, 1600]~GeV \\
        $M_{\tilde{\ell}}$ & [100, 3500]~GeV \\ 
      \hline
        $\delta_{LL}$ & [-0.8, 0.8] \\
        $\delta_{RR}^u$ & [-0.8, 0.8] \\
        $\delta_{RR}^d$ & [-0.8, 0.8] \\
        $\delta_{LR}^u$ & [-0.5, 0.5] \\
        $\delta_{RL}^u$ & [-0.5, 0.5] \\
        $\delta_{LR}^d$ & [-0.05, 0.05] \\
        $\delta_{RL}^d$ & [-0.05, 0.05]
    \end{tabular}
\caption{Supersymmetric and Higgs sectors of our NMFV MSSM parameter space,
  as well as varying Standard Model parameters.
  ${\cal N}(\mu,\sigma)$ denotes a Gaussian profile of mean $\mu$ and width $\sigma$.}
\label{tab:ParamRanges}
\end{center}
\end{table}

\subsection{Indirect constraints on general squark mixing}

The masses and flavour-violating mixings of the superpartners can be indirectly
probed by numerous flavour physics constraints, the anomalous moment of the muon
as well as by the properties of the recently discovered Standard-Model-like
Higgs boson. In our MCMC scanning procedure, we additionally impose the lightest
superpartner to be the lightest neutralino, so that it could be a
phenomenologically viable dark matter candidate. We dedicate the rest of this
section to a brief description of all observables that have been considered in
the scan and that are summarised in Table~\ref{tab:PLMs}.

\begin{table}
\renewcommand{\arraystretch}{1.2}
  \begin{center}
    \begin{tabular}{c|c|c}
        Observable & Experimental result & Likelihood function\\
      \hline
      \hline
        ${\rm BR}(B \rightarrow X_s\gamma) $ 
          & $(3.43 \pm 0.22) \times 10^{-4}$ \cite{Amhis:2014hma}
          & Gaussian\\
        ${\rm BR}(B_s \rightarrow \mu \mu)$
          & $(2.8 \pm 0.7)\times 10^{-9}$ \cite{CMS:2014xfa}
          & Gaussian \\
        ${\rm BR}(B \rightarrow K^* \mu \mu)_{q^2 \in [1,6] ~\text{GeV}^2}$
          & $(1.7 \pm 0.31) \times 10^{-7}$ \cite{Aaij:2013iag}
          & Gaussian \\
        ${\rm AFB}(B \rightarrow K^* \mu \mu)_{q^2 \in [1.1,6] ~\text{GeV}^2}$
          & $(-0.075 \pm 0.036) \times 10^{-7} $ \cite{LHCb:2015dla}
          & Gaussian \\
        ${\rm BR}(B \rightarrow X_s \mu \mu)_{q^2 \in [1,6] ~\text{GeV}^2}$
          & $(0.66 \pm 0.88) \times 10^{-6}$ \cite{Lees:2013nxa}
          & Gaussian \\
        ${\rm BR}(B \rightarrow X_s \mu \mu)_{q^2 > 14.4 ~\text{GeV}^2}$ 
          & $(0.60 \pm 0.31) \times 10^{-6}$ \cite{Lees:2013nxa}
          & Gaussian \\
        ${\rm BR}(B_u \rightarrow \tau \nu)/{\rm BR}(B_u \rightarrow \tau \nu)_{\rm SM} $
          & $1.04\pm 0.34$~\cite{Beringer:1900zz}
          & Gaussian \\
        $\Delta M_{B_s}$
          & $(17.719 \pm 3.300)\; {\rm ps}^{-1}$~\cite{Beringer:1900zz}
          & Gaussian \\
      \hline
        $\epsilon_K$
          & $(2.228 \pm 0.29^{\rm th}) \times 10^{-3}$~\cite{Beringer:1900zz}
          & Gaussian\\
        ${\rm BR}(K^0 \rightarrow \pi^0 \nu \nu)$
          & $  \leq 2.6 \times 10^{-8}$~\cite{Beringer:1900zz}
          & 1 if yes, 0 if no \\
        ${\rm BR}(K^+ \rightarrow \pi^+ \nu \nu)$
          & $ 1.73^{+1.15}_{-1.05} \times 10^{-10}$~\cite{Beringer:1900zz}
          & Two-sided Gaussian \\
      \hline
        $\Delta a_\mu$
          & $(26.1 \pm 12.8)\times 10^{-10}$ $[e^+e^-]$~\cite{Beringer:1900zz}
          & Gaussian \\
      \hline
        $m_h$
          & $125.5\pm2.5$ GeV~\cite{Aad:2012tfa,Chatrchyan:2012ufa} & 1 if yes, 0 if no \\
      \hline
        Lightest supersymmetric particle & Lightest neutralino & 1 if yes, 0 if no 
    \end{tabular}
    \caption{Experimental constraints imposed in our scan of the NMFV MSSM parameter space.}
  \label{tab:PLMs}
  \end{center}
\end{table}

Non-minimal flavour-violating squark mixing involving third generation squarks
is by construction very sensitive to constraints arising from $B$-physics
observables. In particular, $B$-meson rare decays and oscillations are expected
to play an important role as the Standard Model contributions are
loop-suppressed. Although we only
consider squark mixing between the second and third generations, we also
include constraints arising from observables related to the kaon sector. Even if
not present at the scale at which we calculate the supersymmetric spectrum
(\textit{i.e.} the electroweak symmetry breaking scale), squark mixings with the
first generation are induced by the non-vanishing CKM matrix and
renormalisation-group running so that kaon physics observables (calculated
at a different scale) are also relevant for
extracting constraints on the NMFV MSSM parameter space.

We focus on the branching ratios associated with the rare
$B \rightarrow X_s\gamma$, $B \rightarrow K^* \mu \mu$,
$B \rightarrow X_s \mu \mu$ and $B_u \rightarrow \tau \nu$ decays, as well as
on the forward-backward asymmetry (AFB) arising in
$B \rightarrow K^* \mu \mu$ decays. The associated predictions are calculated with
the \superiso\ package~\cite{Mahmoudi:2007vz,Mahmoudi:2008tp}.
In addition, we compute the neutral $B$-meson mass
difference $\Delta M_{B_s}$, the branching ratio associated with the $B_s\to\mu^+\mu^-$,
$K^0 \rightarrow \pi^0 \nu \nu$ and $K^+ \rightarrow \pi^+ \nu \nu$ decays and
the kaon parameter $\epsilon_K$ with the \spheno~code~\cite{Porod:2003um,Porod:2011nf}.
We furthermore employ \spheno\ for
the estimation of the supersymmetric contributions to the anomalous magnetic
moment of the muon $a_\mu$\footnote{Imposing predictions for $a_\mu$ to agree with
the related measured values leads to a preference for a lighter
slepton mass spectrum. This can indirectly imply constraints on the NMFV MSSM
parameters \textit{via} the flavour observables that involve sleptons.} and for
a calculation of the lightest Higgs boson
mass $m_h$. The data transfer between \spheno\ and \superiso\
is achieved through the Flavour Les Houches Accord standard~\cite{Mahmoudi:2010iz}, and
the Wilson coefficients for all hadronic observables are calculated in \spheno\ (at the
scale $Q = 160$~GeV) and \superiso\ (at a scale $Q=m_W$) 
from the values of the running coupling constants and the supersymmetric masses
and parameters that have been evaluated with \spheno.

We now briefly collect all references where the formulas that have been employed for the
calculation of the considered observables can be found,
and we indicate which NMFV MSSM parameters are mainly constrained by each of
these observables. The calculation of the
$B \to X_s \gamma$ branching ratio is mainly based on the results of
Refs.~\cite{Misiak:2006zs,Misiak:2006ab,Misiak:2010tk}, while those of the
$B \to X_s \mu^+\mu^-$ and $B \to K^* \mu^+\mu^-$ branching ratios
respectively follow Refs.~\cite{Dai:1996vg,Ghinculov:2003qd,Huber:2005ig,%
Huber:2007vv} and Refs.~\cite{Beneke:2001at,Beneke:2004dp,Kruger:2005ep,%
Egede:2008uy,Egede:2010zc,Khodjamirian:2010vf}. All three decays are sensitive
to the left-left and to the left-right squark mixing parameters.

In the case of the $B_s \to \mu^+ \mu^-$ branching ratio, the formulas of
Ref.~\cite{Dreiner:2012dh} have been used. For large values of $\tan\beta$, the
pseudoscalar Higgs boson contribution gives a sizeable deviation from the
Standard Model expectation so that when non-minimal flavour violation in the squark
sector is allowed, this observable mainly restricts left-right mixing
parameters~\cite{Buras:2002vd}. Additionally, it is also sensitive to $\delta_{LL}$
and $\delta^d_{RR}$ when the gluino is not too heavy~\cite{Isidori:2002qe}.
For the $B$-meson oscillation parameter $\Delta M_{B_s}$, we use the formulas of
Refs.~\cite{Baek:2001kh,Buras:2002vd} with the hadronic parameters
$\bar{P}^{LR}_1 = -0.71$, $\bar{P}^{LR}_2 = -0.9$, $\bar{P}^{SLL}_1 = -0.37$
and $\bar{P}^{SLL}_1 = -0.72$.
The NMFV contributions are mainly sensitive to the $\delta_{LL} \delta^d_{RR}$,
$\delta^d_{LR} \delta^d_{RL}$ and $\delta^u_{LR} \delta^u_{RL}$ products,
the relative suppression and enhancement of their
contributions being driven by the ratio of the chargino over the gluino
mass~\cite{Buras:2002vd,Hurth:2009ke}.

In the kaon sector, the $\epsilon_K$ observable is estimated by combining the
formulas of Refs.~\cite{Buras:2001ra,Buras:2002vd}, the loop-contributions
being evaluated with $\eta_{tt} = 0.5$,
$\eta_{ct} = 0.47$ and $\eta_{cc} = 1.44$~\cite{Herrlich:1996vf}.
In addition, we fix all hadronic parameters at the scale $Q=2$~GeV as
$B^{VLL}_1 = 0.61, B^{SLL}_1 = 0.76, B^{SLL}_2 = 0.51, B^{LR}_1 = 0.96$
and $B^{LR}_2 = 1.2$~\cite{Buras:2001ra},
and we set the decay constant $f_K$ to 155.8~MeV. The quantity $\epsilon_K$ is
not directly sensitive to a single NMFV MSSM parameter but will allow us to
constrain $\delta_{LL,13}\delta^d_{RR,23}$ and $\delta_{LL,23} \delta^d_{RR,13}$
products (recalling that first generation squark mixings are
generated by renormalisation-group running). On
different lines, the branching ratios associated with the rare
$K^+\to \pi^+ \nu\nu$ and
$K_L\to \pi^0 \nu\nu$ decays are calculated from the formulas given in
Ref.~\cite{Buras:2004qb} with $\kappa_L = 2.13 10^{-11}$,
$\kappa_+ = 5.16 10^{-11}$ and $P_c=0.39$. These observables mainly constrain
the product $\delta^u_{LR,13} \delta^{u*}_{LR,23}$ as well as higher-order
combinations of $\delta$-parameters that in particular appear in
gluino/down-type squark box-contributions~\cite{Colangelo:1998pm,Buras:2004qb}.

In all the calculations described above, we have used
the results of Ref.~\cite{Crivellin:2011jt} for calculating
chirally-enhanced interaction strengths that include, \textit{e.g.} the
resummation of loop-induced holomorphic coupling effects when $\tan\beta$ and/or
the sfermion-Higgs trilinear couplings are large.

As we allow for relatively light sleptons, charginos and neutralinos, we
calculate the supersymmetric contributions to the anomalous magnetic moment of
the muon by using the formulas of Ref.~\cite{Ibrahim:1999hh}. The related impact
in terms of constraints on our NMFV MSSM parameter space depends on the
higgsino/gaugino nature of the lighter charginos and neutralinos.

Finally, the calculation of the Higgs boson mass includes the complete one-loop
contribution that embeds all possible flavour structures and that is obtained by
extending the formulas of Ref.~\cite{Pierce:1996zz}. For the two-loop
corrections, we have made use of the formulas of Refs.~\cite{Degrassi:2001yf,%
Brignole:2001jy,Brignole:2002bz,Dedes:2003km,Dedes:2002dy,Allanach:2004rh} where
generation mixing is neglected so that
only third generation mass parameters in the super-CKM basis are used
as input parameters. Although flavour effects can shift the Higgs mass by
a few GeV at the one-loop level, in particular when the product $\delta^u_{LR}
\delta^u_{RL}$ is large~\cite{AranaCatania:2011ak,Bartl:2012tx,Kowalska:2014opa}, the two-loop effects are expected
to be of one order of magnitude smaller so that ignoring
the associated flavour mixing is expected to be a reasonable approximation.


\section{Results and discussion \label{sec:Results}}

The analysis of the results of the Markov Chain Monte Carlo scan presented in
Section~\ref{sec:Model} gives us information on the regions of parameter space
that are favoured by the experimental data shown in Table \ref{tab:PLMs}.
The influence of a specific experimental result on a given parameter can be
studied by comparing its theoretical prior distribution to the posterior
one that is derived after imposing the related constraint. The prior
distributions of all parameters are obtained from a uniform random scan in which
we ignore scenarios that exhibit tachyons, where the electroweak symmetry is not
successfully broken and where the lightest neutralino is not the lightest
supersymmetric particle. We hence include about $1.5 \times 10^6$ theoretically
accepted setups. The posterior distributions are then computed
on the basis of our MCMC scan in which all the experimental constraints,
except the one on the Higgs boson mass, are imposed. This scan consists of 100 chains
of 6000 scenarios in which the first 900 ones (the burn-in length) are removed.
The constraint on the Higgs boson mass is eventually imposed and the final
posterior distributions include about $100\,000$ points.
To estimate the importance of each observable separately, we have run a separate MCMC scan
consisting of 100 chains of 2000 scenarios
for each observable. After removing the burn-in length, $170\,000$ points remain.
For each scan, the convergence test of Gelman and Rubin has been verified~\cite{gelman1992}.

Although we are mainly interested in the non-minimally flavour-violating parameters defined
in Eq.~\eqref{eq:NMFVparams}, we first also discuss for completeness the flavour-conserving
parameters of our model description.

\subsection{Flavour-conserving parameters \label{sec:ResultsFC}}

\begin{figure}
	\begin{center}
	\includegraphics[width=0.3\textwidth]{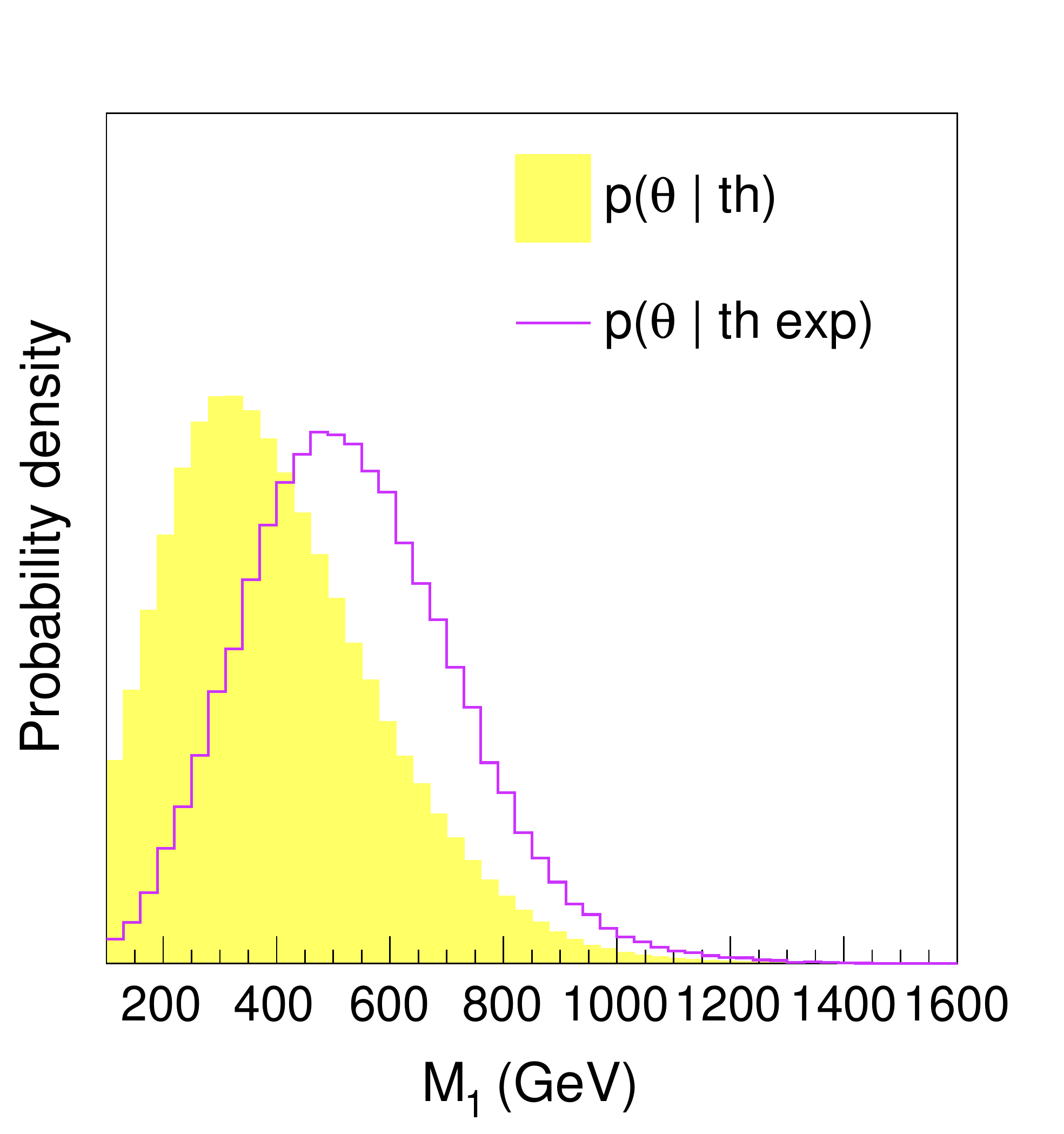} 
	\includegraphics[width=0.3\textwidth]{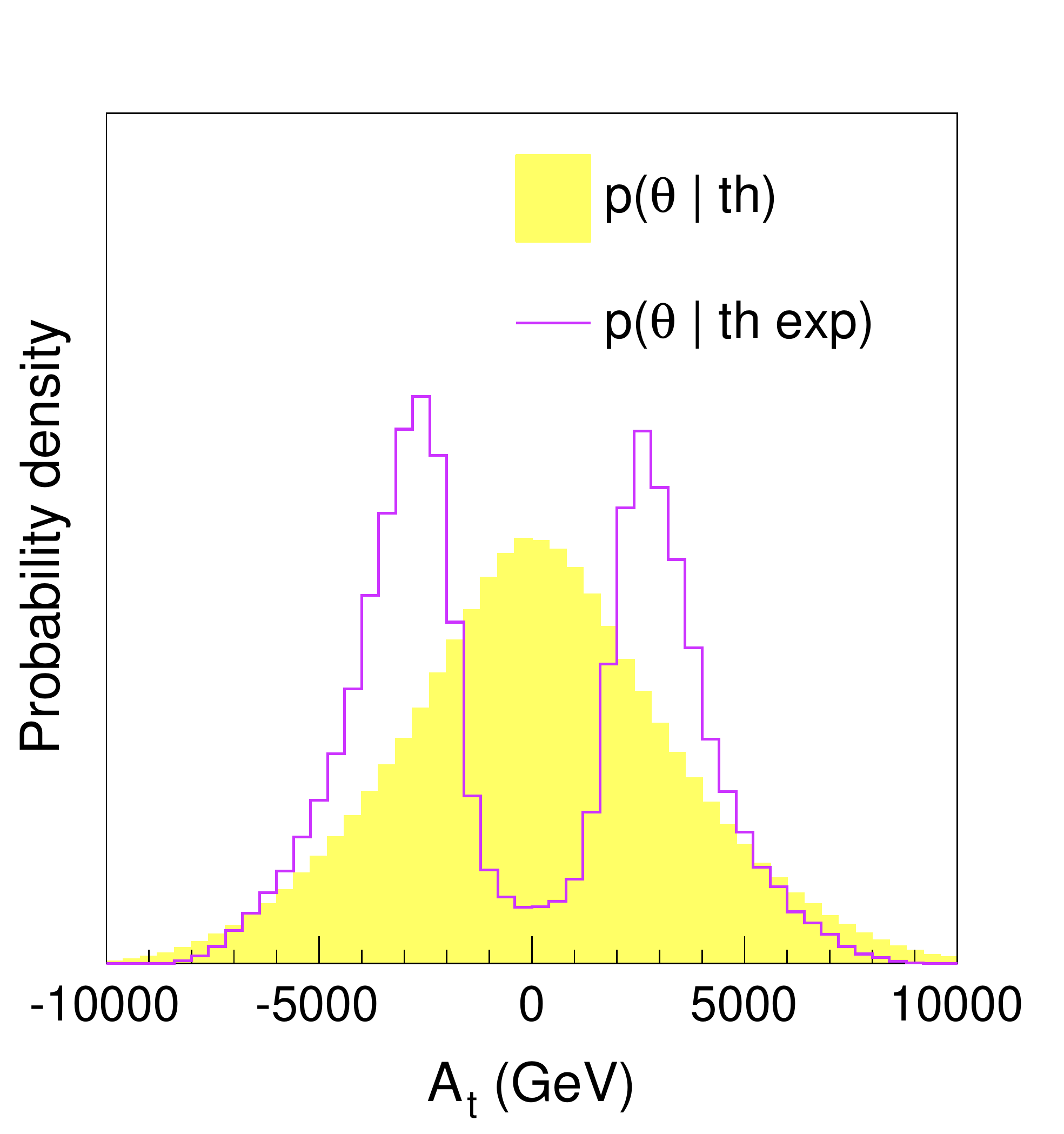} 
	\includegraphics[width=0.3\textwidth]{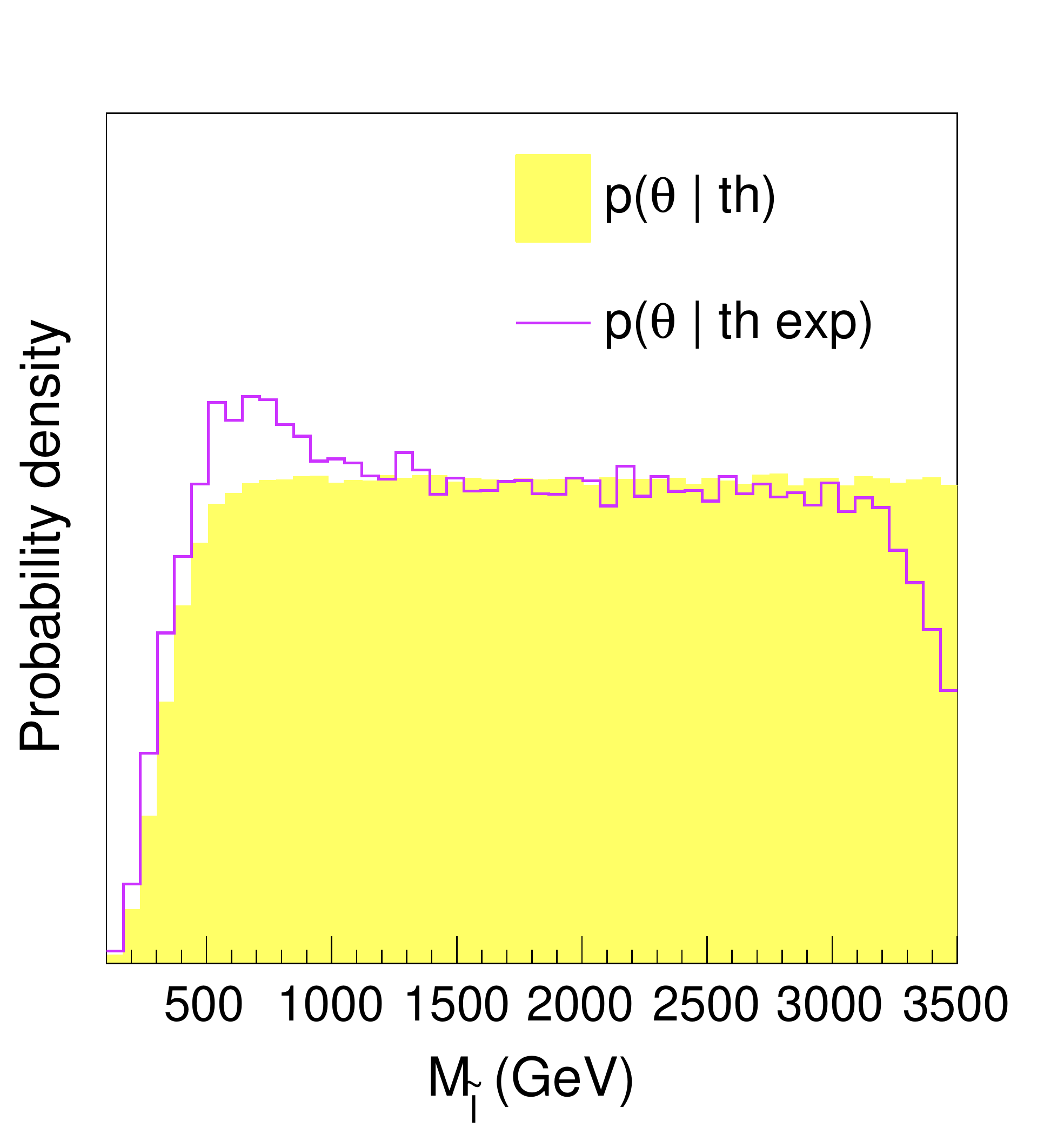} 
	\includegraphics[width=0.3\textwidth]{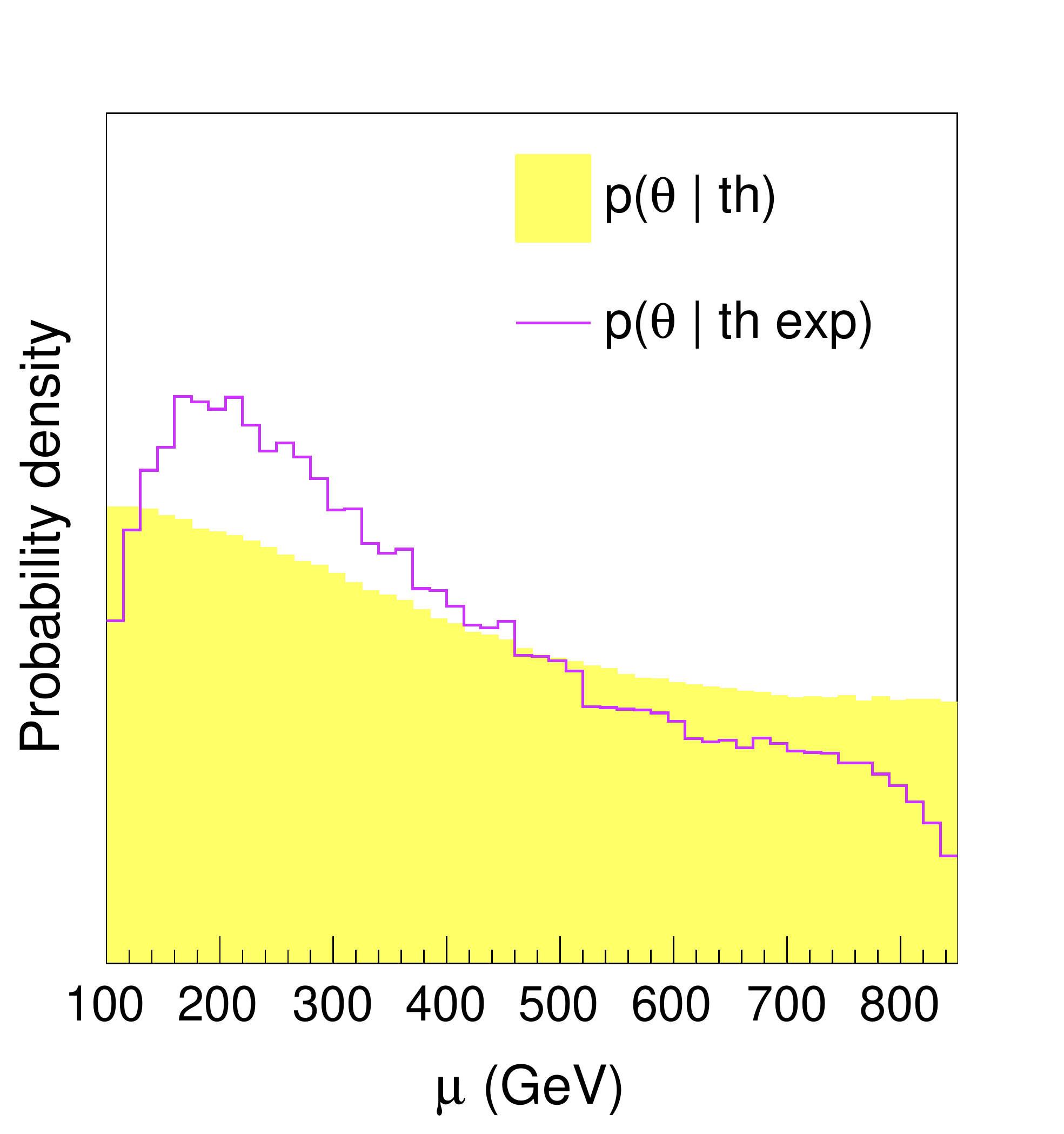} 
	\includegraphics[width=0.3\textwidth]{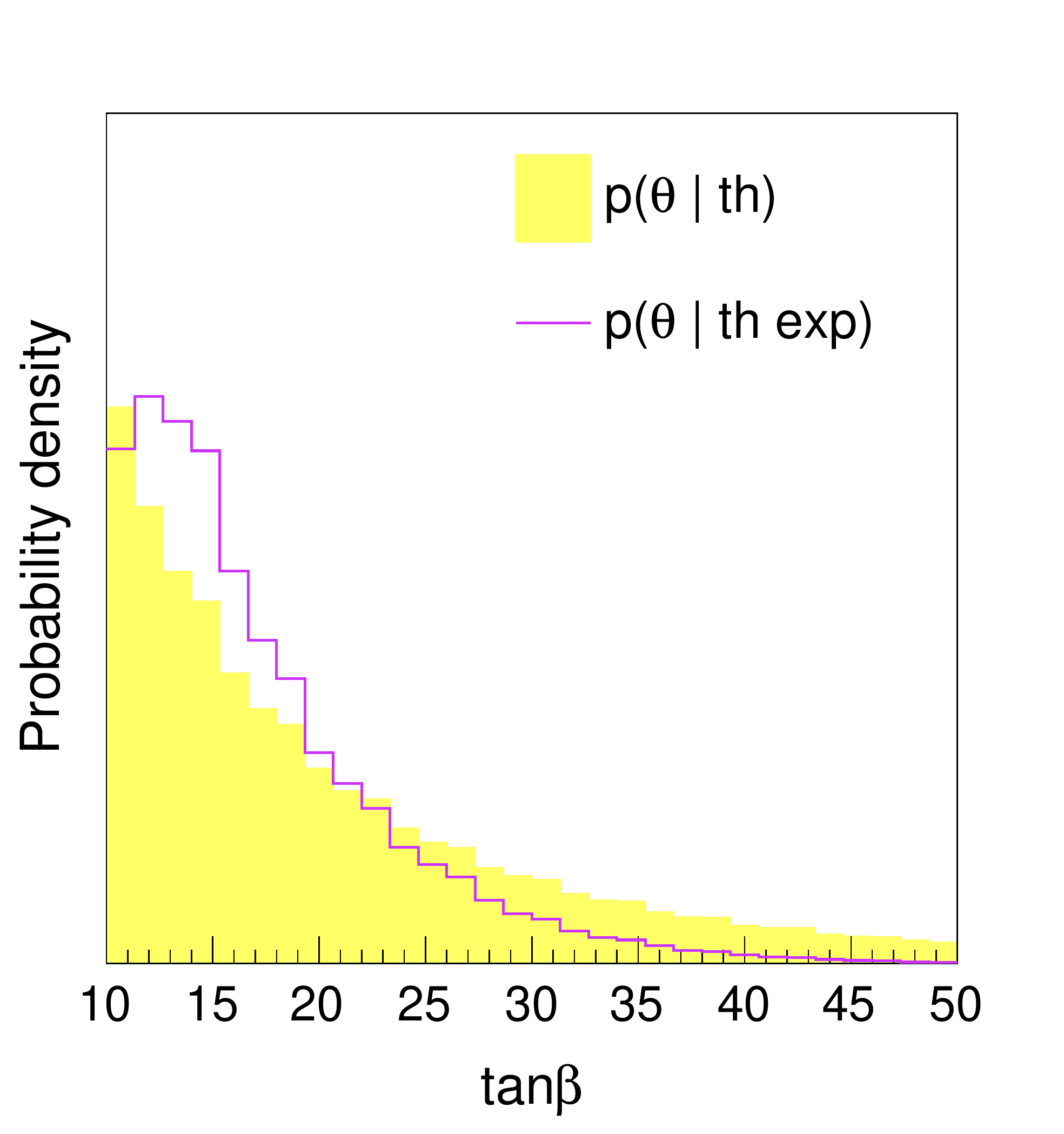} 
	\includegraphics[width=0.3\textwidth]{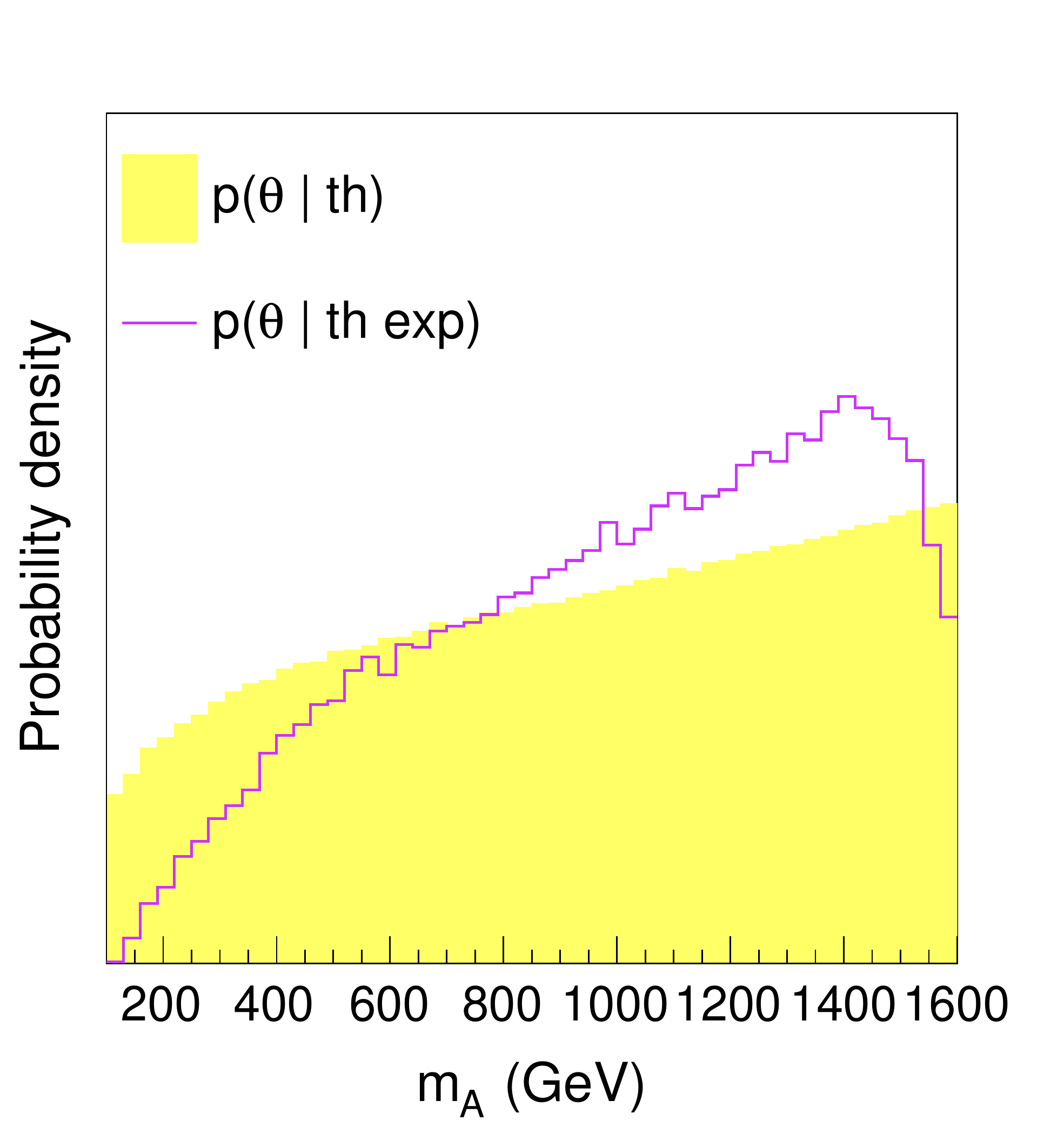} 
	\includegraphics[width=0.3\textwidth]{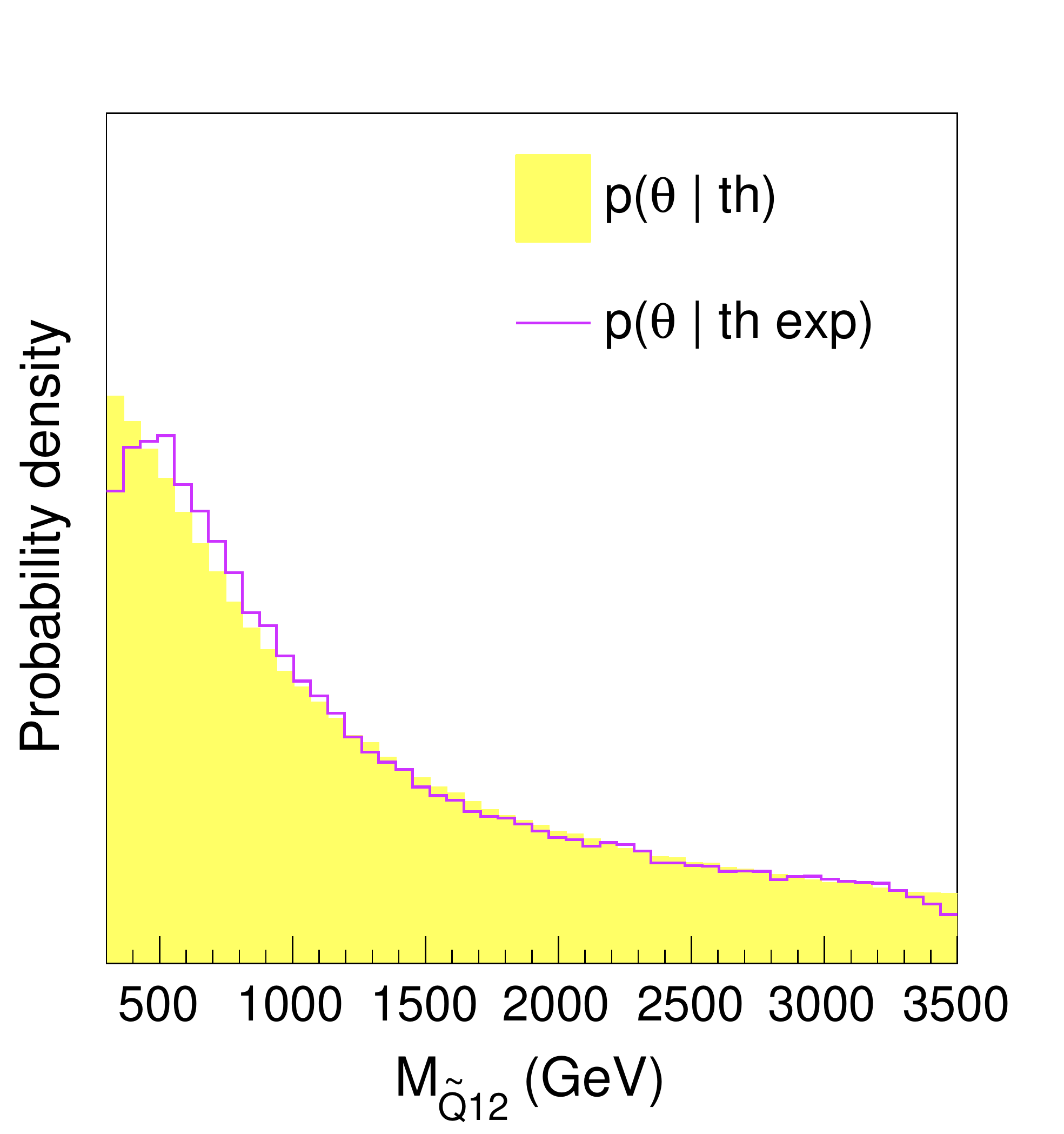} 
	\includegraphics[width=0.3\textwidth]{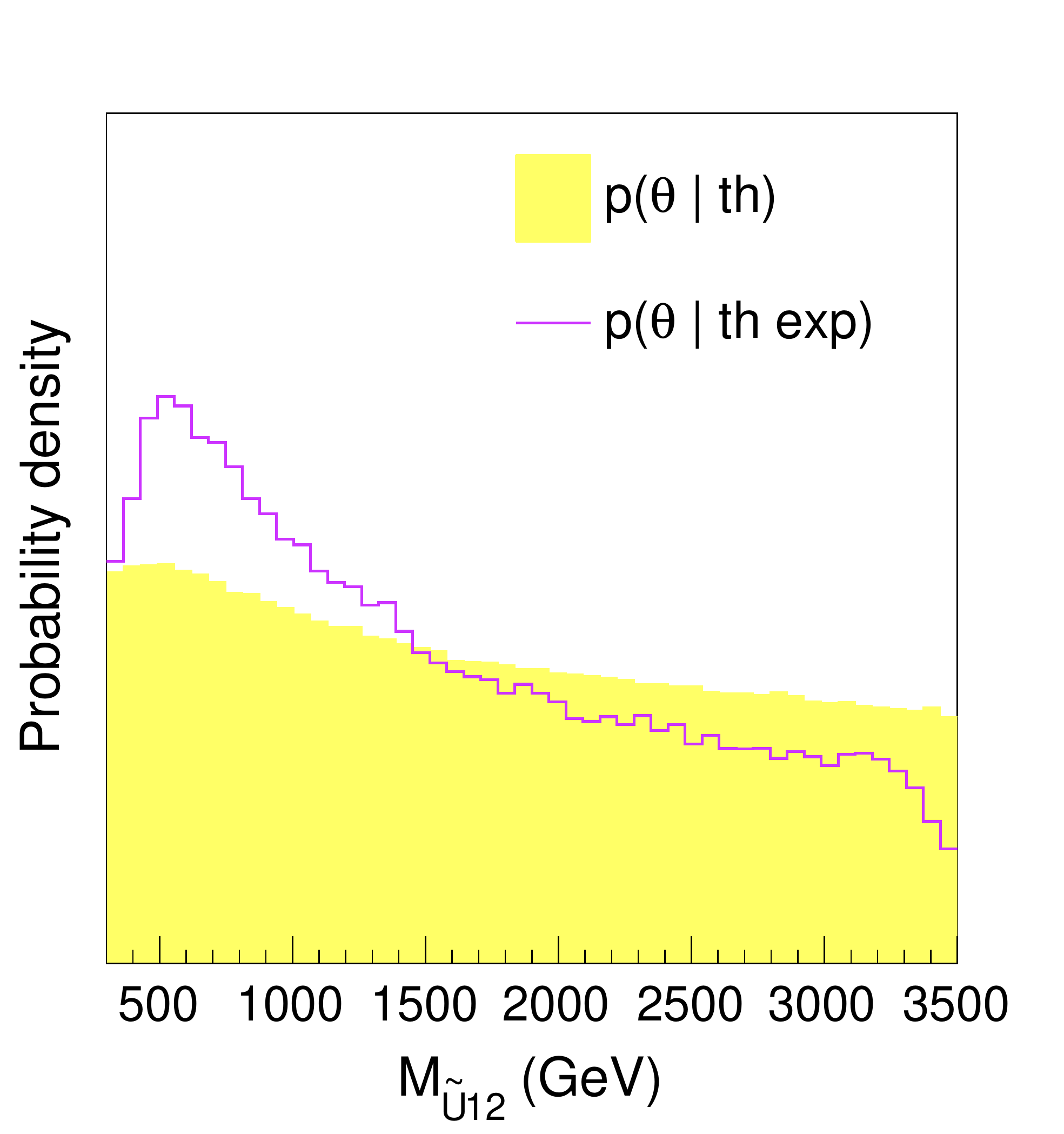} 
	\includegraphics[width=0.3\textwidth]{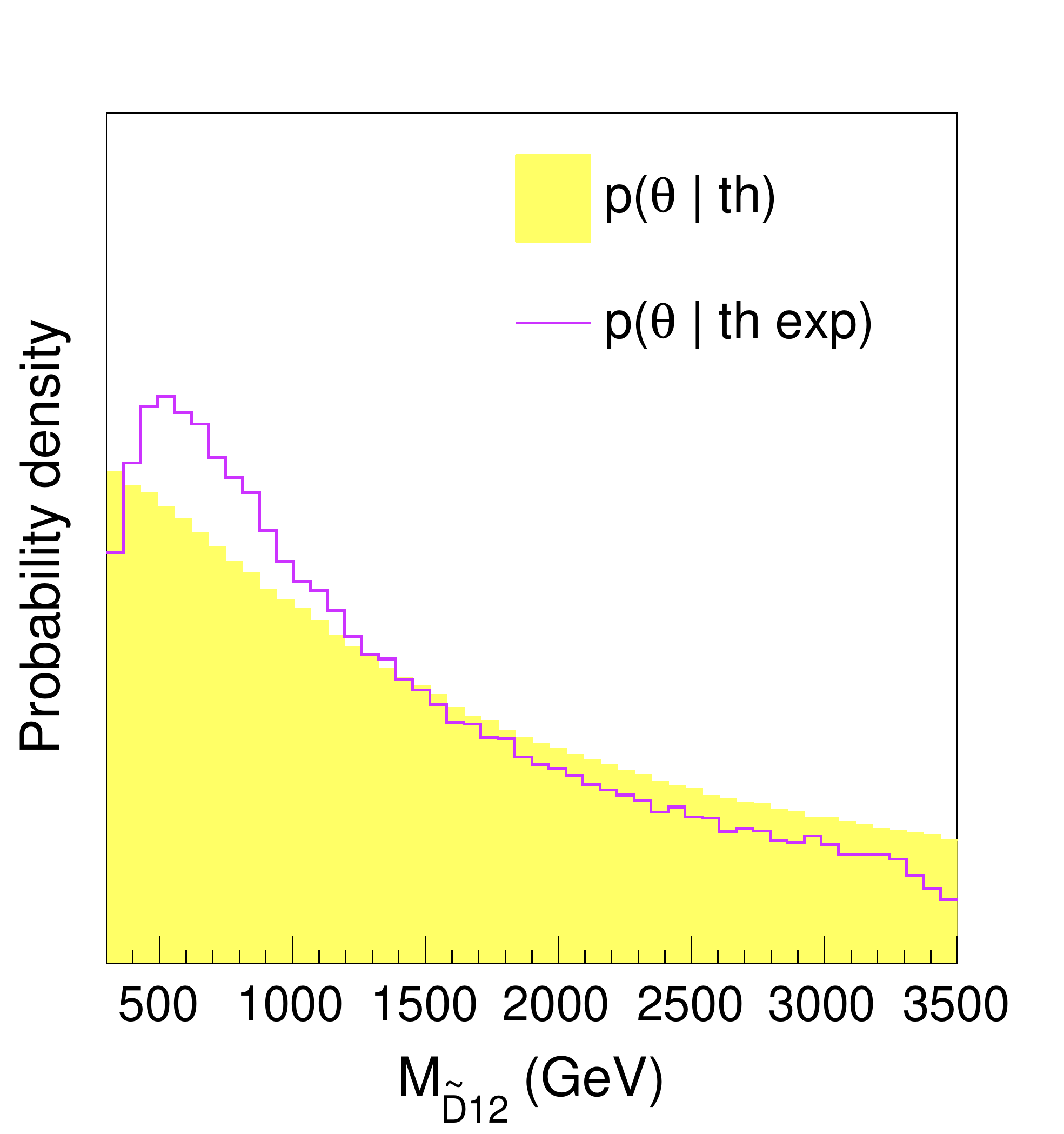} 
	\includegraphics[width=0.3\textwidth]{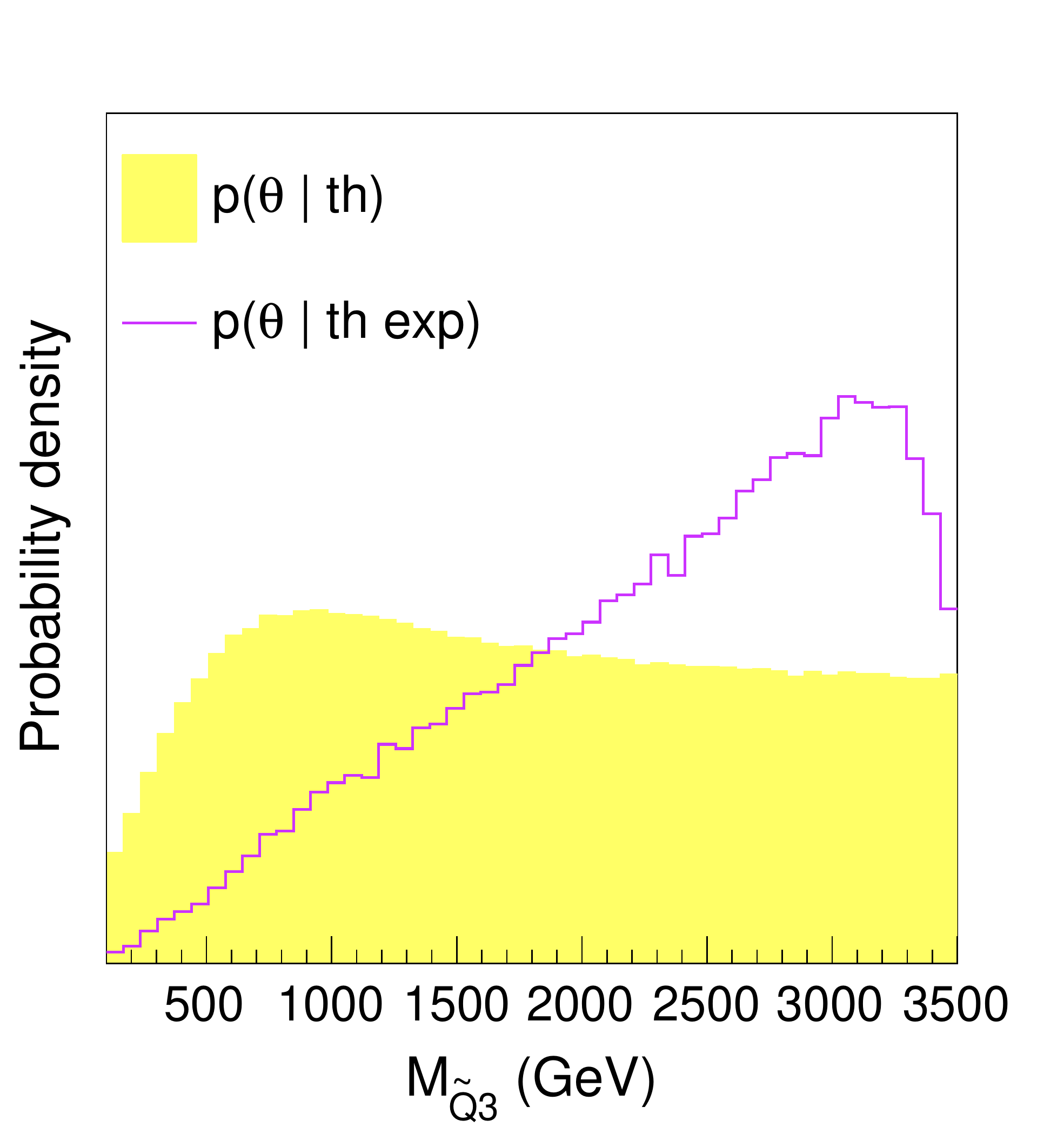} 
	\includegraphics[width=0.3\textwidth]{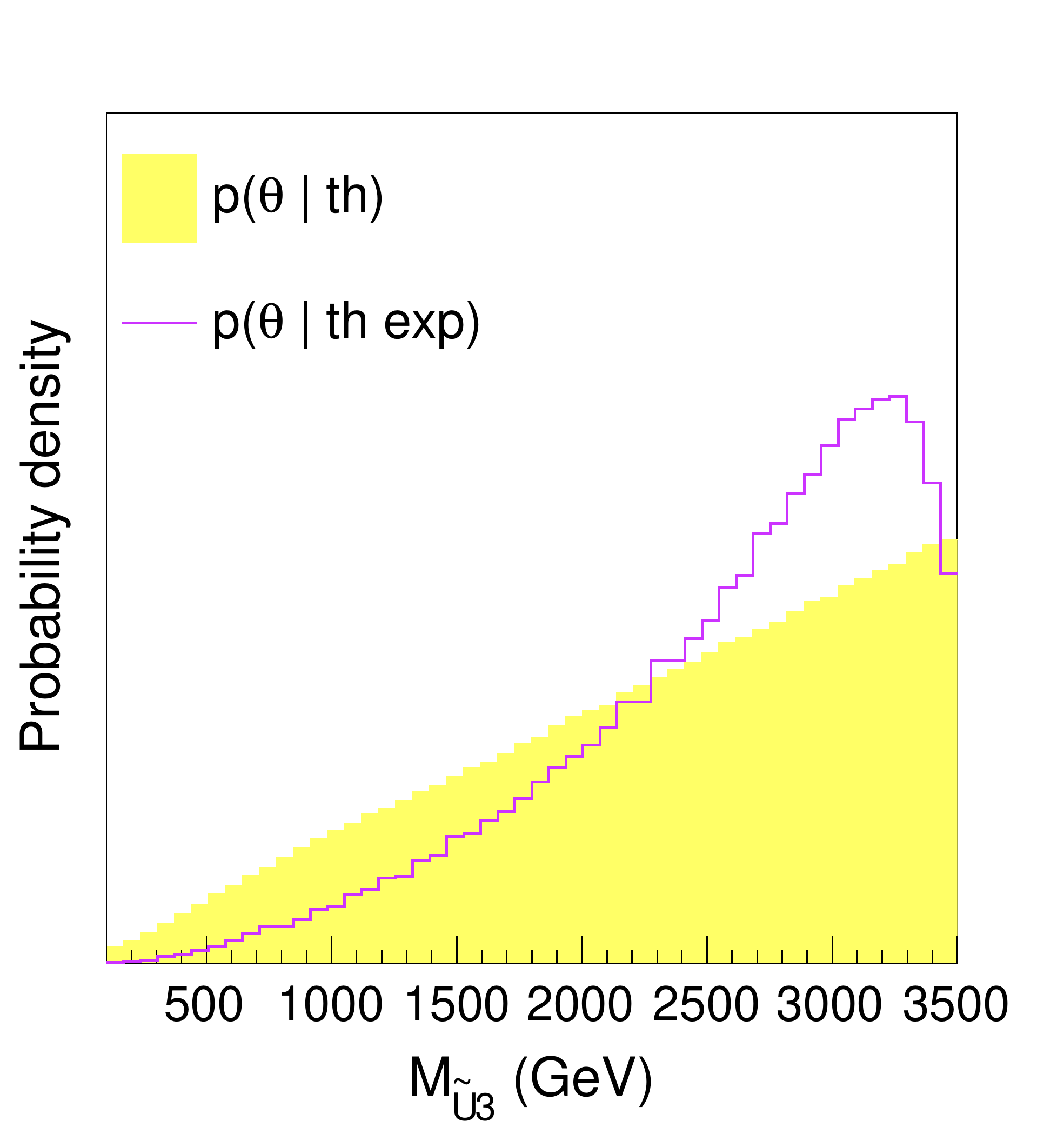} 
	\includegraphics[width=0.3\textwidth]{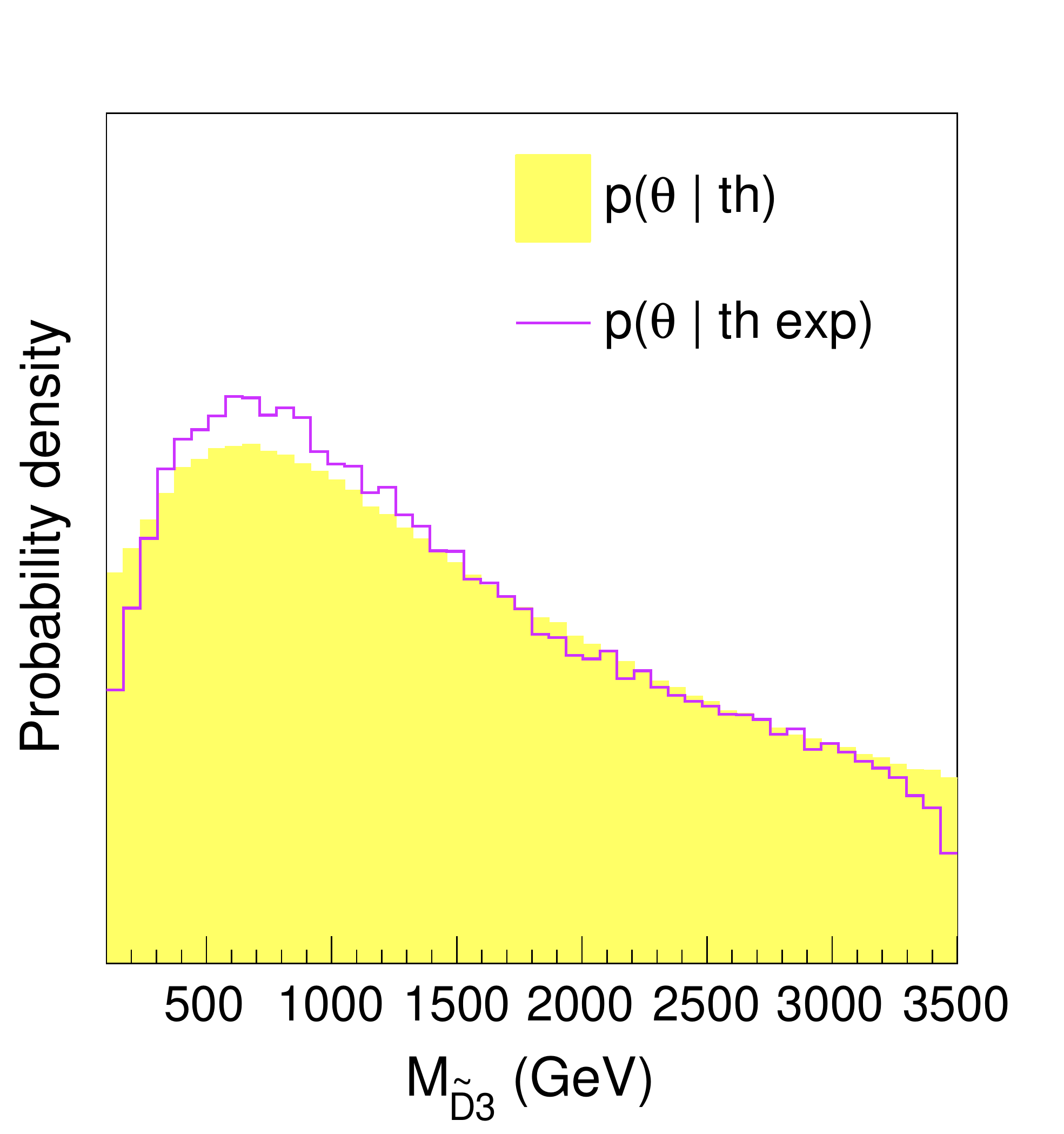} 
	\end{center}
	\caption{The one-dimensional prior (yellow histogram) and posterior
  (violet curve) distributions of the parameters of our NMFV MSSM description. The
  prior only incorporates theoretical inputs while the posterior distribution
  shows the impact of all experimental observations listed in Table~\ref{tab:PLMs}.}
	\label{fig:1Dpmssm}
\end{figure}

We start the discussion with the twelve flavour-conserving parameters of our
NMFV MSSM description. Figure \ref{fig:1Dpmssm} shows their probability density
distributions over the respective parameter ranges. In each panel, we display
the theoretical prior (yellow area) as well as the posterior distribution (solid
line), which shows the impact of all constraints given in Table~\ref{tab:PLMs}
together.

The prior distribution of the gaugino mass parameter $M_1$ is centred at
relatively low values of $M_1 \sim 400$~GeV and may reach values ranging up to
about 1000~GeV. When imposing all considered
experimental constraints, the distribution is shifted by about 100~GeV to higher
values. This feature can be traced to the
chargino contributions to the $B_s \to \mu\mu$ branching ratio and to the neutral
$B$-meson mass difference $\Delta M_{B_s}$, as we have fixed the ratios of the
gaugino mass parameters $M_1$ and $M_2$ so that chargino
effects are connected to $M_1$.

For the trilinear coupling parameter $A_f$, the prior distribution is centred
around zero. Large values of $A_f$ are indeed often rejected since they can induce
a large left-right squark mixing implying tachyonic states. Imposing
the experimental constraints drastically changes the shape of the distribution
and leads to two peaks corresponding to $|A_f| \sim 3000$~GeV. This feature is induced
by the Higgs boson mass requirement that necessitates, in order to be satisfied, a relatively large
splitting of the masses of the squarks exhibiting the largest stop component
$m_{\tilde q_1}$ and $m_{\tilde q_2}$.
More precisely, the flavour-conserving formula for the leading contributions to the Higgs mass,
\begin{equation}
	m^2_h ~=~ m^2_Z \cos^2 2\beta + \frac{3 g^2 m_t^4}{8\pi m^2_W} \left[ \log\frac{M^2_{\rm SUSY}}{m^2_t} 
		+ \frac{X_t^2}{M^2_{\rm SUSY}} \left( 1 - \frac{X_t^2}{12 M^2_{\rm SUSY}} \right) \right] \ ,
\end{equation}
where $X_t = A_t - \mu/\tan\beta$ and
$M^2_{\rm SUSY} = m_{\tilde{q}_1} m_{\tilde{q}_2}$, stays approximatively
valid in the NMFV regime, so that peaks defined by $|X_t| \sim \sqrt{6} M_{\rm SUSY}$ are
expected (see, \textit{e.g.}\ Ref.~\cite{Allanach:2004rh} and references
therein).

Moving on with the slepton mass parameter $M_{\tilde{\ell}}$, we observe a peak
centred at around $600$~GeV after imposing all experimental constraints. This
is mainly inferred by the anomalous magnetic moment of the muon requirement that
strongly depends on the slepton sector properties.
Turning to the Higgs sector (second line of Figure~\ref{fig:1Dpmssm}), the prior
distribution of the $\mu$-parameter shows a preference for low values while
its posterior distribution slightly peaks around
$\mu\sim 200$~GeV due to the $B_s \rightarrow \mu \mu$,
$\Delta a_\mu$ and $\Delta M_{B_s}$ constraints which all depend on the chargino
and neutralino sector. Next, the $\tan\beta$ parameter tends towards lower values
both in its prior and posterior distributions, the
favourite values being pushed to satisfy $12 \lesssim \tan\beta \lesssim 18$.
Finally, the posterior distribution of the mass of the pseudoscalar Higgs boson $m_A$
is shifted towards higher values with respect to its prior distribution. This results
from the interplay of most considered observables for which low values of $m_A$ would
induce too large Higgs contributions.

The last two lines of Figure~\ref{fig:1Dpmssm} concern the soft squark mass parameters.
Low values are preferred for the first and second generation
squark masses $M_{\tilde{Q}_{1,2}}$, $M_{\tilde{U}_{1,2}}$ and
$M_{\tilde{D}_{1,2}}$, a feature that is mostly caused by the Higgs boson.
This behaviour can be understood from the limiting case in which
$M^2_{\tilde{Q}_{1,2}} \simeq M^2_{\tilde{U}_{1,2}} \simeq M^2_{\tilde{D}_{1,2}}
\equiv \tilde m^2$. The one-loop corrections to $m_h$ that are proportional to
$\delta^u_{LR}$ are here approximately given by~\cite{Kowalska:2014opa}
\begin{equation}
\Delta m^2_h = \frac{3 v_u^4}{8 \pi^2(v^2_d+v^2_u)} 
\left[ \frac{(T_u)_{23}^2}{\tilde m^2} \left( \frac{Y^2_t}{2} - \frac{(T_u)_{23}^2}{12 \tilde m^2} \right)\right] \ ,
\end{equation}
while the corresponding contributions of down-type squarks are obtained by replacing
$T_u$ by $T_d$, $Y_t$ by $Y_b$ and by exchanging $v_u$ and $v_d$.
In our parameterisation,
\begin{equation}
	\left( T_u \right)_{23} ~=~ \frac{\sqrt{2}}{v_u}\delta^u_{LR} M_{\tilde{Q}_{1,2}} M_{\tilde{U}_3}
	\label{eq:TUdLR}
\end{equation}
so that for non-zero $\delta^u_{LR}$, the Higgs boson becomes tachyonic
if $\tilde m^2$ is too large.
Similarly, the requirement of a physical solution for the
electroweak vacuum also favours lower values for $M_{\tilde{Q}_{1,2}}$.
The distributions of the third-generation
mass parameters $M_{\tilde{Q}_3}$ and $M_{\tilde{U}_3}$ prefer in contrast
larger values due to $\Delta M_{B_s}$
and the mass of the Higgs boson constraints.
Finally, both the prior and posterior distributions of the right-handed
down-type squark mass $M_{\tilde{D}_3}$ prefer lower values and are
in this case very similar.

\subsection{Flavour-violating parameters}

\begin{figure}
	\begin{center}
	\includegraphics[width=0.32\textwidth]{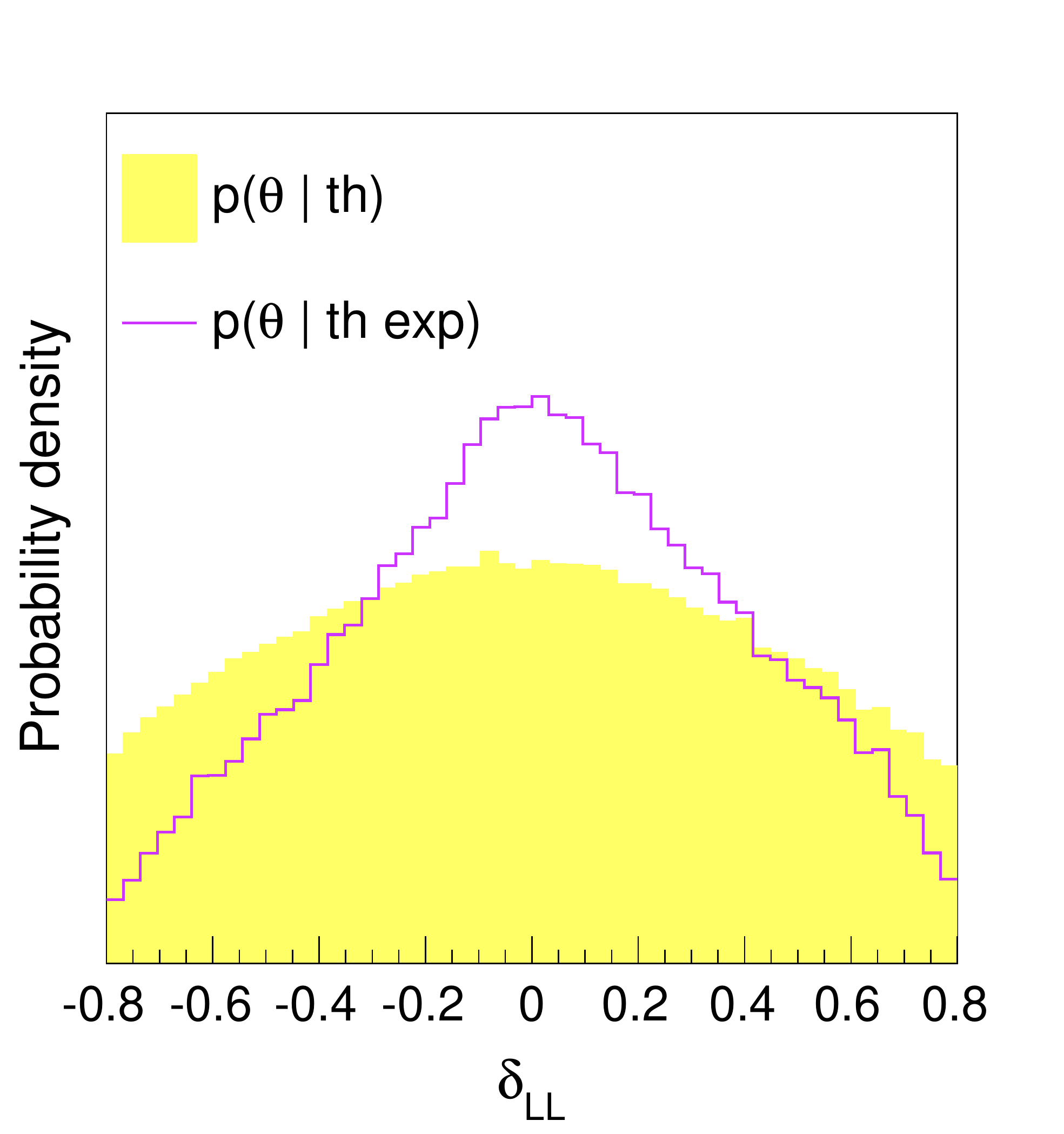} \\
	\includegraphics[width=0.32\textwidth]{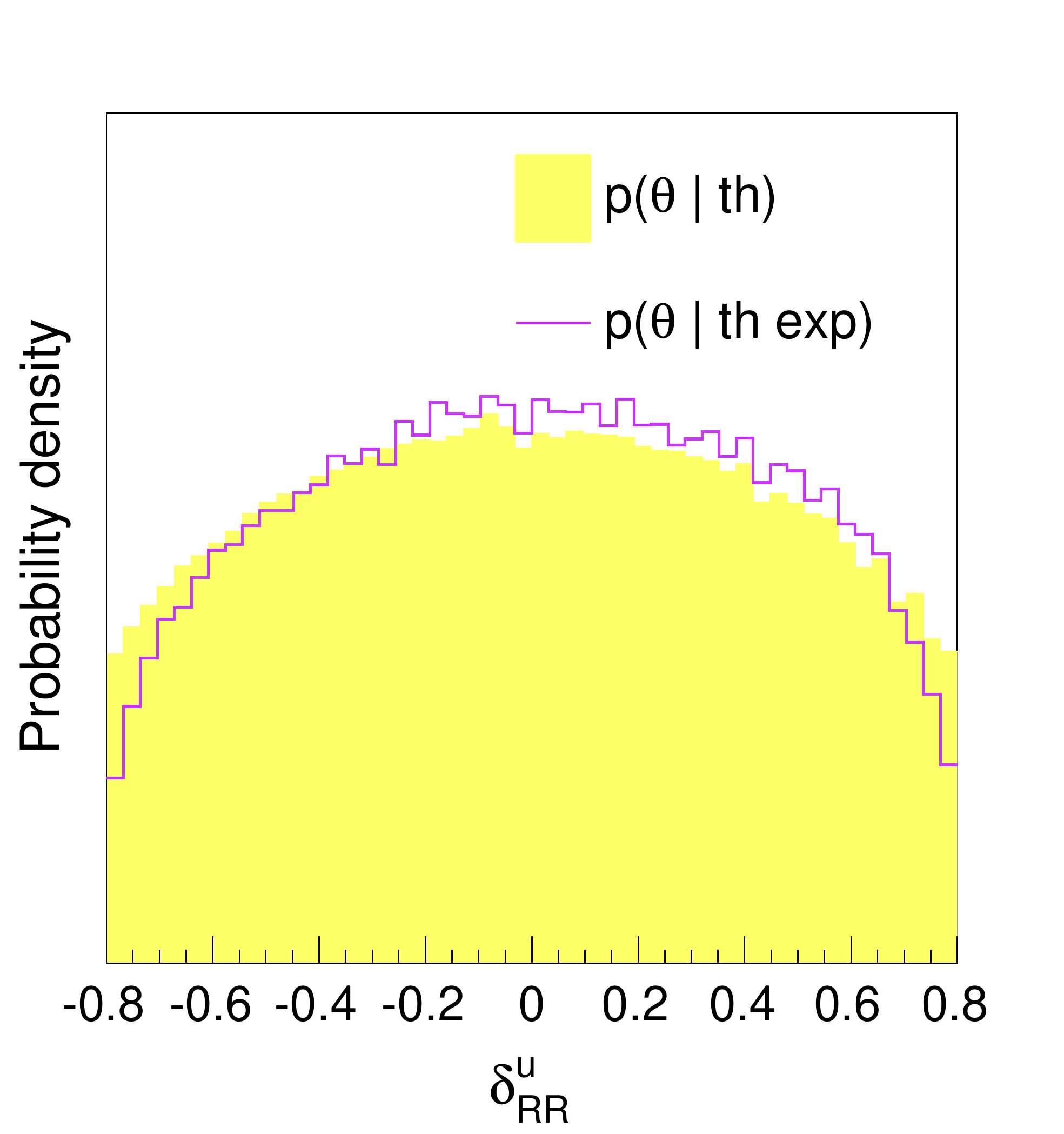} 
	\includegraphics[width=0.32\textwidth]{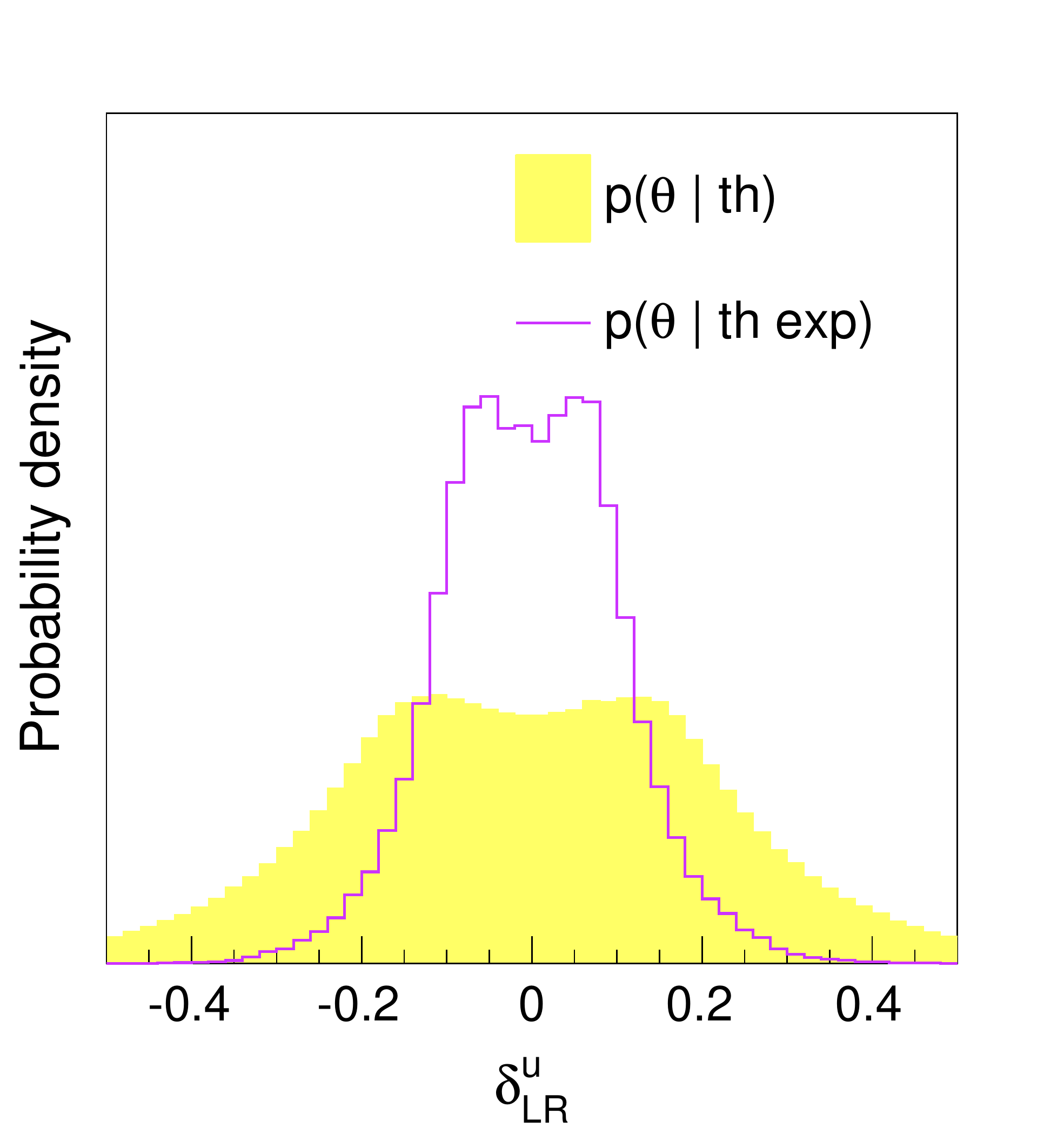} 
	\includegraphics[width=0.32\textwidth]{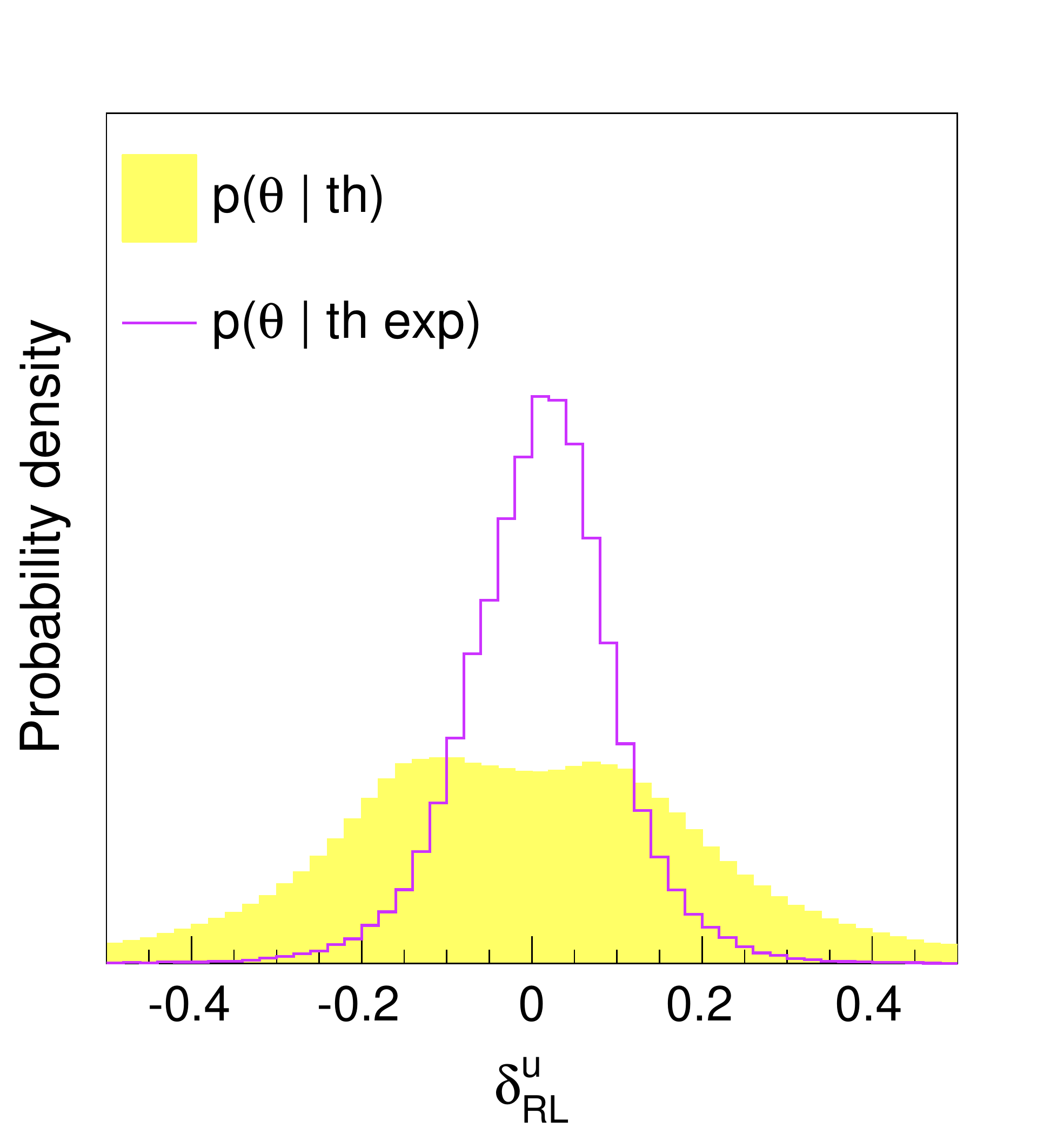} 
	\includegraphics[width=0.32\textwidth]{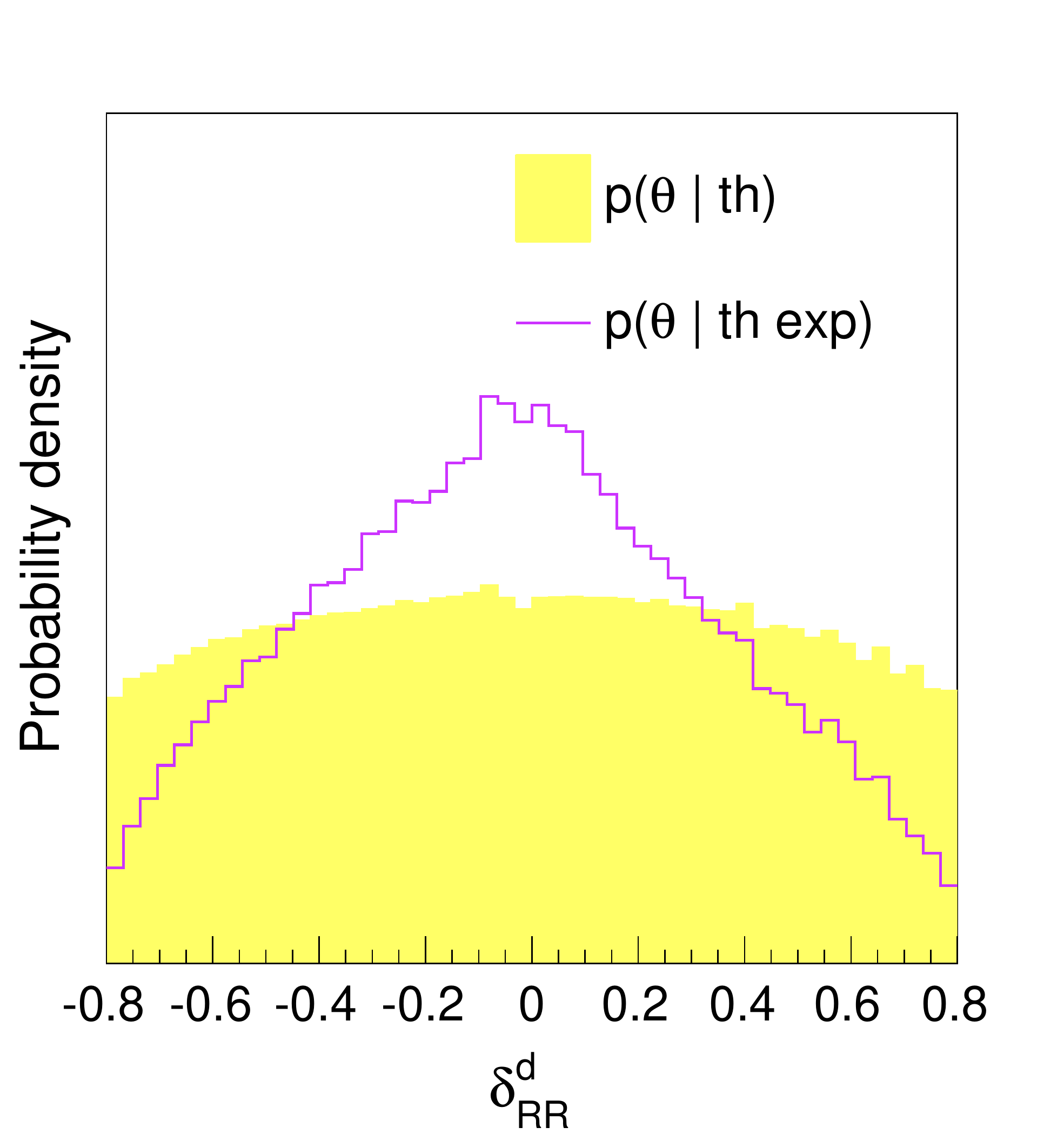} 
	\includegraphics[width=0.32\textwidth]{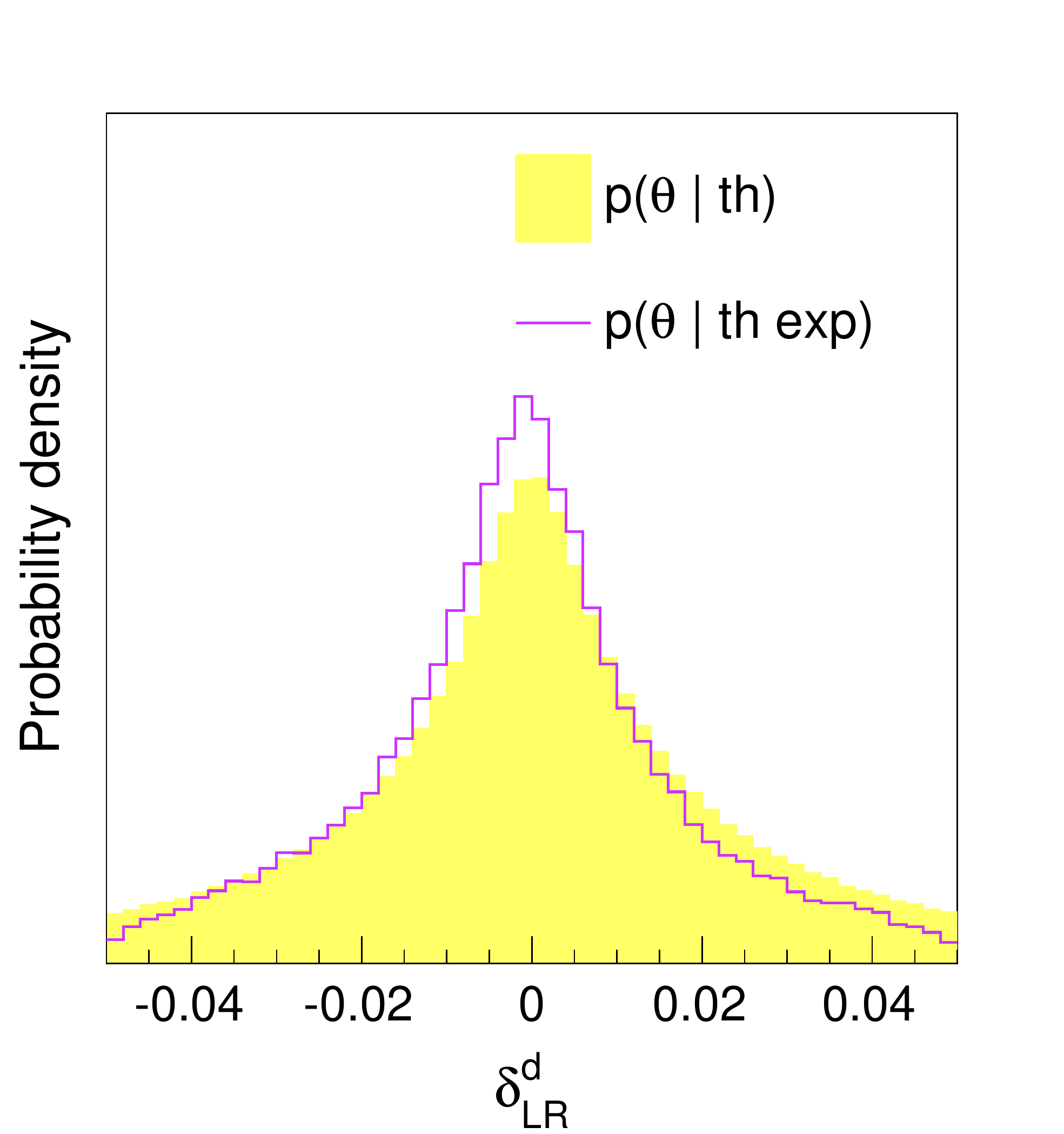} 
	\includegraphics[width=0.32\textwidth]{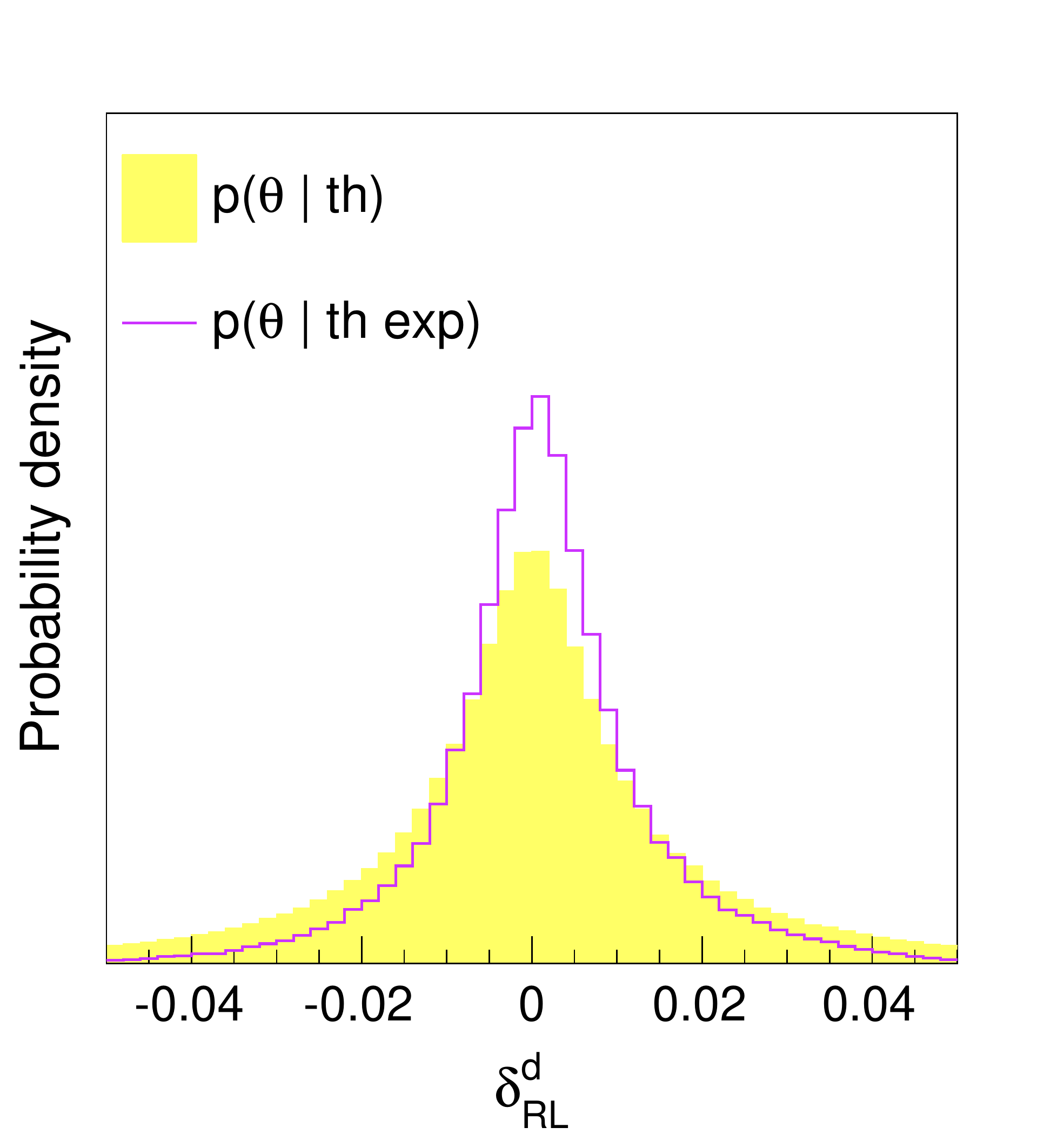} 
	\end{center}
	\caption{Same as Figure~\ref{fig:1Dpmssm} in the case of the flavour-violating
  input parameters of our NMFV MSSM description.}
	\label{fig:1Dnmfv}
\end{figure}

\begin{figure}
	\begin{center}
	\includegraphics[width=0.32\textwidth]{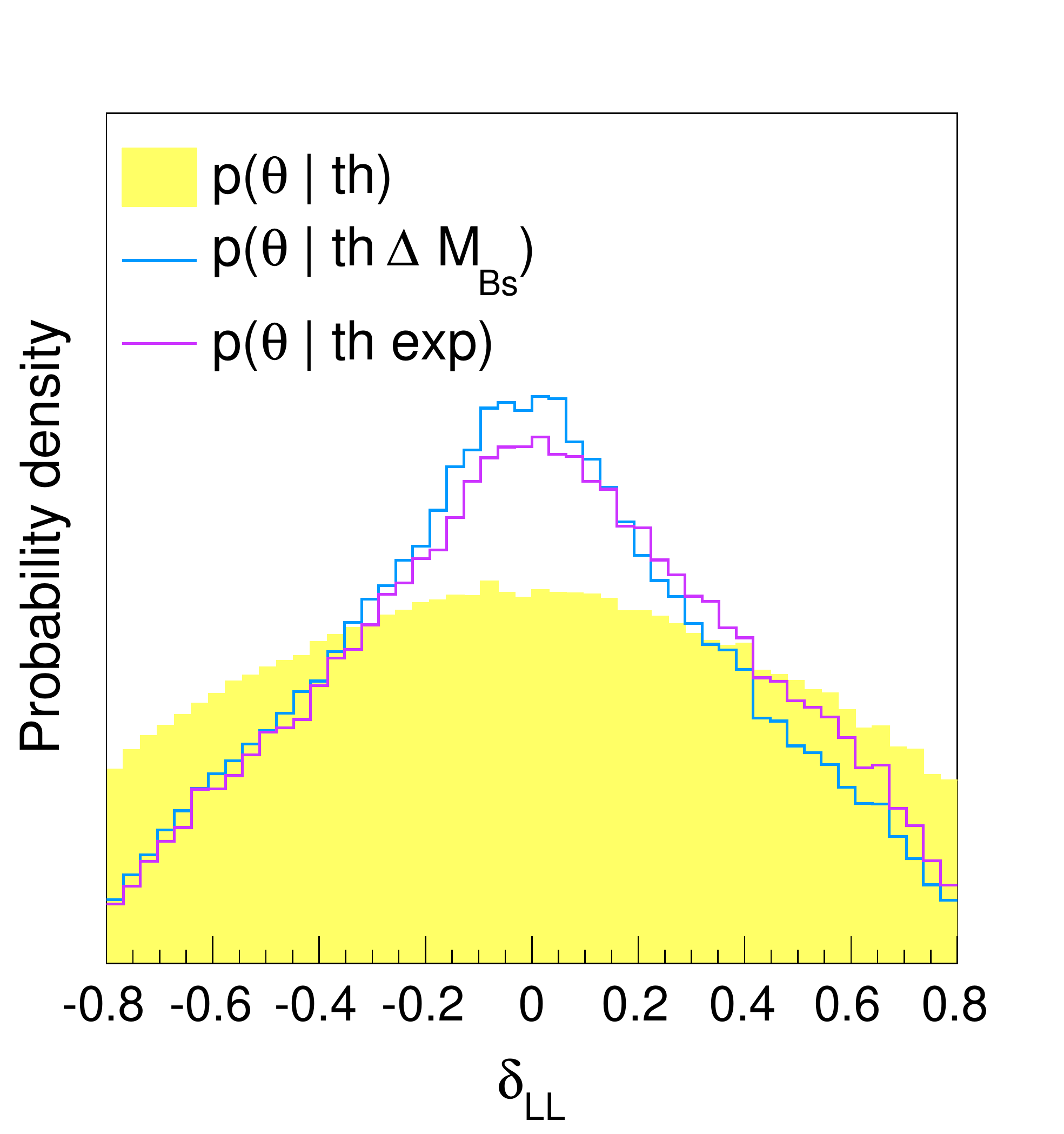} 
	\includegraphics[width=0.32\textwidth]{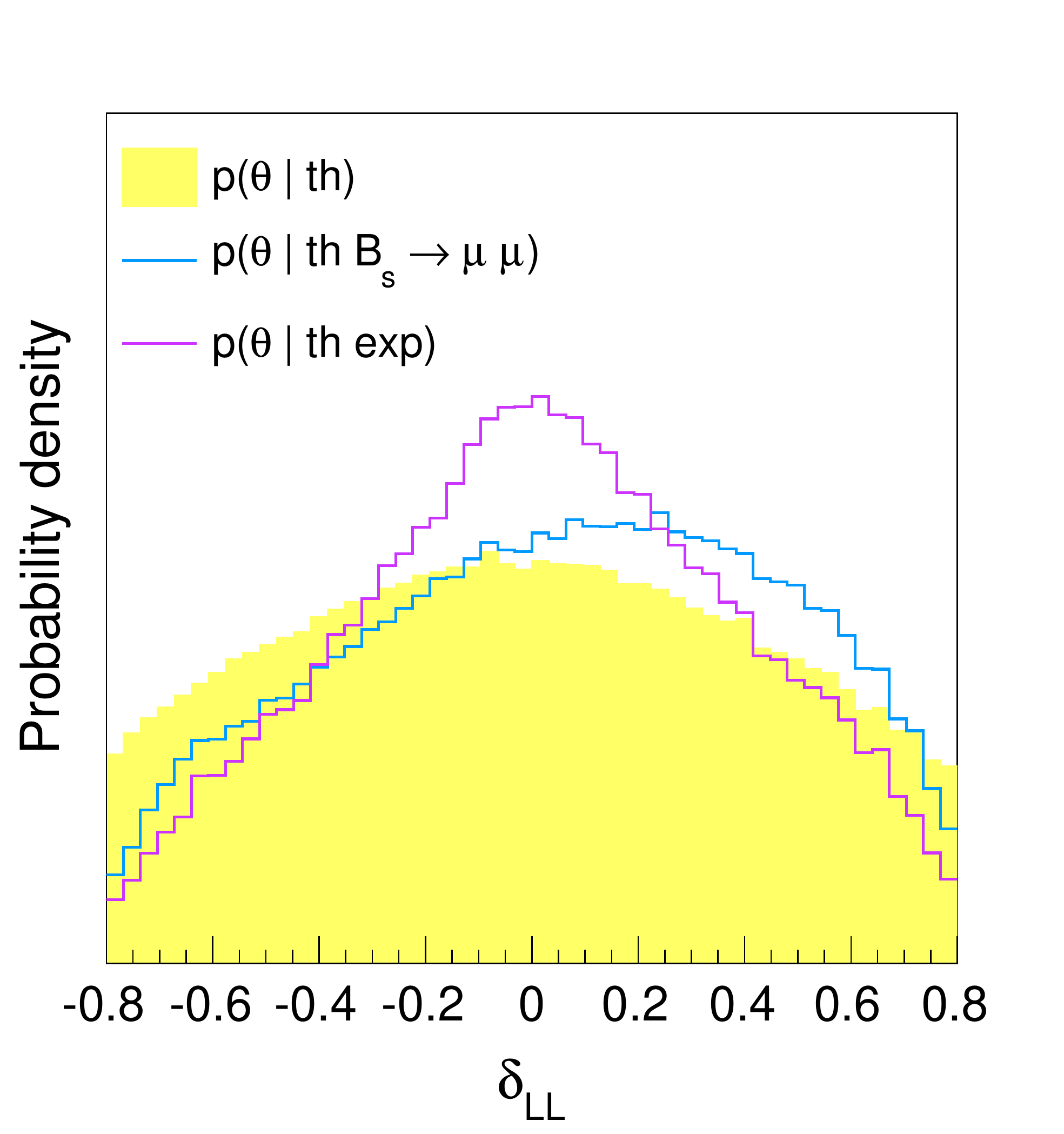}
	\\
	\includegraphics[width=0.32\textwidth]{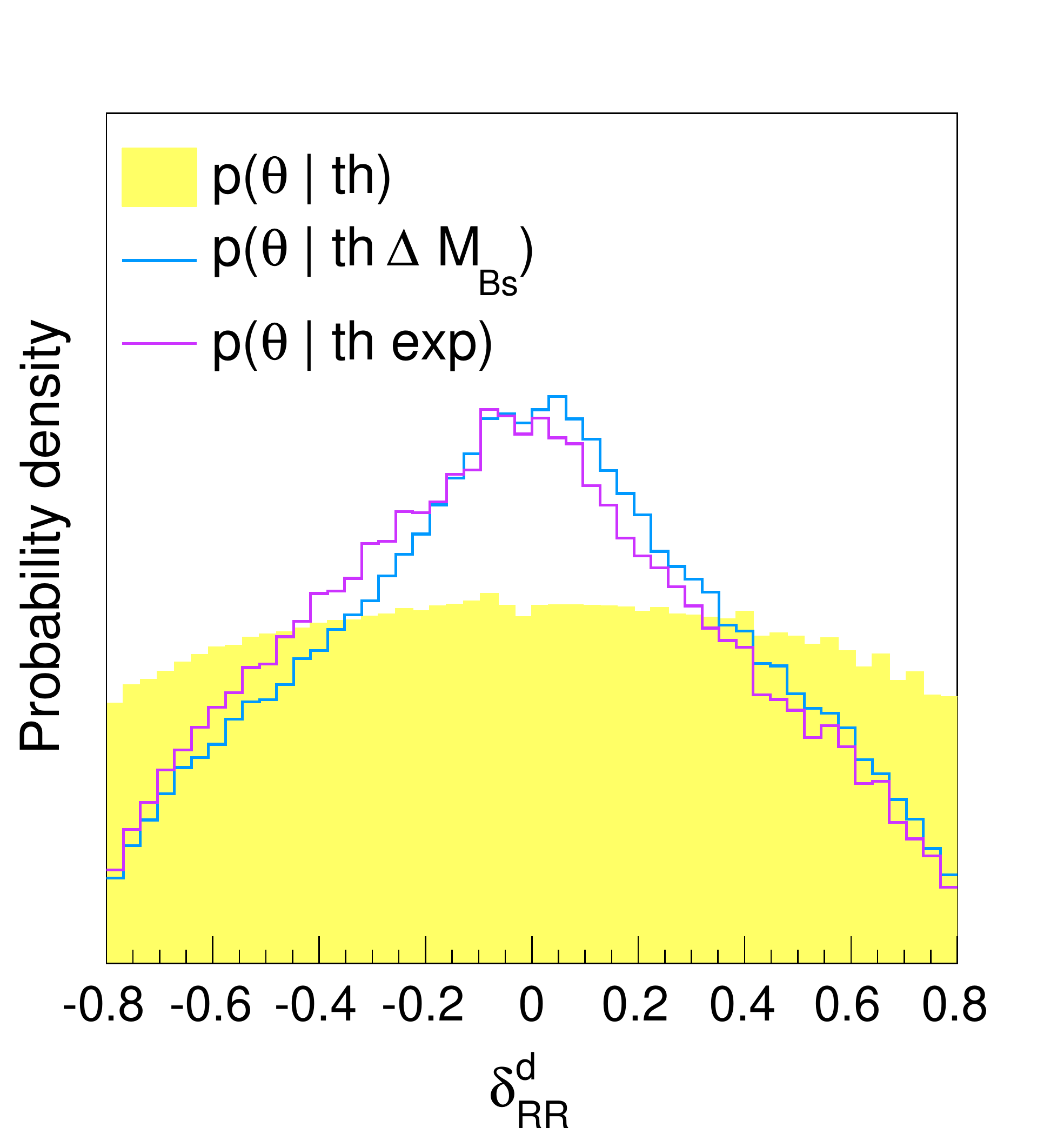}
	\includegraphics[width=0.32\textwidth]{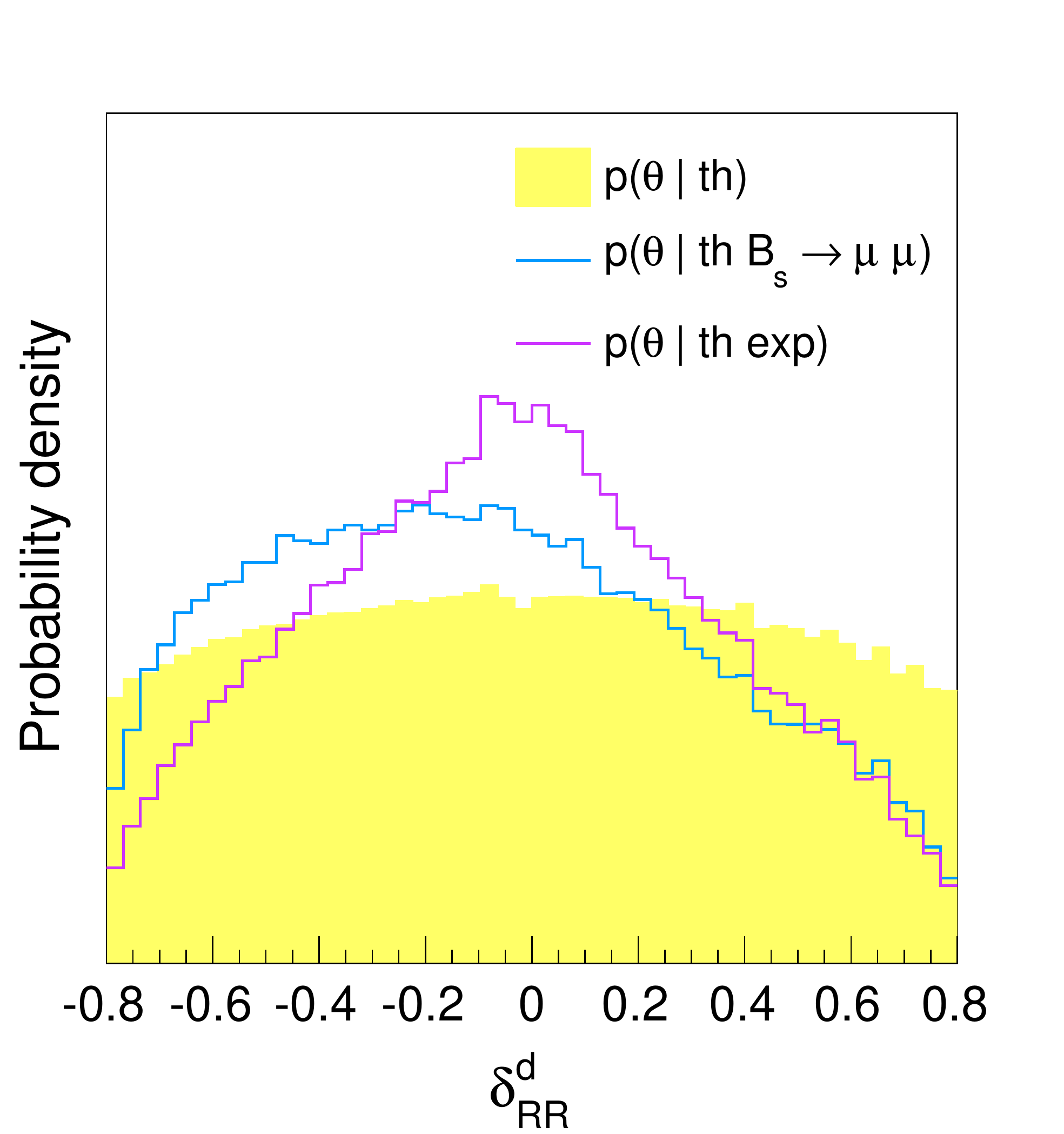}
	\end{center}
	\caption{Most relevant observables constraining the
    $\delta_{LL}$ (upper panel) and $\delta_{RR}^d$ (lower panel) parameters.
	}
	\label{fig:1DLLRR}
\end{figure}

\begin{figure}
	\begin{center}
	\includegraphics[width=0.32\textwidth]{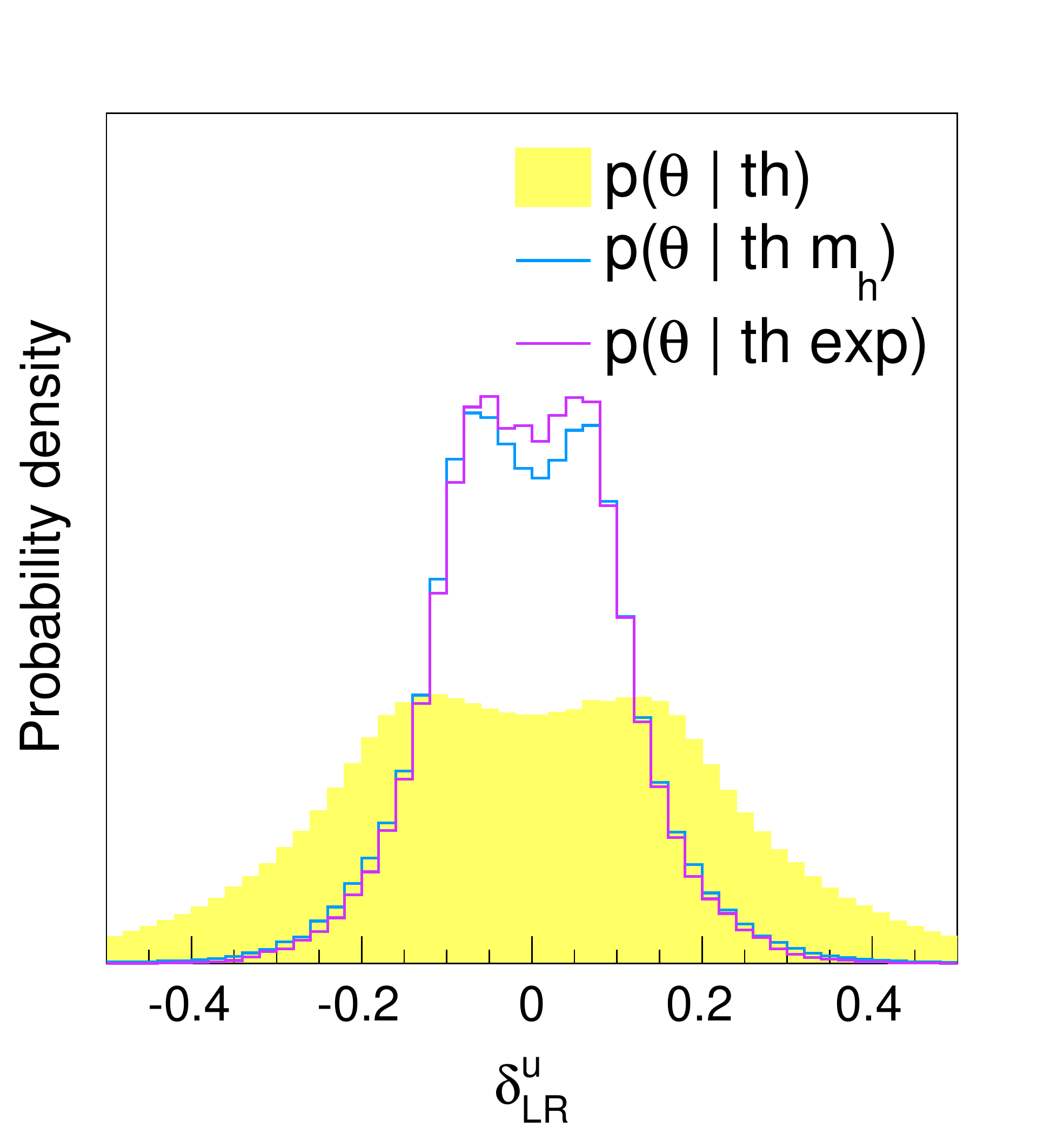}
	\includegraphics[width=0.32\textwidth]{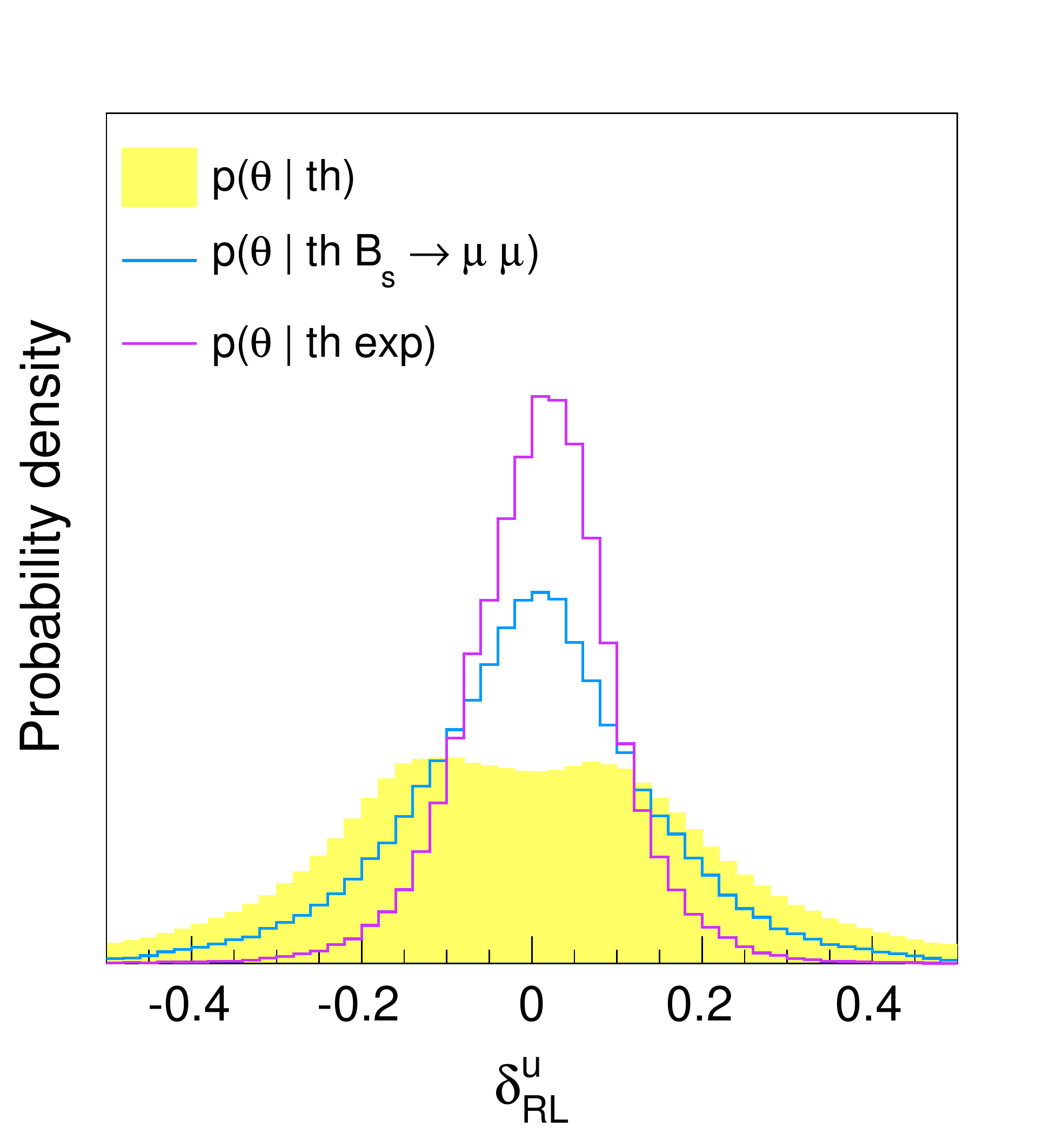}
	\includegraphics[width=0.32\textwidth]{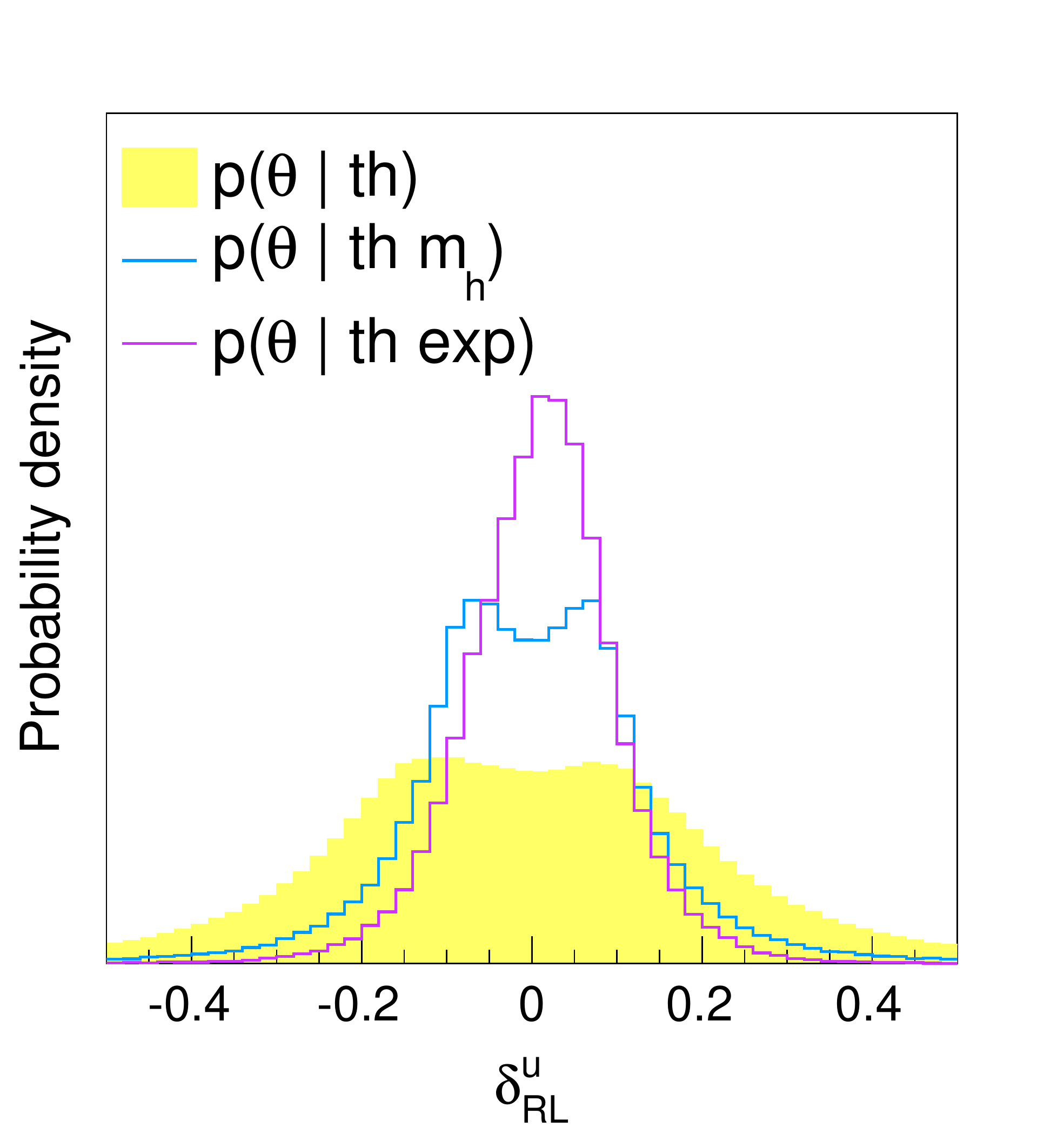}
	\end{center}
	\caption{Most relevant observables constraining the
   $\delta^u_{LR}$ (left panel) and $\delta^u_{RL}$ (centre and right panel) parameters.
   }
	\label{fig:1DuLRRL}
\end{figure}

\begin{figure}
	\begin{center}
	\includegraphics[width=0.32\textwidth]{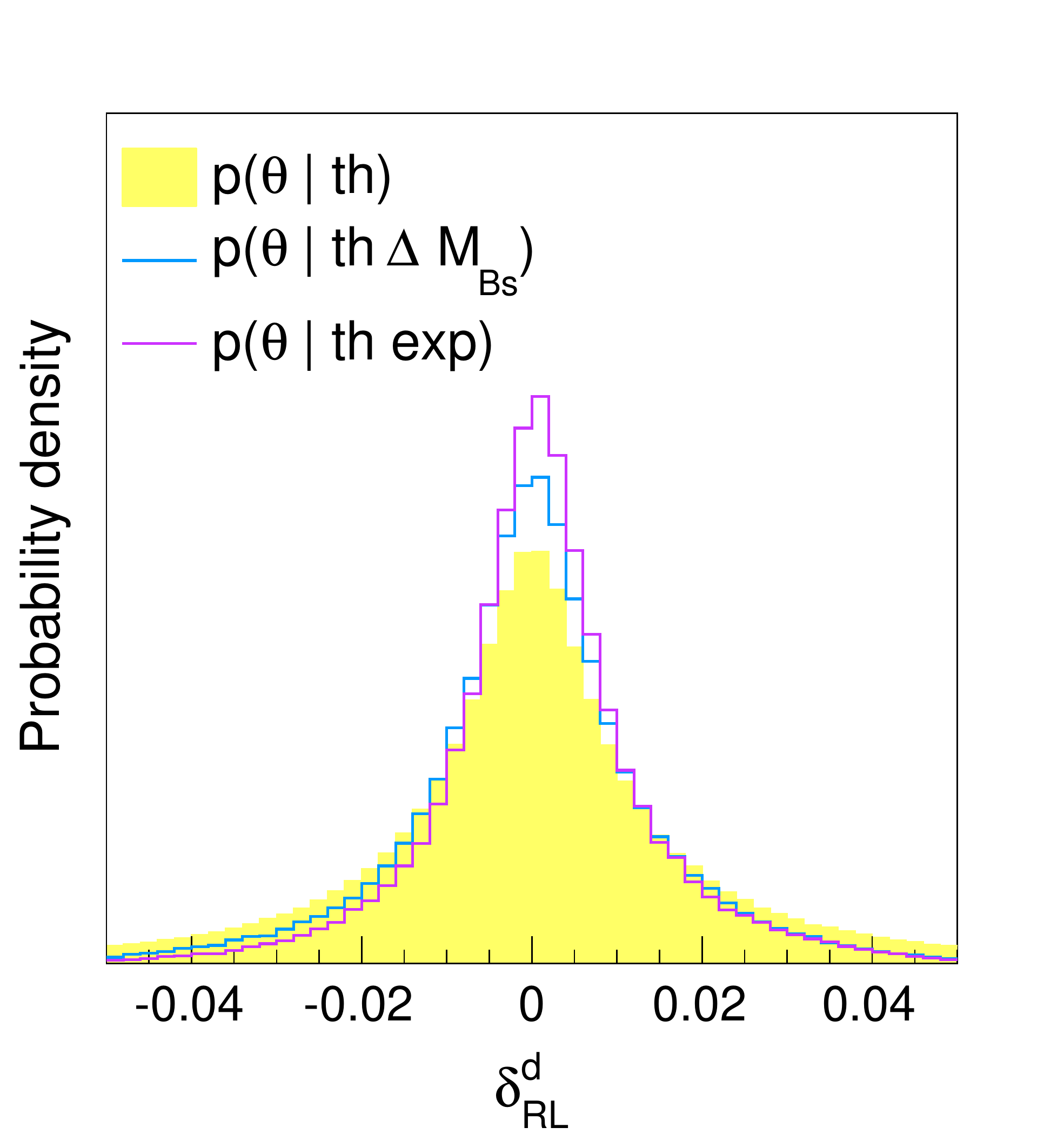}
	\includegraphics[width=0.32\textwidth]{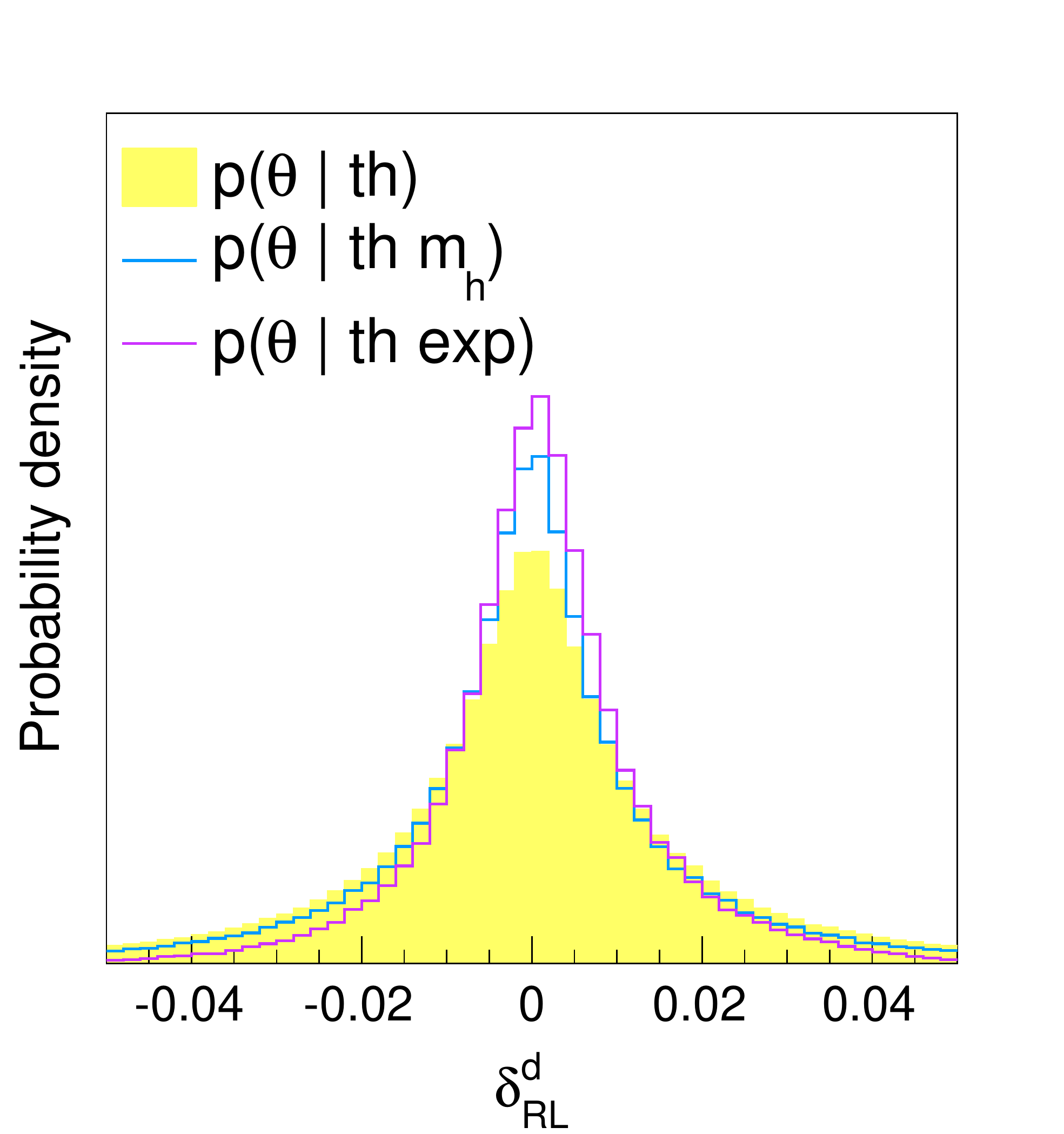}
	\end{center}
	\caption{Most relevant observables constraining the $\delta^d_{RL}$ parameter.}
	\label{fig:1DdLRRL}
\end{figure}

We now turn to the analysis of the constraints that are imposed on the seven
non-minimally flavour-violating parameters $\delta^{q}_{\alpha\beta}$ that are
at the centre of interest of the present analysis. The corresponding prior and
posterior distributions are displayed in Figure~\ref{fig:1Dnmfv}, and we detail
the impact of the most important observables on Figure~\ref{fig:1DLLRR},
Figure~\ref{fig:1DuLRRL} and Figure~\ref{fig:1DdLRRL}.

The theoretical constraints on any additional stop-scharm mixing in the left-left
sector ($\delta_{LL}$) are relatively mild such that an almost flat behaviour is
observed (see Figure~\ref{fig:1Dnmfv}). The $\delta_{LL}$
parameter is then mainly constrained by the $B$-meson oscillation parameter
$\Delta M_{B_s}$ (which favours smaller absolute values of $\delta_{LL}$)
and the branching ratio for the $B_s \rightarrow \mu \mu$ decay (which causes a slight
preference to positive values), as shown in Figure~\ref{fig:1DLLRR}.
Values ranging up to $|\delta_{LL}| = 0.8$ can nevertheless be
reached, but this simultaneously requires large values for other $\delta$ quantities so that
cancellations between the different contributions to the considered observables
occur (see Section~\ref{sec:correlations}). In a similar way, the prior distributions of the parameters $\delta^u_{RR}$
and $\delta^d_{RR}$ show a mild preference for low absolute values.
The posterior distribution of the $\delta^u_{RR}$ parameter does not differ much from
its prior distribution so that $\delta^u_{RR}$ is not sensitive to the
experimental constraints under study. In contrast,
the $B$-meson oscillation parameter $\Delta M_{B_s}$ restricts the posterior distribution of
$\delta^d_{RR}$ to be narrower while the $B_s \rightarrow \mu \mu$ branching ratio
implies a preference to negative values (see Figure~\ref{fig:1DLLRR}).
However, the full explored range of
$-0.8 \lesssim \delta^{u,d}_{RR} \lesssim 0.8$ stays accessible in the context
of both right-right mixing parameters.

The flavour-violating left-right and right-left elements of the up-type squark
mass matrix ($\delta^u_{LR}$ and $\delta^u_{RL}$) turn out to be
mainly constrained by the necessity to incorporate a Higgs boson with a
mass of about $125$~GeV, as can be seen in Figure~\ref{fig:1DuLRRL}. 
The posterior distribution of $\delta^u_{LR}$ exhibits two peaks at 
$|\delta^u_{LR}|\sim 0.5$ and is restricted to
$-0.15 \lesssim \delta^u_{LR} \lesssim 0.15$.
Theoretically, this behaviour is expected from Eq.~\eqref{eq:TUdLR}.
The $\delta^u_{RL}$ parameter however receives extra
constraints stemming from the ${\rm BR}(B_s \to \mu\mu)$ observable (see
Figure~\ref{fig:1DuLRRL}) so that the posterior distribution peaks around zero
and has a maximal value of $|\delta^u_{RL}| \sim 0.2$. We recall
that the two parameters $\delta^u_{LR}$ and $\delta^u_{RL}$ are independent and
induce different mixing patterns. More precisely, $\delta^u_{LR}$ describes a
$\tilde{c}_L$--$\tilde{t}_R$ mixing, while $\delta^u_{RL}$ corresponds to mixing
between the $\tilde{c}_R$ and $\tilde{t}_L$ eigenstates. The impact of the
constraints and the resulting distributions are therefore different and directly
related to the structure of the
chargino-squark-quark and neutralino squark-quark interactions.

In the down-type squark sector, the prior distributions of the
$\delta^d_{LR}$ and $\delta^d_{RL}$ mixing parameters show a 
clear peak for values close to zero. Large values are often discarded as
they imply large off-diagonal terms in the $M_{\tilde d}$ mass matrix
so that the resulting spectrum likely contains tachyons.
Both parameters are hardly constrained by any of the observables under
consideration and we only observe minor effects. The posterior distribution
of $\delta^d_{LR}$ slightly prefers negative values, and the
posterior distribution of $\delta^d_{RL}$ is slightly narrower, when both distributions
are compared to their respective prior. This mostly results from an interplay of
all observables, although but for the $\delta^d_{RL}$ case, the $B$-meson oscillation
observable $\Delta M_{B_s}$ and the Higgs boson mass requirement
play a non-negligible role (see Figure~\ref{fig:1DdLRRL}).

\begin{figure}
	\begin{center}
	\includegraphics[width=0.32\textwidth]{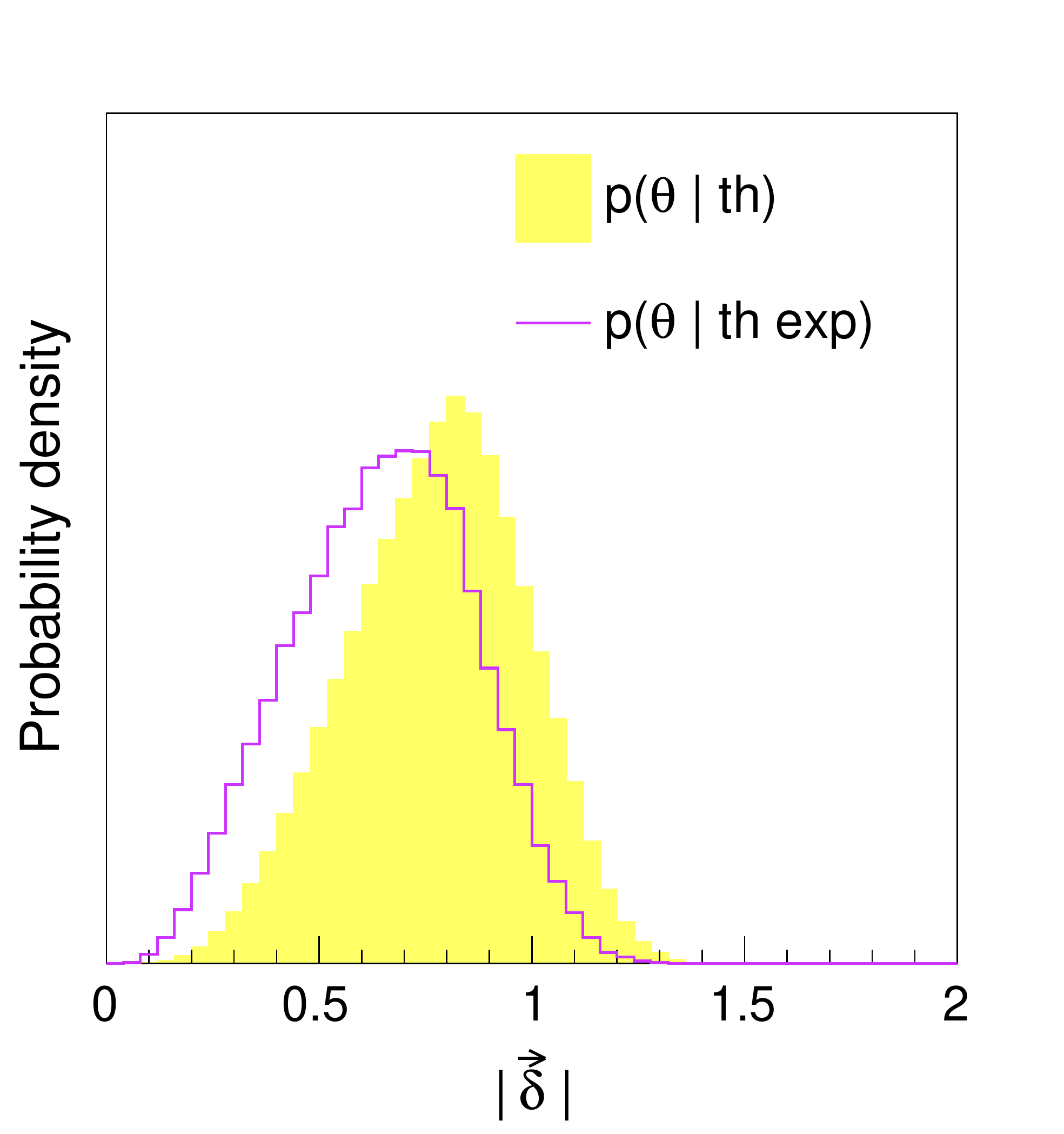} \qquad \qquad
	\includegraphics[width=0.32\textwidth]{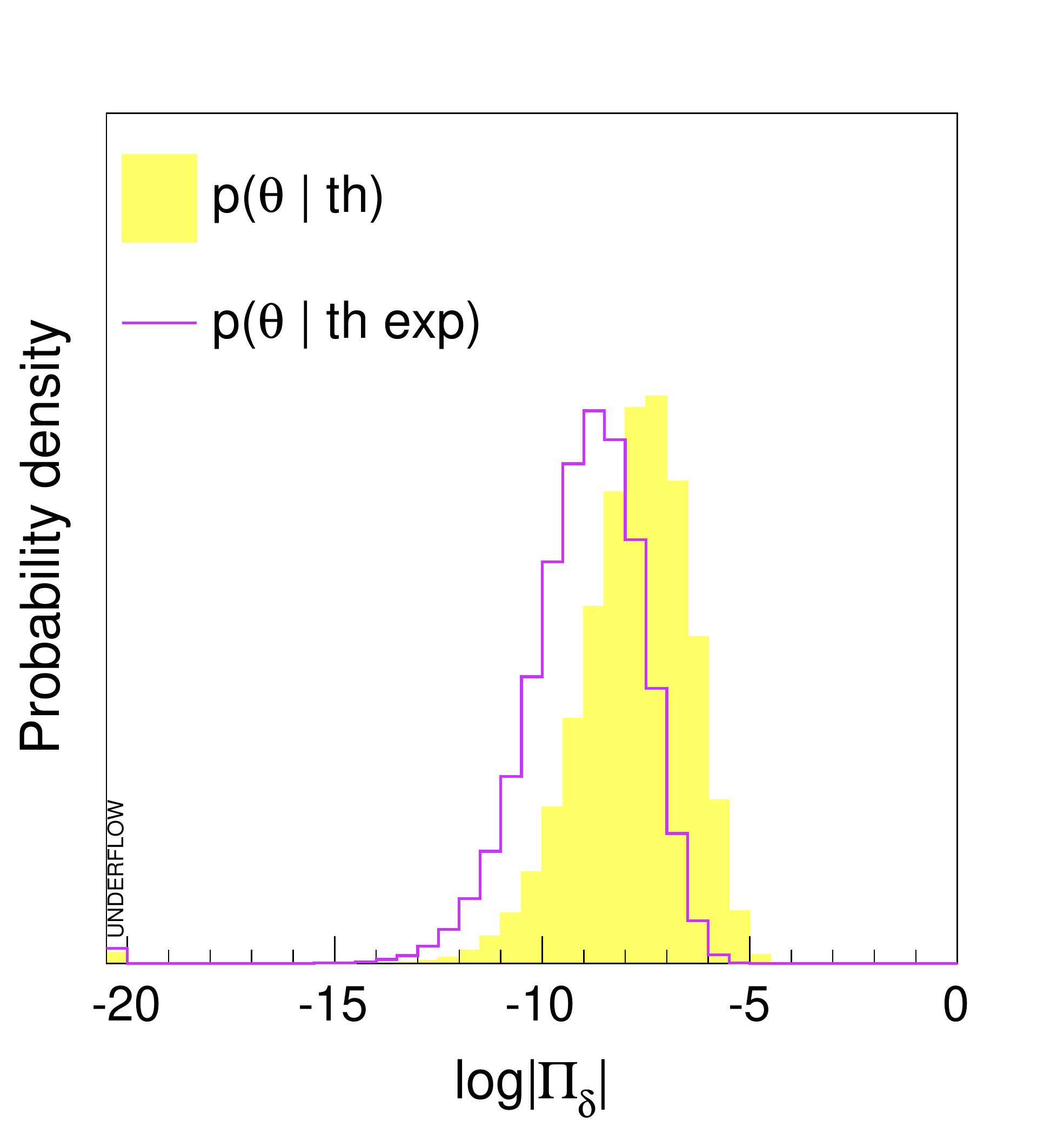}
	\end{center}
	\caption{The one-dimensional prior (yellow histogram) and posterior (violet curve) distributions of the quantities $|\vec{\delta}|$ and $\log|\Pi_{\delta}|$ defined in Eq.~\eqref{eq:norms}.}
	\label{fig:alldelta}
\end{figure}

We now illustrate the global distribution of all NMFV
parameters. To this end, we introduce the quantities
\be\bsp
	|\vec{\delta}| &= \left[ \big( \delta_{LL} \big)^2 + \big( \delta_{RR}^u \big)^2 + \big( \delta_{RR}^d \big)^2 
		+ \big( \delta_{LR}^u \big)^2 + \big( \delta_{RL}^u \big)^2 
		+ \big( \delta_{LR}^d \big)^2 + \big( \delta_{RL}^d \big)^2 \right]^{1/2}\ , \\
	\log|\Pi_{\delta}| &= \log{ \left| \delta_{LL} \delta_{RR}^u \delta_{RR}^d \delta_{LR}^u \delta_{RL}^u \delta_{LR}^d \delta_{RL}^d \right | }\ .
\esp\label{eq:norms}\ee
The former, $|\vec{\delta}|$, corresponds to the norm of a vector whose
components are the seven NMFV parameters. Its value gives a measure of how far
a given benchmark is situated from the minimally flavour-violating setup where
$|\vec{\delta}| = 0$. The maximum value that can be reached in our scan is
$|\vec{\delta}| \approx 1.56$. The
second quantity, $\log|\Pi_{\delta}|$, corresponds to the logarithm of the
absolute value of the product of the seven NMFV parameters. 
The case where all NMFV parameters are maximum corresponds to
$\log|\Pi_{\delta}| \approx -3.5$. In Figure~\ref{fig:alldelta}, we show the
prior and posterior distributions of these two quantities. All scanned points
feature $|\vec{\delta}| > 0$ so that at least one of the NMFV parameters is
sizeable and non-vanishing. The second quantity is in general large and negative
so that at least one of the NMFV parameters has to be small.
However, since the distribution shows a peak around $\log|\Pi_{\delta}| \approx -7$
it is clear that a large fraction of the scanned points exhibit seven non-vanishing
(with some sizeable) NMFV parameters.

\subsection{Correlations within the flavour-violating parameters}
\label{sec:correlations}

Having discussed the distribution of single parameters, it is interesting to
investigate possible correlations between different NMFV quantities. A correlation indicator
between two parameters $x$ and $y$ can be computed as
\begin{equation}
    r ~=~ \frac{\sum_{i=1}^{n}(x_i - \bar{x})(y_i - \bar{y})}{\sqrt{\sum_{i=1}^{n}(x_i - \bar{x})^2} \sqrt{\sum_{i=1}^{n}(y_i - \bar{y})^2}} \ ,
    \label{eq:correlation}
\end{equation}
where the sum runs over all the points $(x_i, y_i)$ of the sample and $\bar{x}$
and $\bar y$ are the mean values of the two parameters. The correlation factor
is vanishing when there is no correlation, while $r=\pm1$ indicates a linear
correlation with the exception of the case in which the sampled points lie on a
line parallel to one of the $x$ and $y$ axes. As the study of the correlations only
makes sense when the parameter ranges cover the entire distribution spreads, we
restrict our analysis to the NMFV parameters. The correlation indicators have been computed
for any pair out of the seven NMFV parameters and the results are shown in
Table~\ref{tab:correlations} when one only accounts for the theoretical prior
(second column) and after imposing the full set of constraints (last column). The
correlations are found not particularly pronounced with all $r$-values being close to
zero.

\begin{figure}[]
	\begin{center}
		\includegraphics[width=0.32\textwidth,clip=true,trim=285 0 0 0]{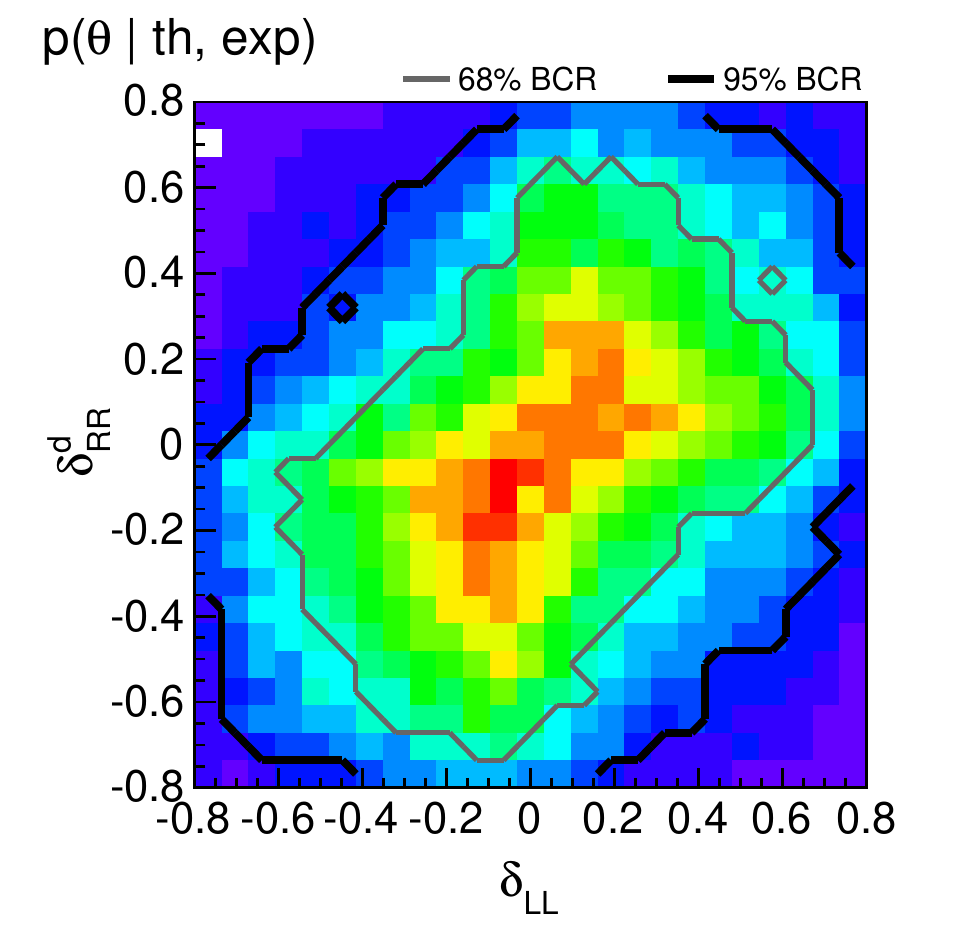}
		\includegraphics[width=0.32\textwidth,clip=true,trim=285 0 0 0]{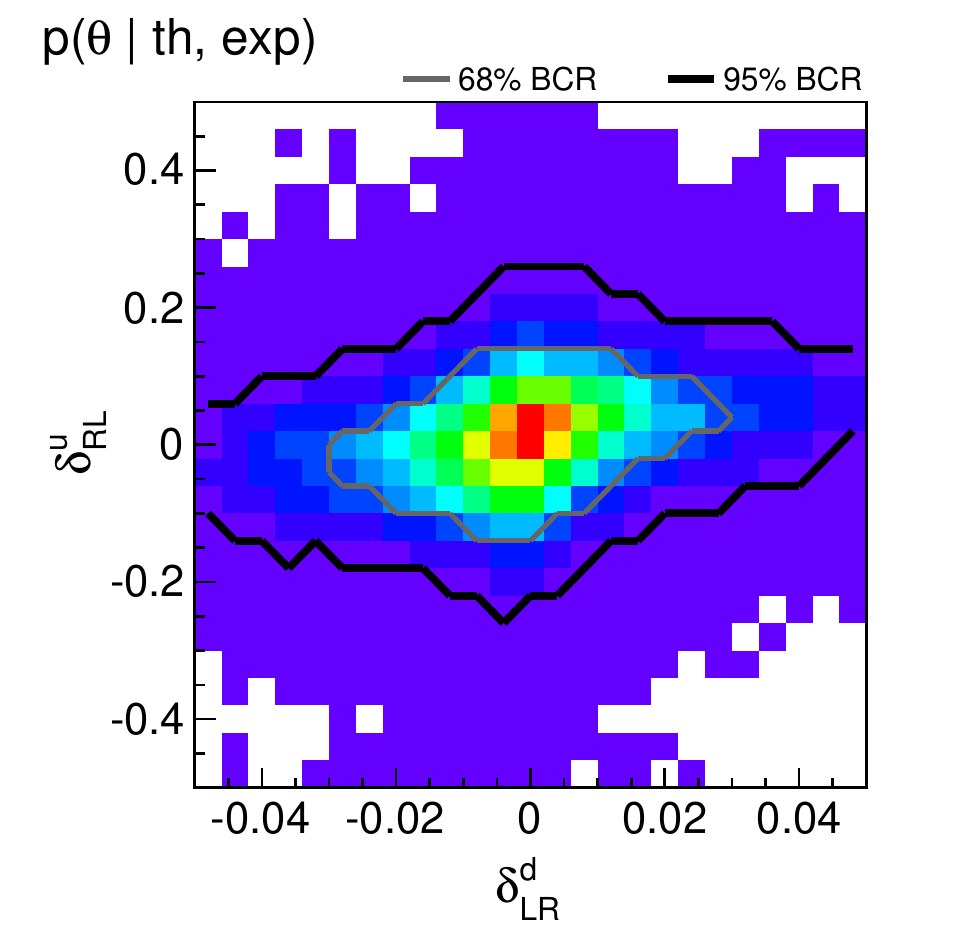}
		\includegraphics[width=0.32\textwidth,clip=true,trim=285 0 0 0]{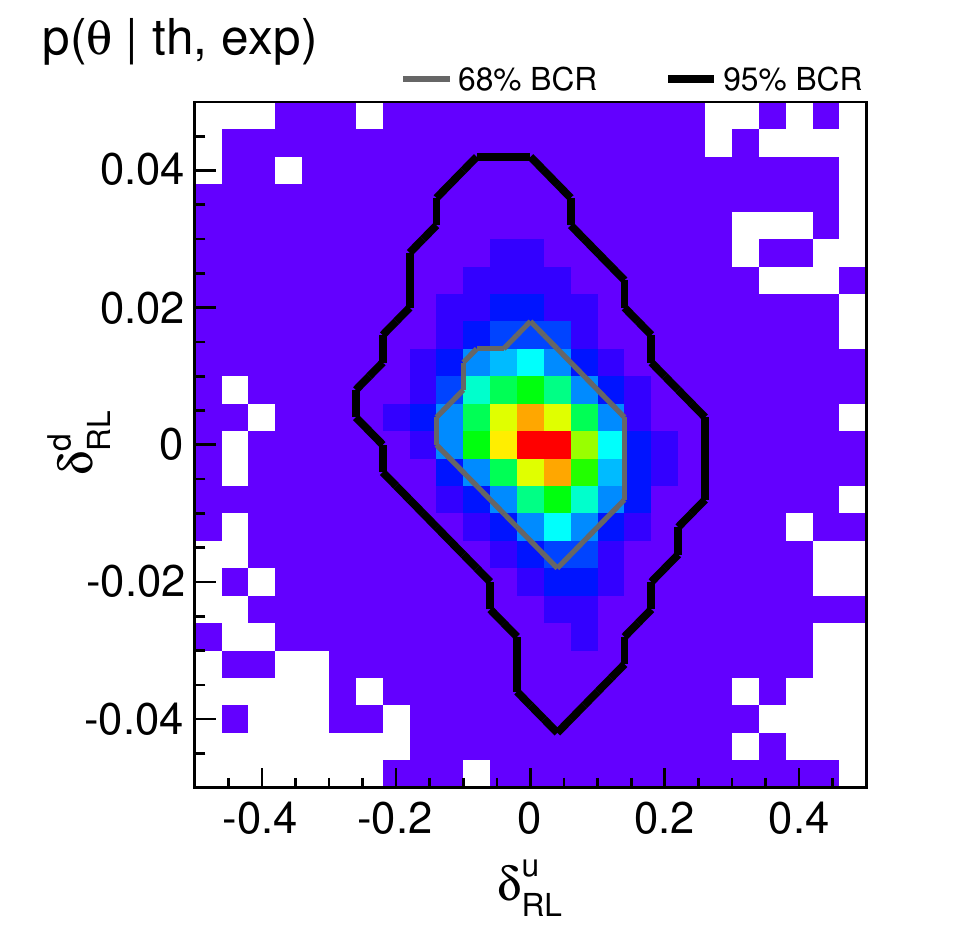}
	\end{center}
	\vspace*{-0.5cm}
	\caption{Two-dimensional distributions of the mostly correlated pairs of NMFV
    parameters after including all constraints.}
	\label{fig:correl}
\end{figure}

We illustrate the correlations between different NMFV parameters on
Figure~\ref{fig:correl}. We however only focus on cases where the correlation
indicator is above $|r|>0.25$, namely on the $(\delta_{LL},\delta_{RR}^{d})$,
$(\delta_{LR}^{d},\delta_{RL}^{u})$ and $(\delta_{RL}^{u},\delta_{RL}^{d})$
pairs. This shows that scenarios in which several NMFV
parameters are non-zero (and even significantly large) simultaneously are
still allowed by current low-energy flavour and Higgs data.

\renewcommand{\arraystretch}{1.2}
\begin{table}
        \centering
        \begin{tabular}{l||c|c} 
                Parameters & th & th + exp  \\           
                \hline \hline
                $(\delta_{LL},\delta_{RR}^{d})$ & -0.003 & 0.270 \\ 
                $(\delta_{LR}^{d},\delta_{RL}^{u})$ & 0.007 & 0.267 \\ 
                $(\delta_{RL}^{u},\delta_{RL}^{d})$ & -0.000 & -0.254 \\
                $(\delta_{RR}^{d},\delta_{RL}^{u})$ & -0.002 & 0.185 \\ 
                $(\delta_{LL},\delta_{RL}^{u})$ & 0.009 & -0.158 \\ 
                \hline \hline
                $(\delta_{RR}^{u},\delta_{RL}^{u})$ & 0.003 & -0.037 \\ 
                $(\delta_{LL},\delta_{LR}^{u})$ & 0.002 & -0.031 \\ 
                $(\delta_{RR}^{d},\delta_{LR}^{d})$ & -0.021 & -0.028 \\
                $(\delta_{LR}^{d},\delta_{RL}^{d})$ & -0.001 & 0.027 \\ 
                $(\delta_{LL},\delta_{LR}^{d})$ & -0.002 & 0.023 \\ 
                $(\delta_{LL},\delta_{RL}^{d})$ & -0.024 & 0.013 \\ 
                $(\delta_{LR}^{u},\delta_{LR}^{d})$ & -0.006 & -0.012 \\ 
                $(\delta_{RR}^{u},\delta_{LR}^{u})$ & 0.003 & 0.010 \\ 
                $(\delta_{RR}^{u},\delta_{RL}^{d})$ & -0.000 & -0.010 \\ 
                \hline \hline
                $(\delta_{RR}^{d},\delta_{RL}^{d})$ & -0.002 & -0.008 \\ 
                $(\delta_{LR}^{u},\delta_{RL}^{u})$ & 0.002 & -0.007 \\ 
                $(\delta_{RR}^{u},\delta_{RR}^{d})$ & 0.001 & -0.006 \\ 
                $(\delta_{RR}^{u},\delta_{LR}^{d})$ & 0.000 & -0.003 \\ 
                $(\delta_{RR}^{d},\delta_{LR}^{u})$ & -0.001 & 0.002 \\ 
                $(\delta_{LR}^{u},\delta_{RL}^{d})$ & 0.000 & 0.000 
        \end{tabular} 
         \caption{The correlation coefficient $r$ defined in
  Eq.~\eqref{eq:correlation} for all pairs of NMFV parameters. The parameter
  pairs are ordered by their correlation indicators when taking into account all
  imposed constraints (`th+exp'). We also display the indicator values when
  only the theoretical prior is imposed (`th').}
        \label{tab:correlations}
\end{table}
\renewcommand{\arraystretch}{1.0}

\subsection{Squark masses and flavour decomposition}

\begin{figure}[ht!]
	\begin{center}
		\includegraphics[width=0.3\textwidth]{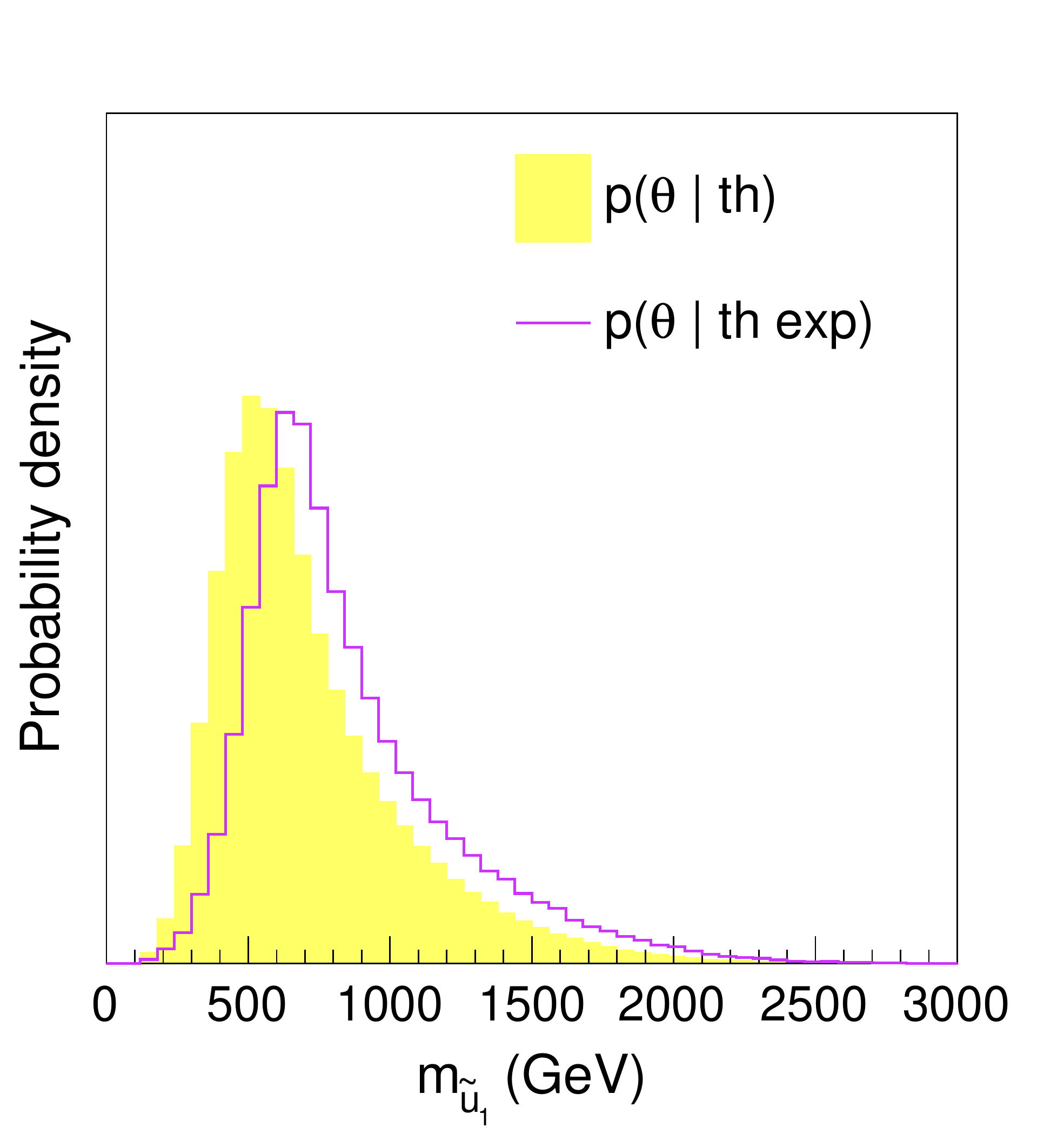} 
		\includegraphics[width=0.3\textwidth]{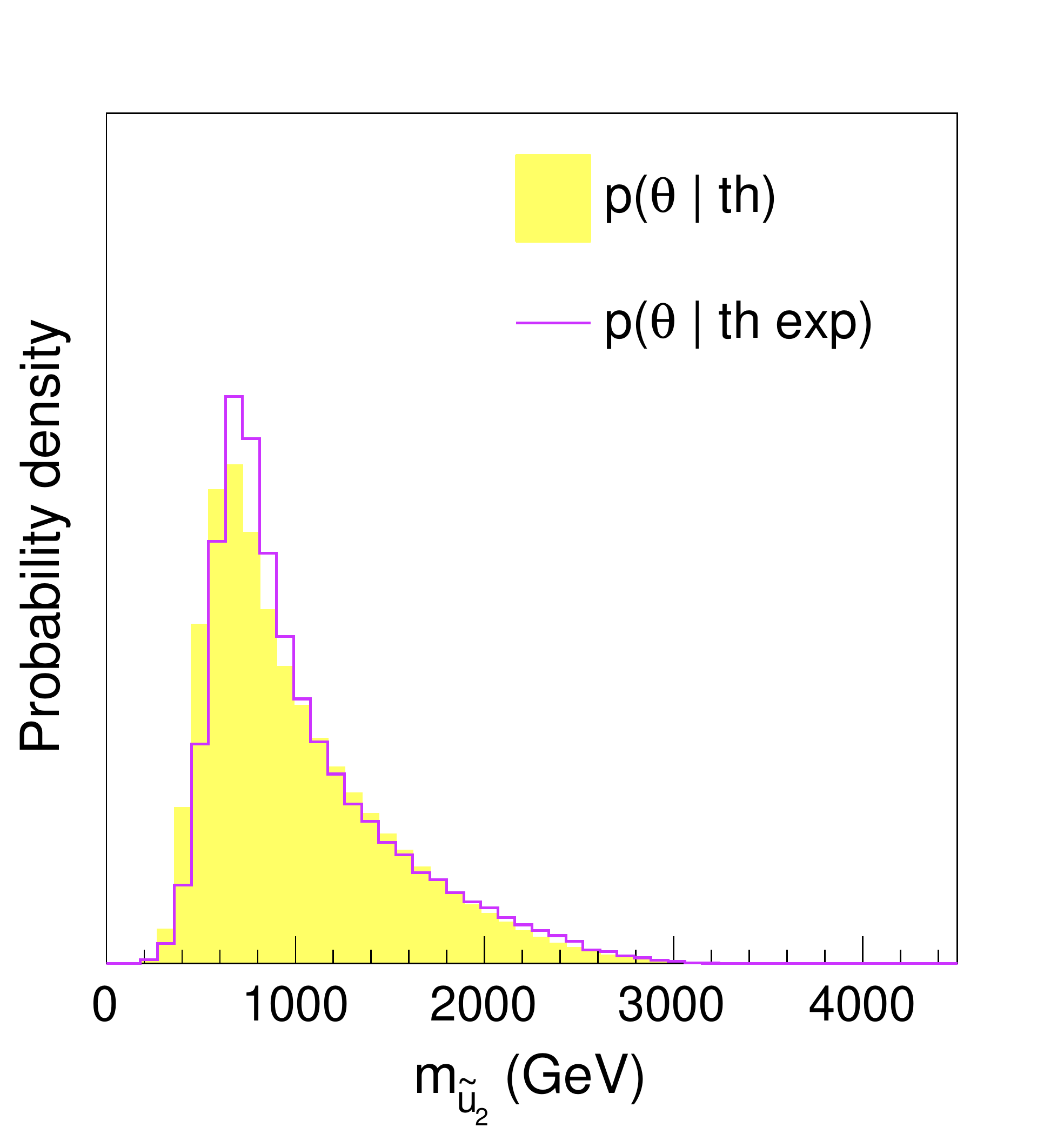} 
		\includegraphics[width=0.3\textwidth]{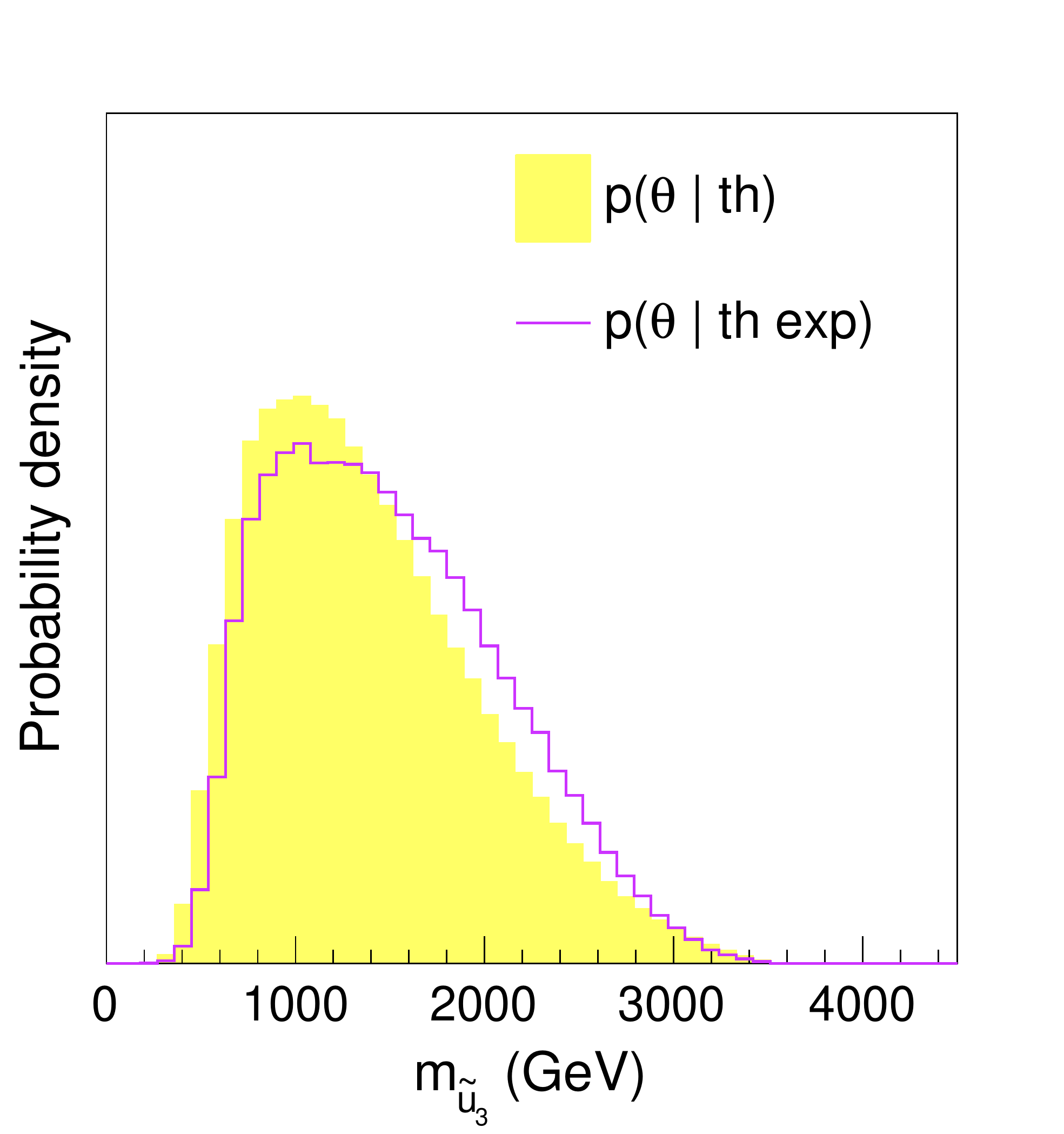} \\
		\includegraphics[width=0.3\textwidth]{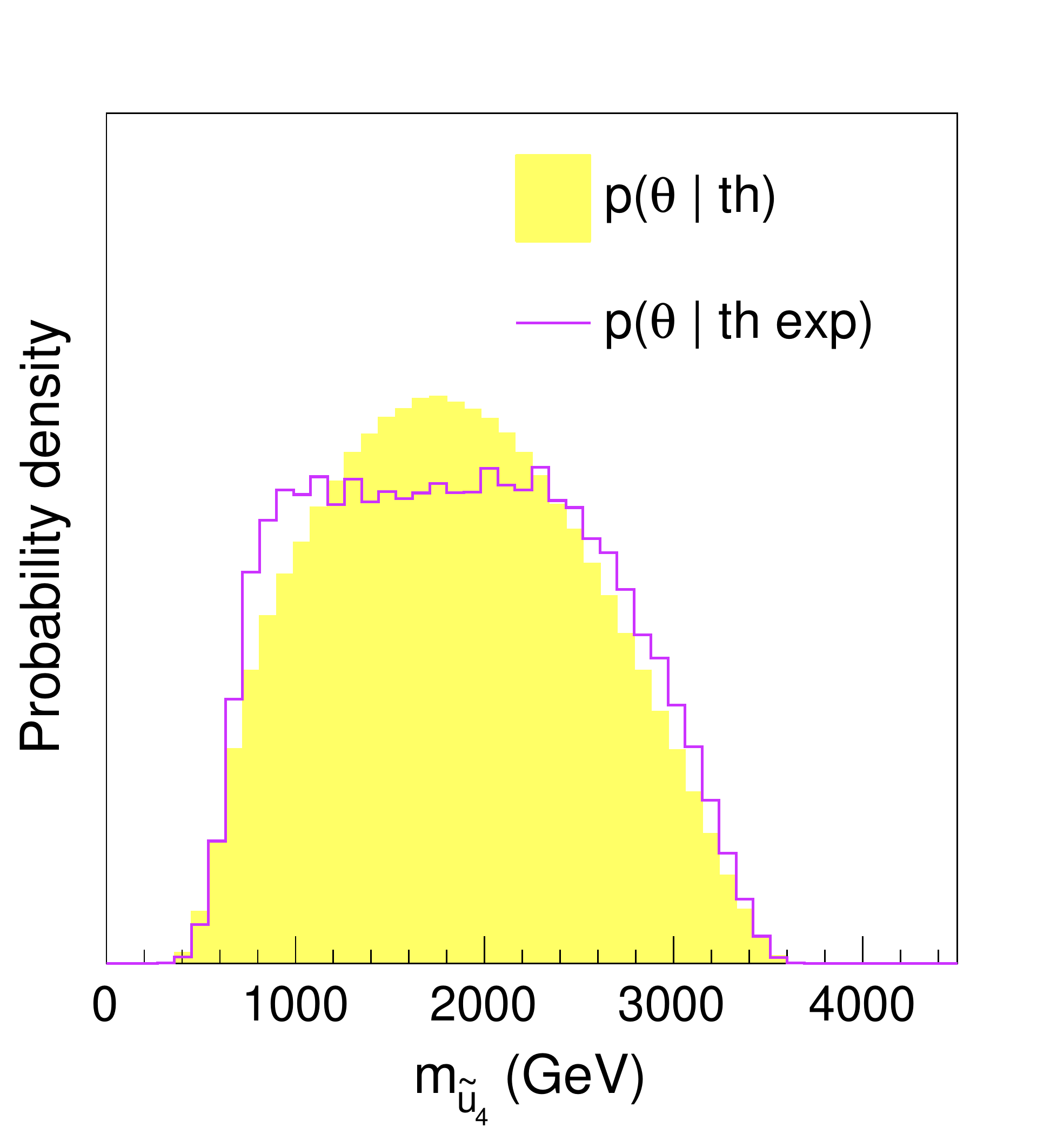} 
		\includegraphics[width=0.3\textwidth]{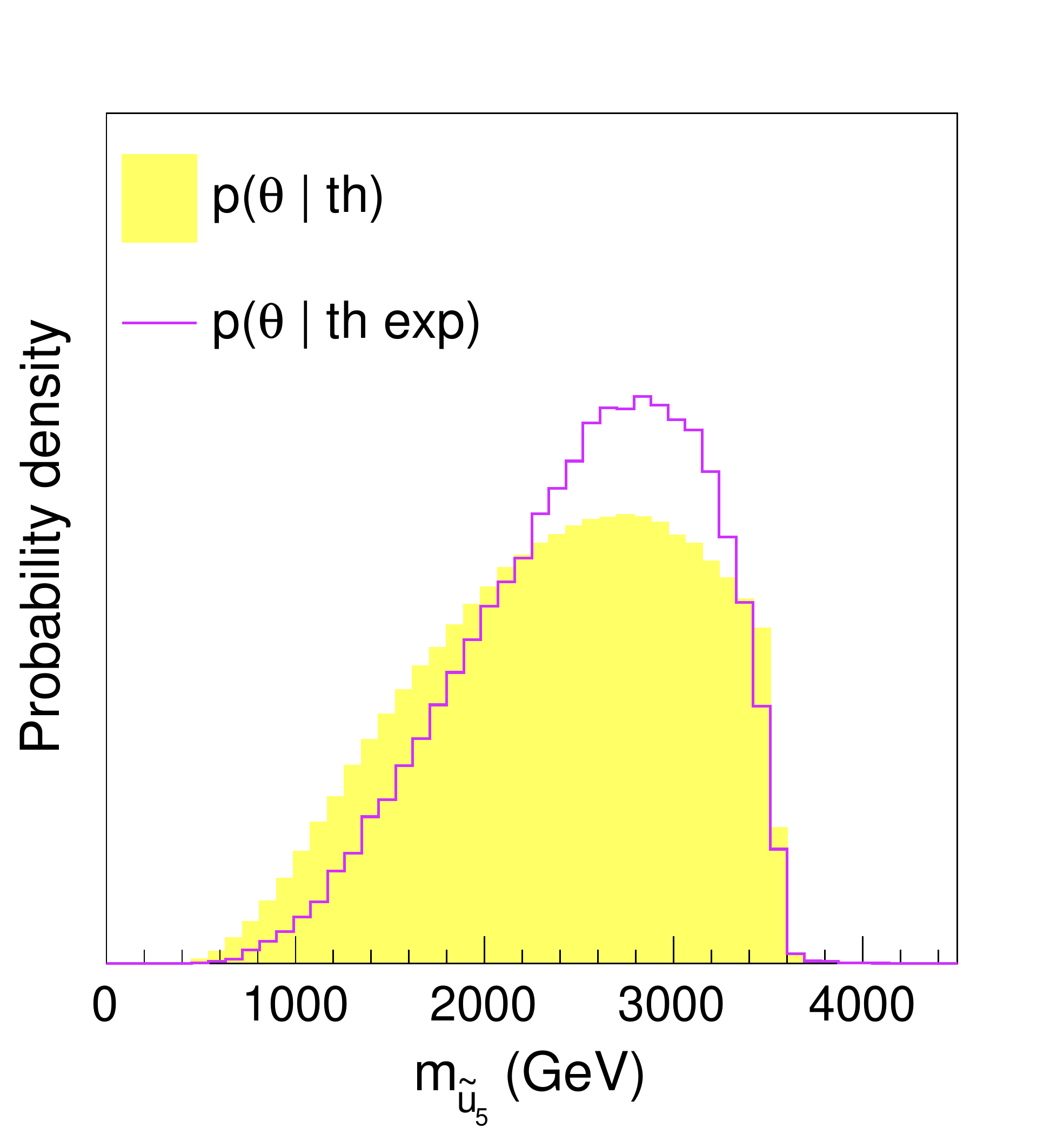} 
		\includegraphics[width=0.3\textwidth]{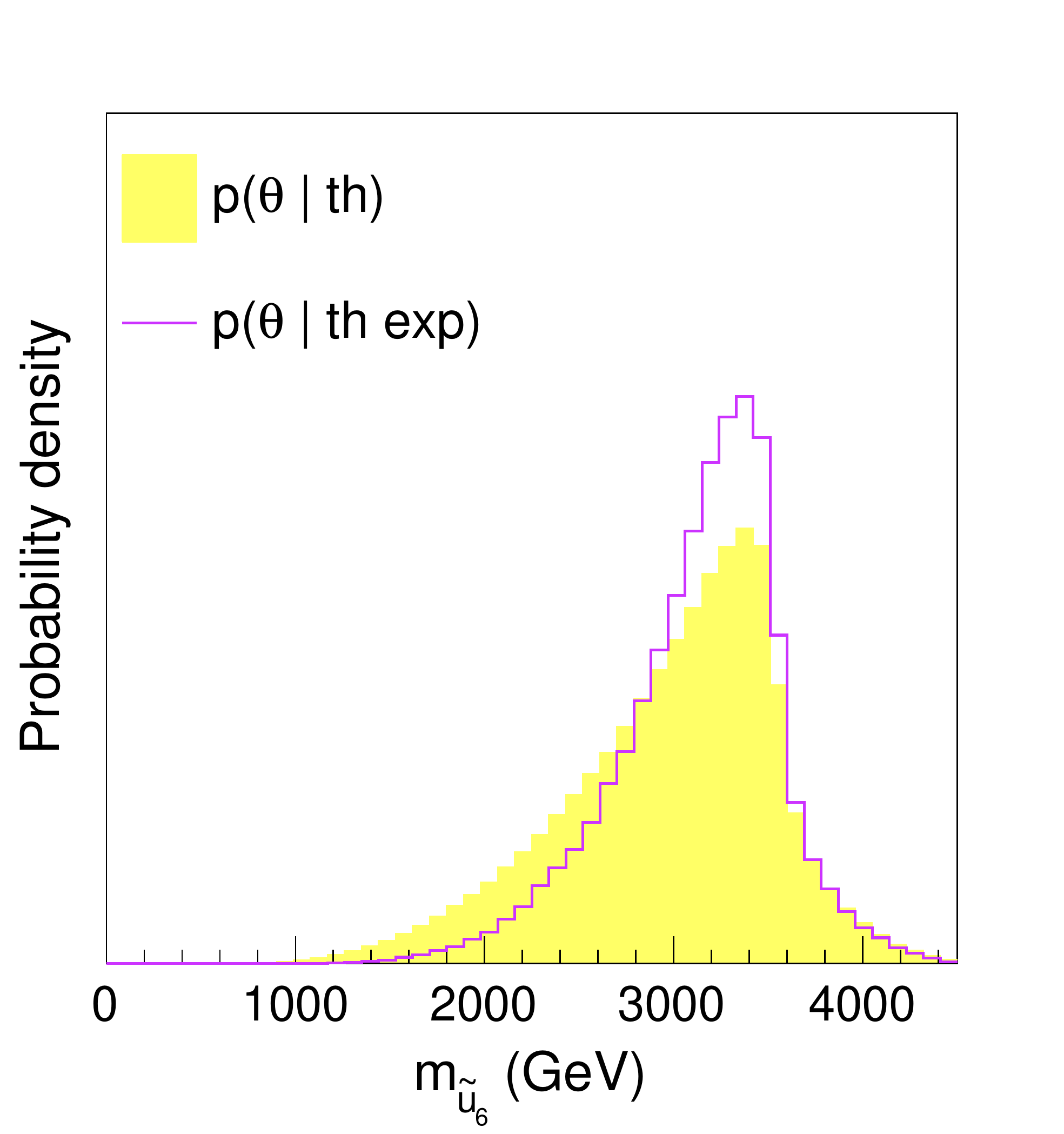}		
	\end{center}
	\vspace*{-0.5cm}
	\caption{One-dimensional prior (yellow histogram) and posterior (violet curve) distributions of the masses of the six up-type squarks.}
	\label{fig:supmasses}
\end{figure}

\begin{figure}[h!]
	\begin{center}
		\includegraphics[width=0.32\textwidth,clip=true,trim=270 0 0 0]{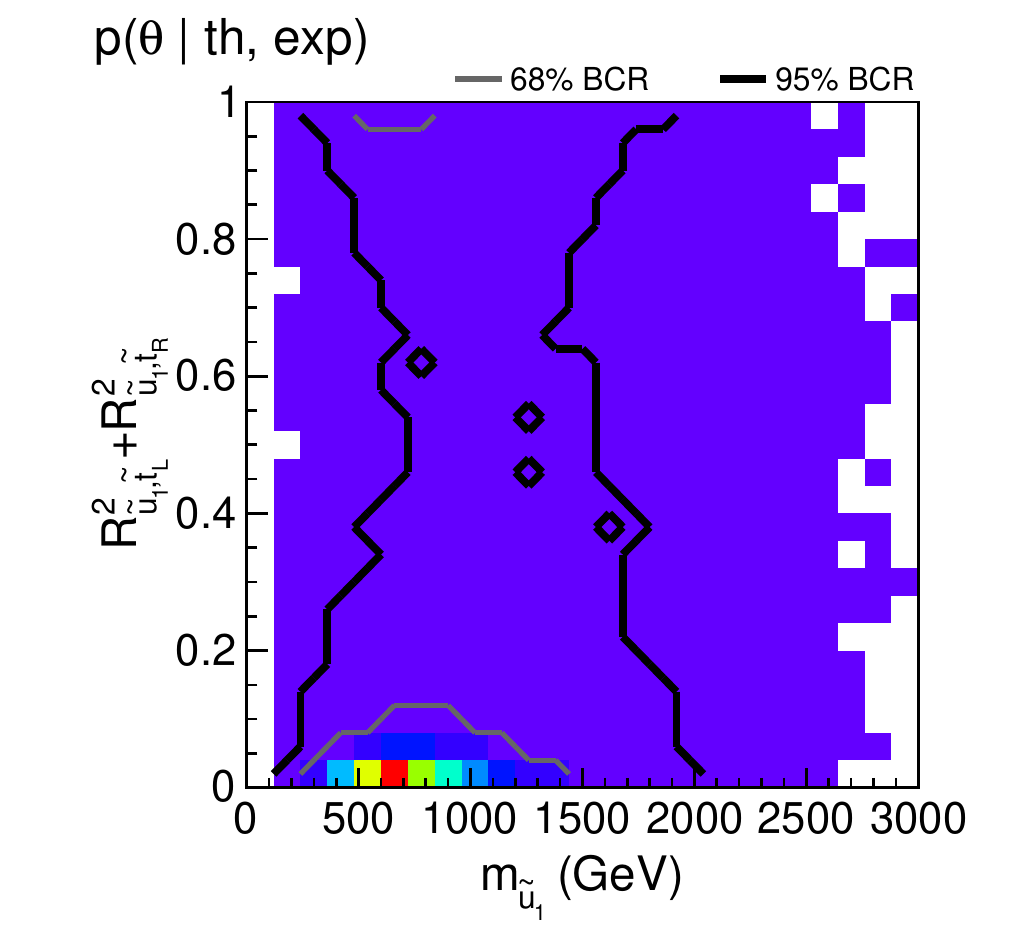}
		\includegraphics[width=0.32\textwidth,clip=true,trim=270 0 0 0]{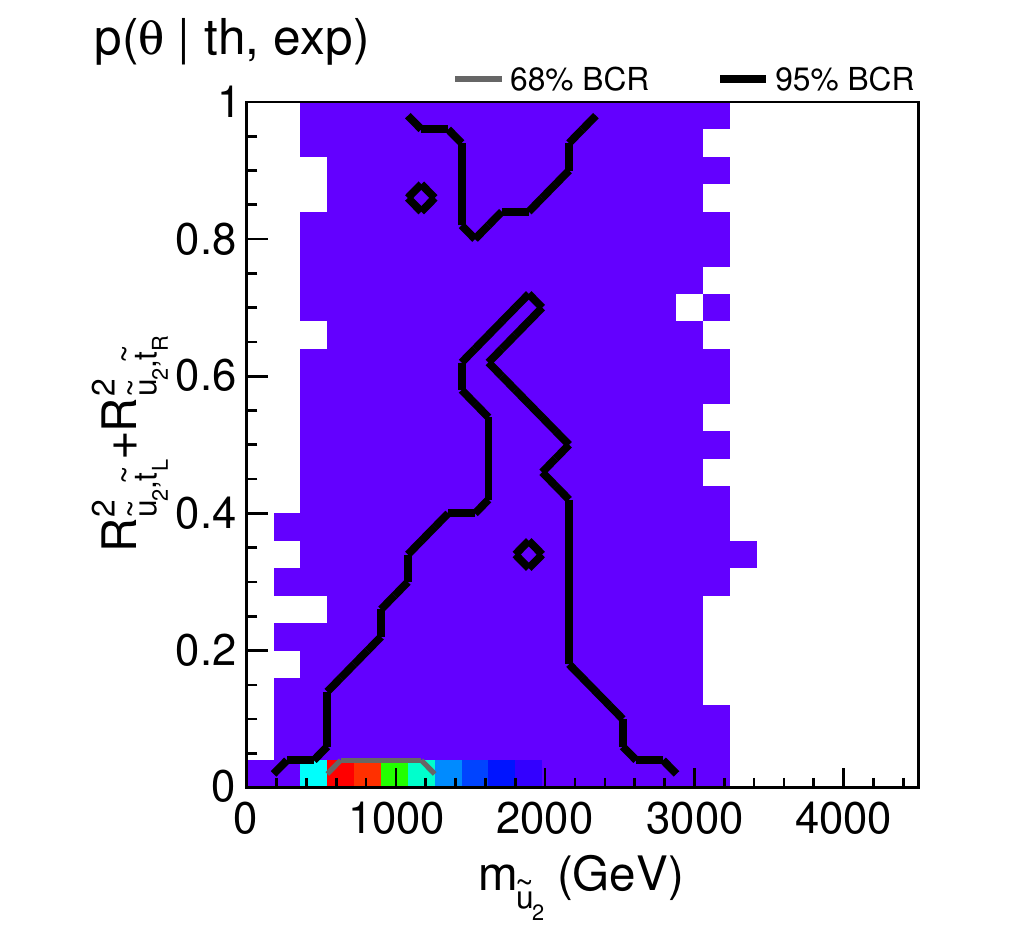}
		\includegraphics[width=0.32\textwidth,clip=true,trim=270 0 0 0]{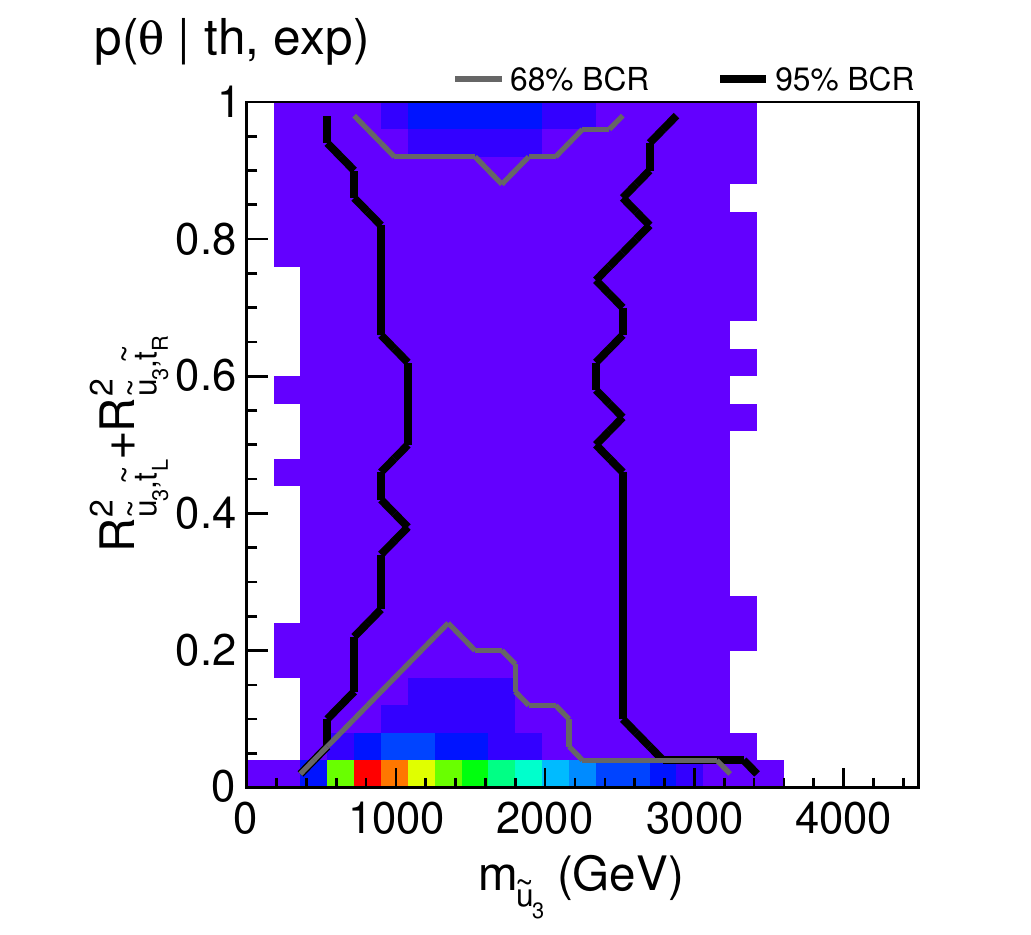}
	\end{center}
	\vspace*{-0.5cm}
	\caption{Resulting correlations between the stop flavour content and the masses
   of the three lightest up-type squarks after imposing all experimental constraints mentioned in Table~\ref{tab:PLMs}. Red colour indicates the highest and dark purple the lowest likelihood.}
	\label{fig:massstop}
\end{figure}

We discuss in this section the distributions of the masses of the squarks,
their flavour decomposition and the mass differences between states
relevant for the LHC phenomenology of NMFV MSSM models.
Figure~\ref{fig:supmasses} shows the prior and posterior distributions for the
up-type squark masses. The shapes of the distributions for the two lightest
states $\tilde{u}_1$ and $\tilde{u}_2$ are very similar and they both peak at
about 800--1000~GeV. The two lightest up-type states $\tilde{u}_1$ and
$\tilde{u}_2$, that are mostly of the first and second generation (see Figure~\ref{fig:massstop}),
are in general relatively close in mass. This is due to the choice
of common mass parameters for the first and second generation squarks.
The heavier $\tilde{u}_3$, $\tilde{u}_4$ and $\tilde{u}_5$ states
exhibit more spread distributions, the masses ranging from 1 to 3.5~TeV. Finally, the
heaviest state $\tilde{u}_6$ is barely reachable at the LHC, with
a mass lying in general above 2~TeV. Although the considered experimental
constraints affect all NMFV supersymmetric parameters, the associated effects on
the mass eigenvalues is at the end only mild, the mass distributions being only
slightly shifted towards higher values.

From a phenomenological point of view, it is interesting to examine the flavour
(in particular the stop) content of the six up-type squarks. 
The posterior distribution
of the stop content of the three lightest up-type squarks is depicted in
Figure~\ref{fig:massstop} and shown in correlation with the respective
squark mass. The lighter states
$\tilde{u}_1$, $\tilde{u}_2$, $\tilde{u}_3$ (and also $\tilde{u}_4$) are mainly not
stop-like, \textit{i.e.}\ they have a significant up or charm component.
Most scanned scenarios indeed exhibit a charm-dominated lightest
$\tilde{u}_1$ squark, while $\tilde{u}_2$ is mostly dominated by its up
component. This contrasts with usual flavour-conserving MSSM setups where the
lightest squark state is typically a stop. This feature can be traced to the
first and second generation soft masses that are driven to lower values as
explained in Section~\ref{sec:ResultsFC}, whilst the third generation squark
masses are pushed towards higher values by the flavour constraints. Furthermore,
even in the presence of large trilinear terms, the lightest states are still
found to be up-like or charm-like.

\begin{figure}
	\begin{center}
		\includegraphics[width=0.3\textwidth]{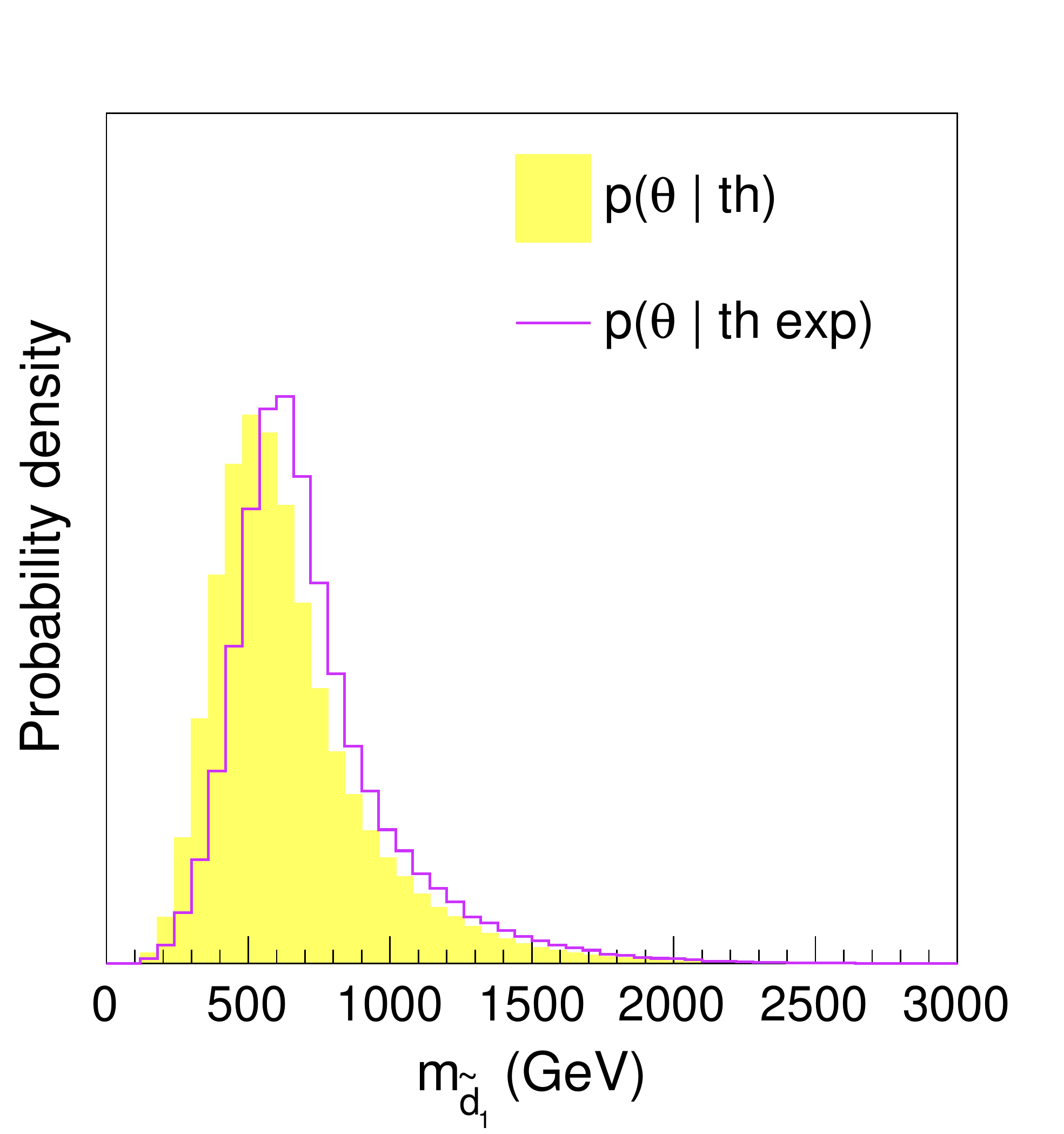} 
		\includegraphics[width=0.3\textwidth]{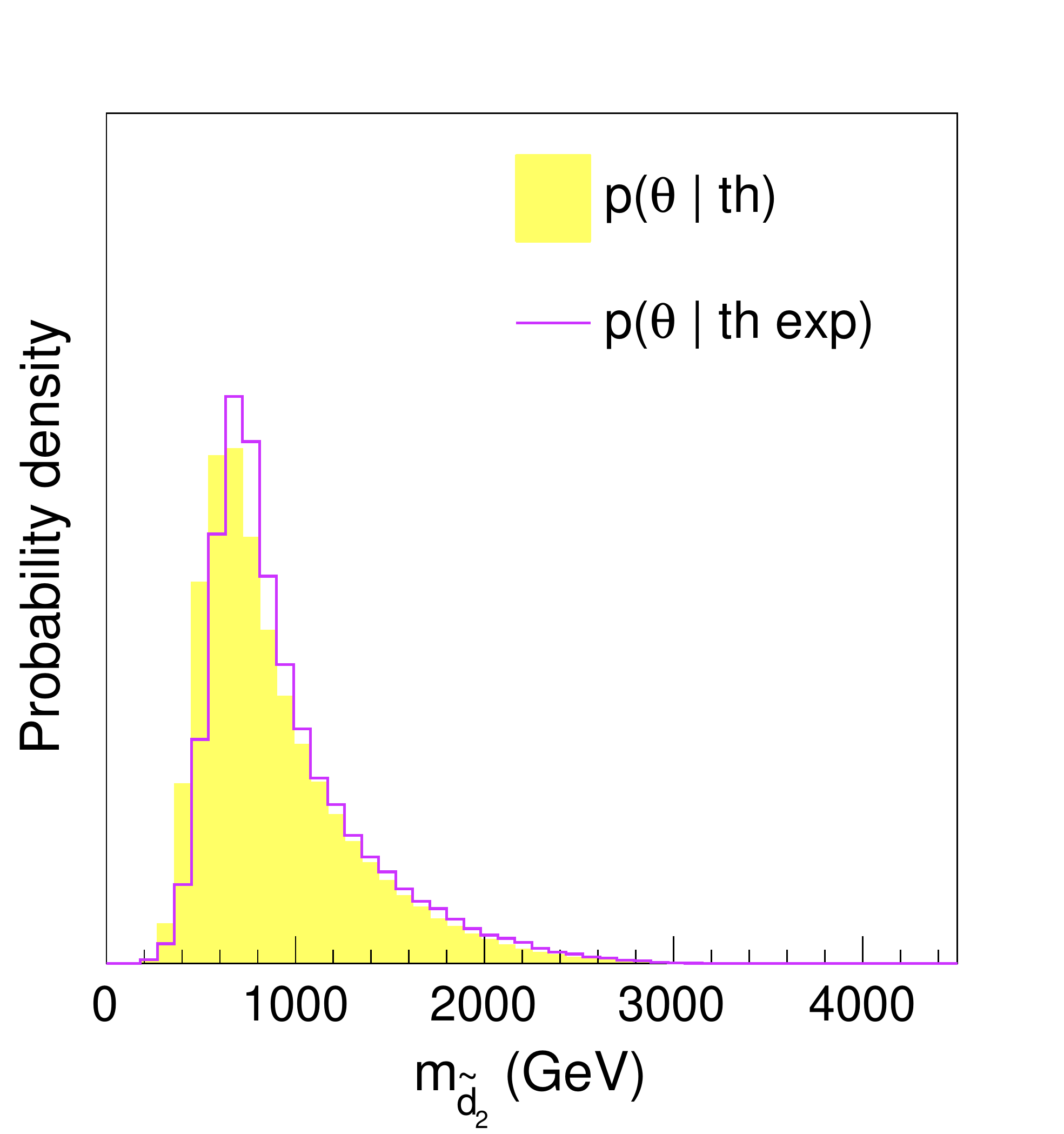} 
		\includegraphics[width=0.3\textwidth]{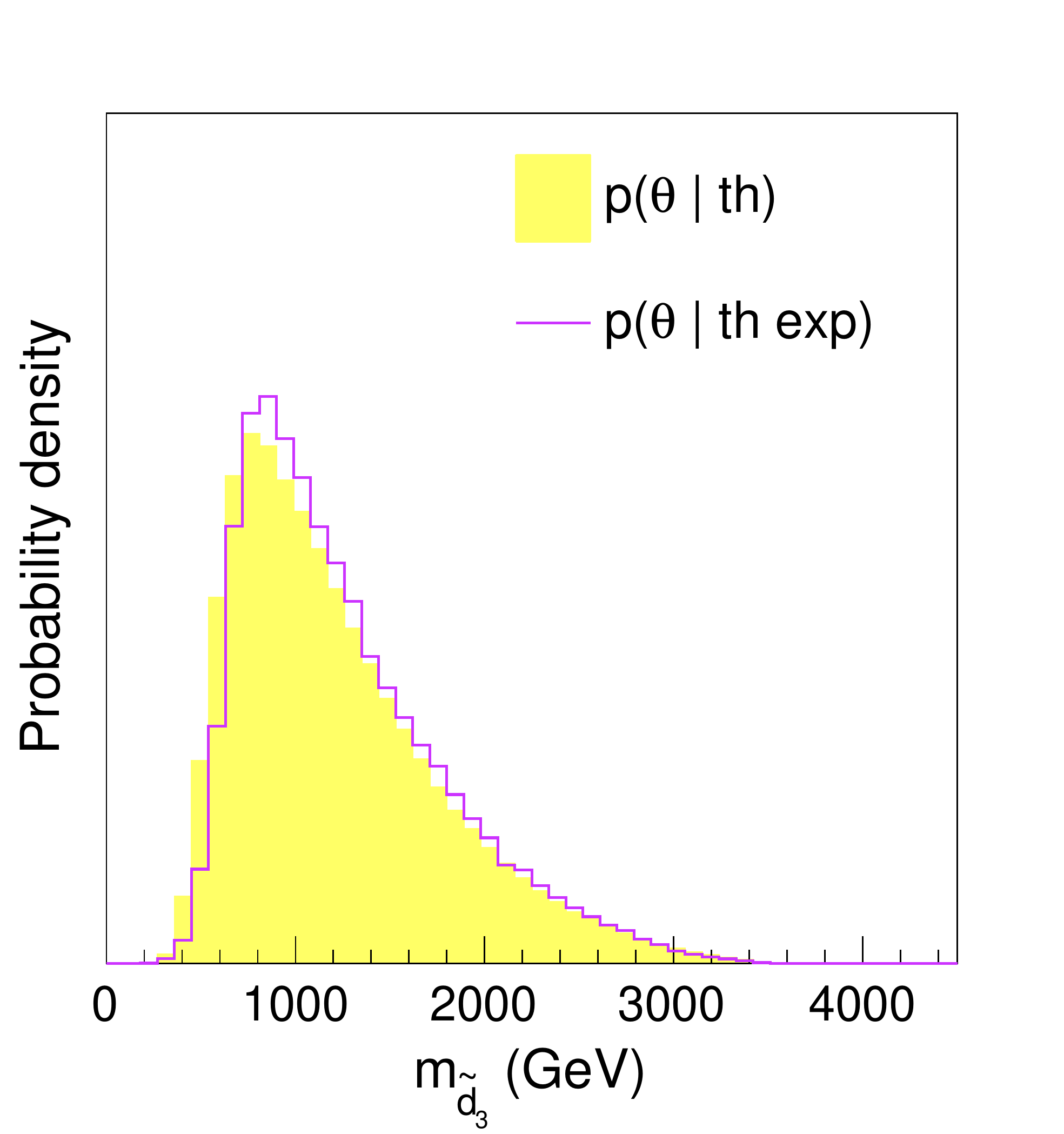} \\
		\includegraphics[width=0.3\textwidth]{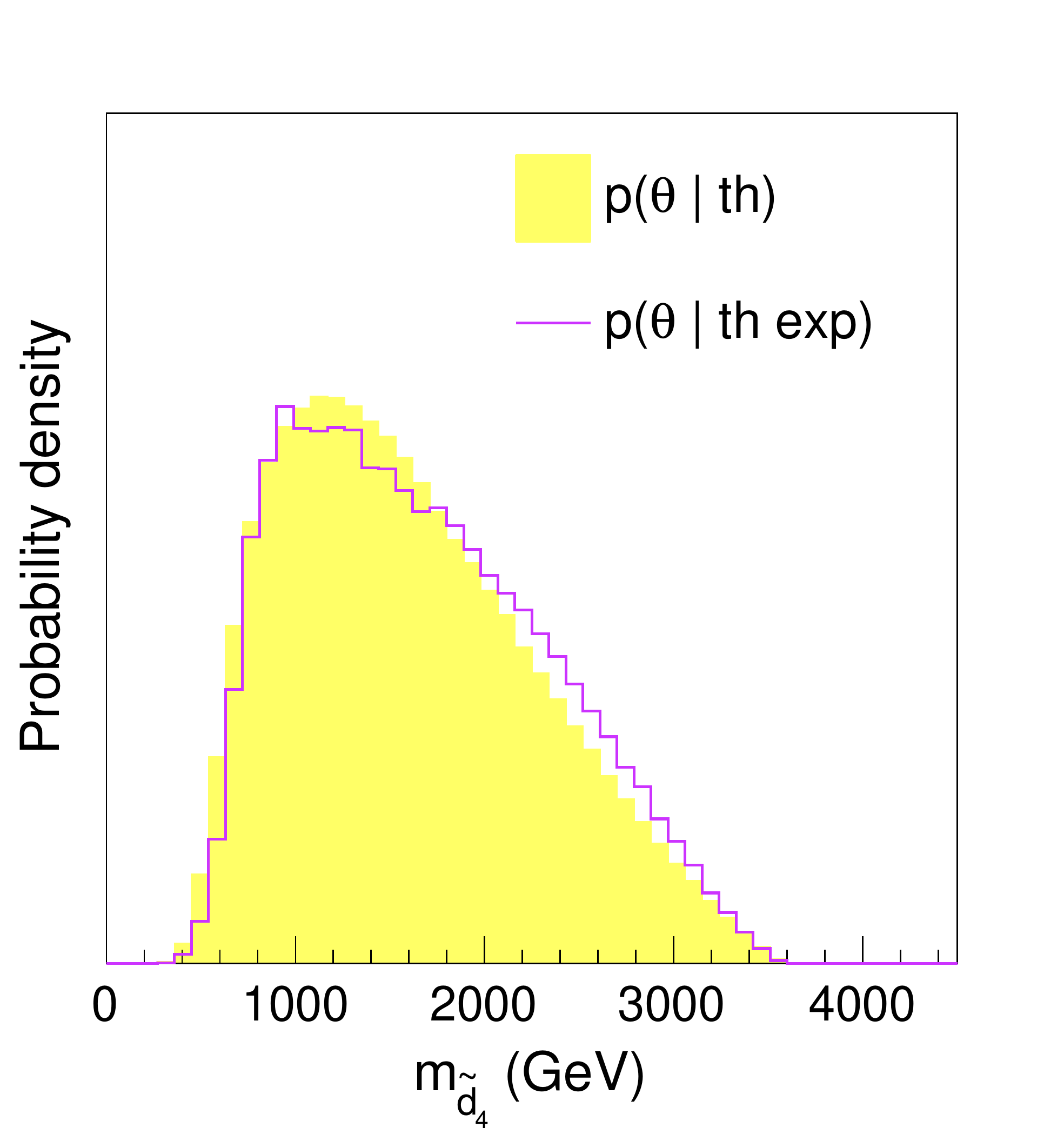} 
		\includegraphics[width=0.3\textwidth]{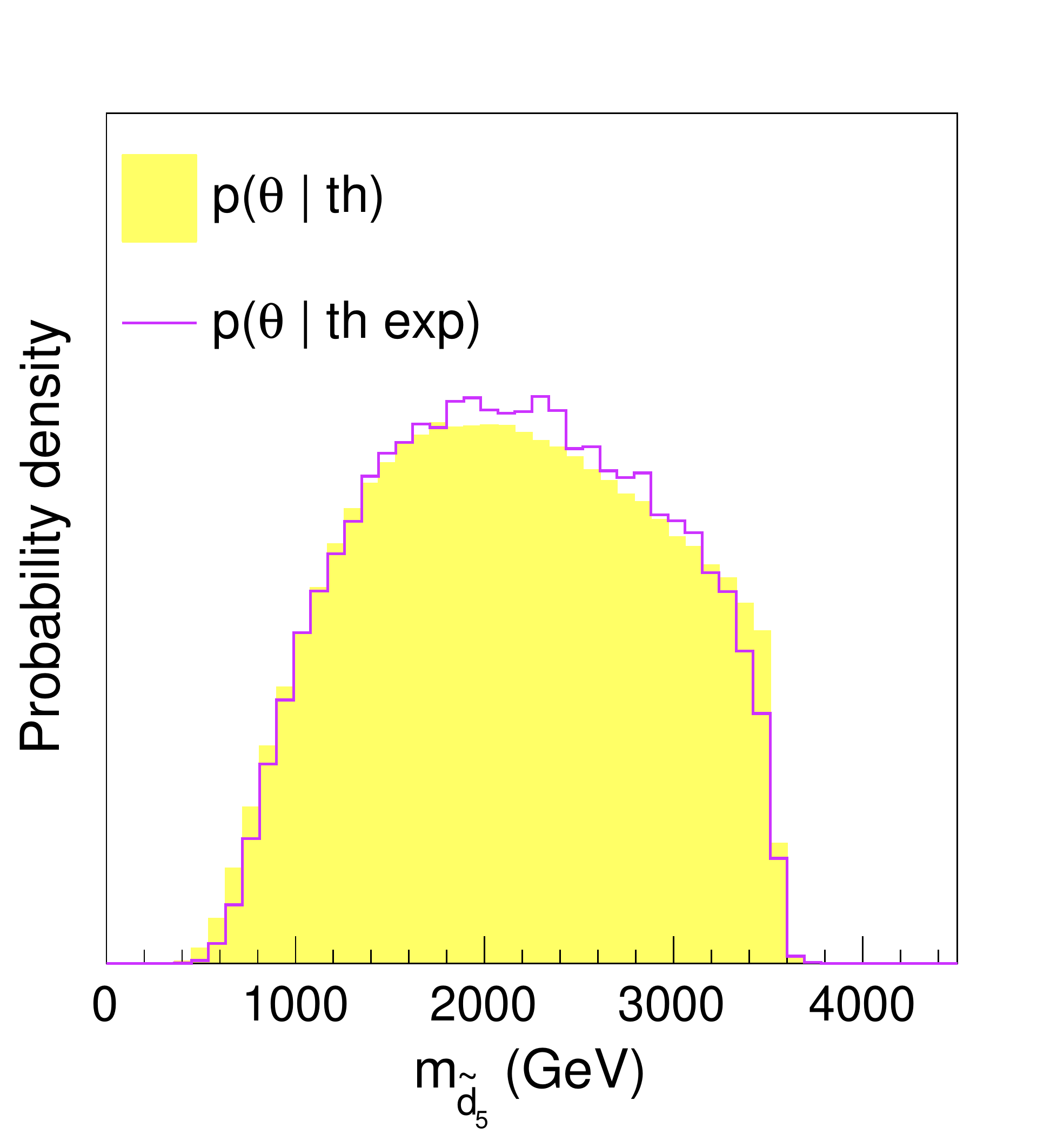} 
		\includegraphics[width=0.3\textwidth]{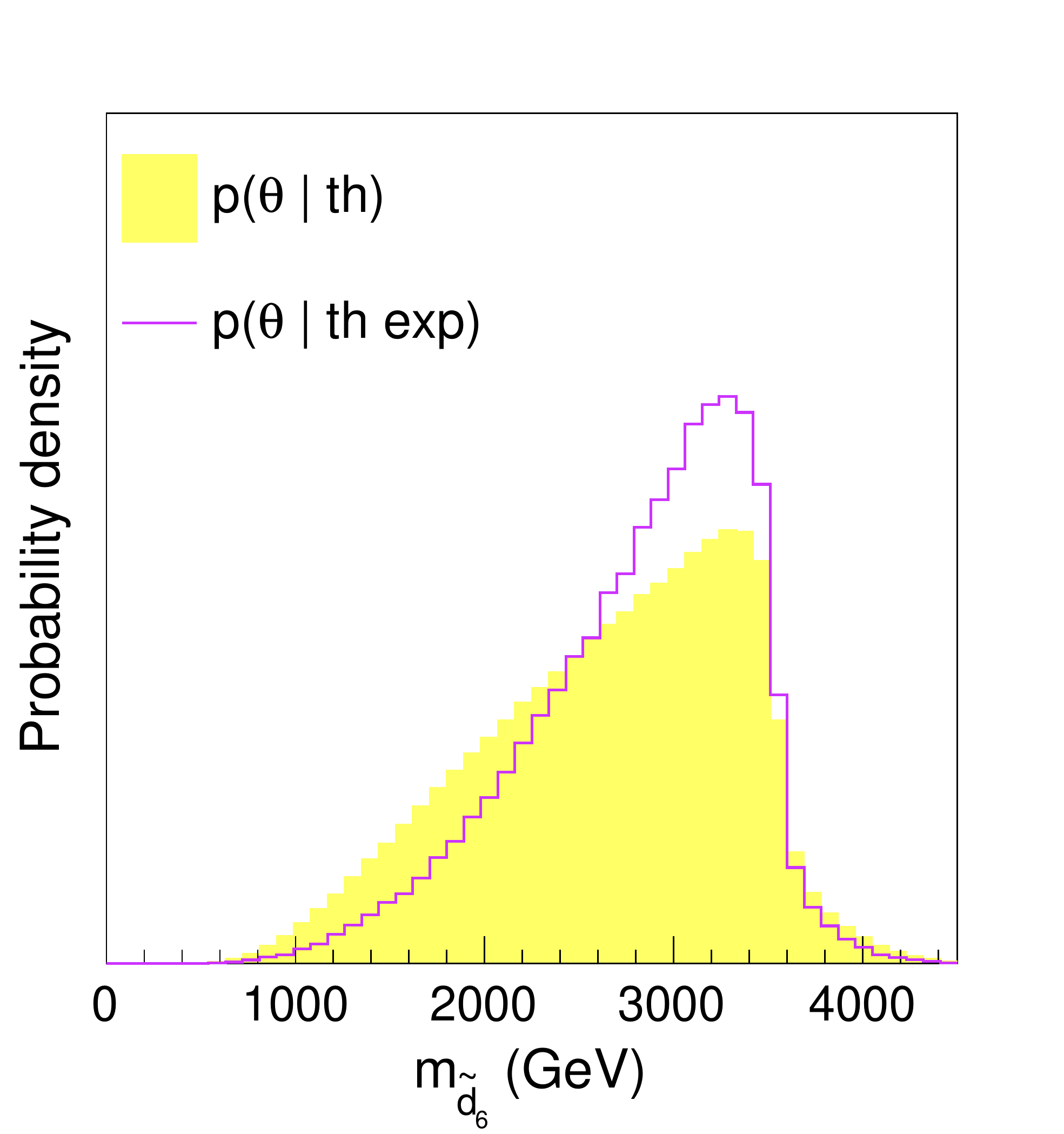}		
	\end{center}
	\vspace*{-0.5cm}
	\caption{Same as Figure~\ref{fig:supmasses} for the down-type squark sector.}
	\label{fig:sdownmasses}
\end{figure}

\begin{figure}
	\begin{center}
		\includegraphics[width=0.32\textwidth,clip=true,trim=270 0 0 0]{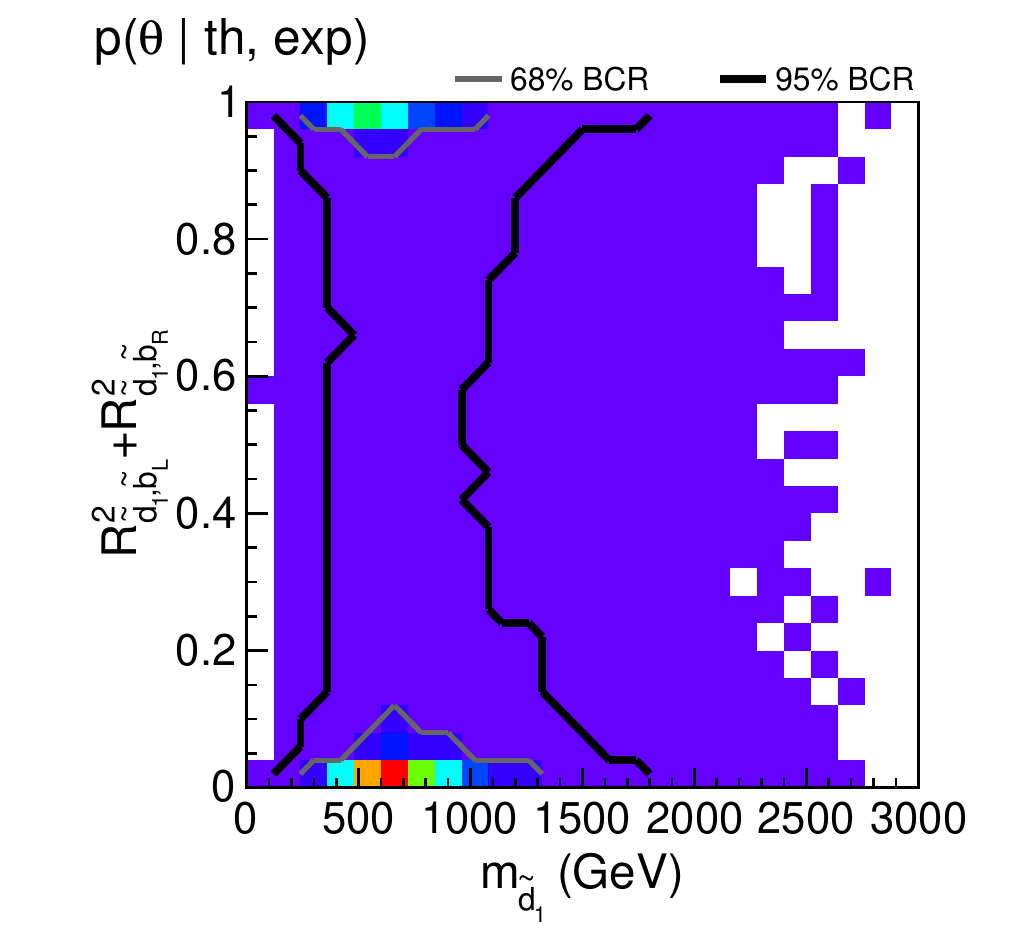}
		\includegraphics[width=0.32\textwidth,clip=true,trim=270 0 0 0]{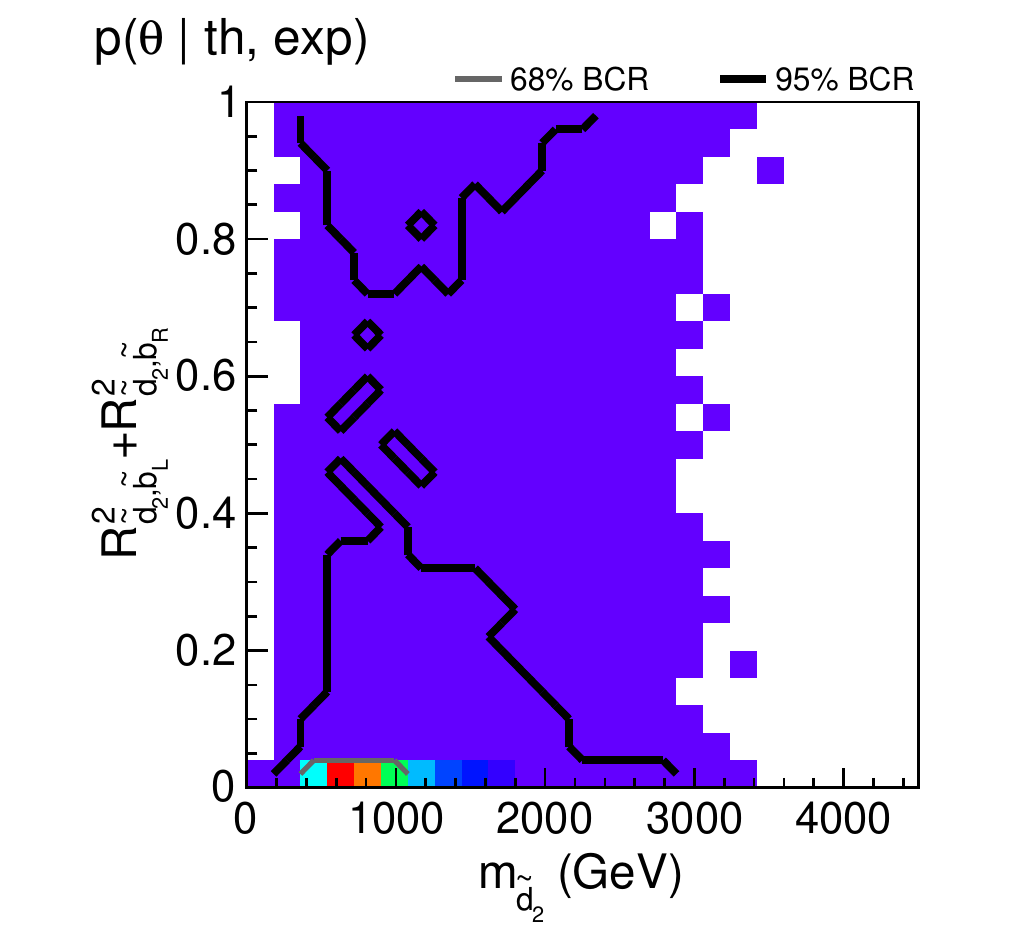}
		\includegraphics[width=0.32\textwidth,clip=true,trim=270 0 0 0]{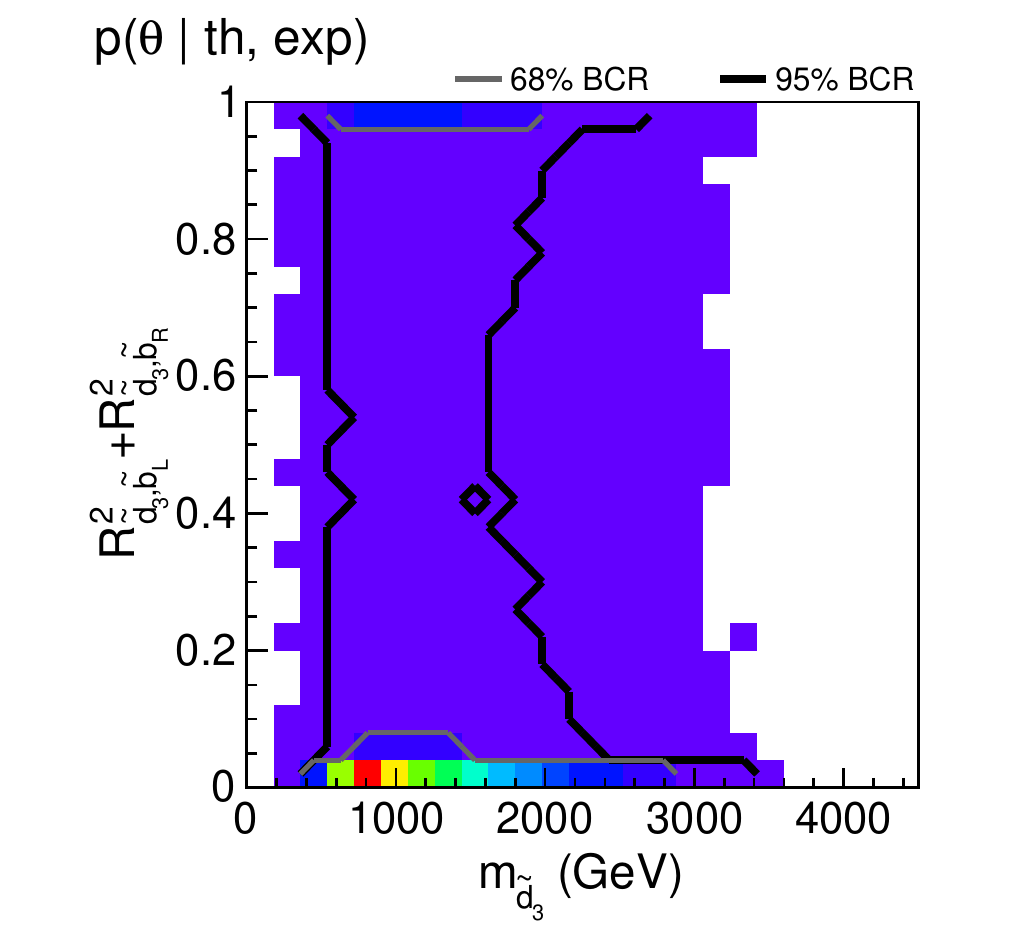}
	\end{center}
	\vspace*{-0.5cm}
	\caption{Resulting correlations between the sbottom flavour content and the masses of the three lightest down-type squarks
  after imposing all experimental constraints mentioned in Table~\ref{tab:PLMs}. Red colour indicates the highest and dark purple the lowest likelihood.}
	\label{fig:masssbot}
\end{figure}

Similar conclusions hold for the sector of the down-type squarks. We show their
masses in Figure~\ref{fig:sdownmasses} and selected flavour decompositions in
Figure~\ref{fig:masssbot}. The three lighter states exhibit comparable
distributions, peaking as for the up-type squarks at about 800--1000~GeV. The
mass distributions of the $\tilde{d}_4$ and $\tilde{d}_5$ states feature
distributions with a larger spread, and the one of heaviest $\tilde{d}_6$ squark
is peaking at about 3~TeV, although masses of about 1~TeV are predicted for a
small subset of scenarios. Flavour mixing in the down-type squark sector is
generally less pronounced than for the up-type squarks, as illustrated on
Figure~\ref{fig:masssbot} where we depict the correlations between the sbottom
content and the masses of the lighter down-type squarks. A majority of
scenarios include light down-like and strange-like squark states and there is only a
small number of parameter points where $\tilde{d}_1$ and $\tilde{d}_2$
contain a sizeable sbottom content.

\begin{figure}
	\begin{center}
		\includegraphics[width=0.4\textwidth,clip=true,trim=270 0 0 0]{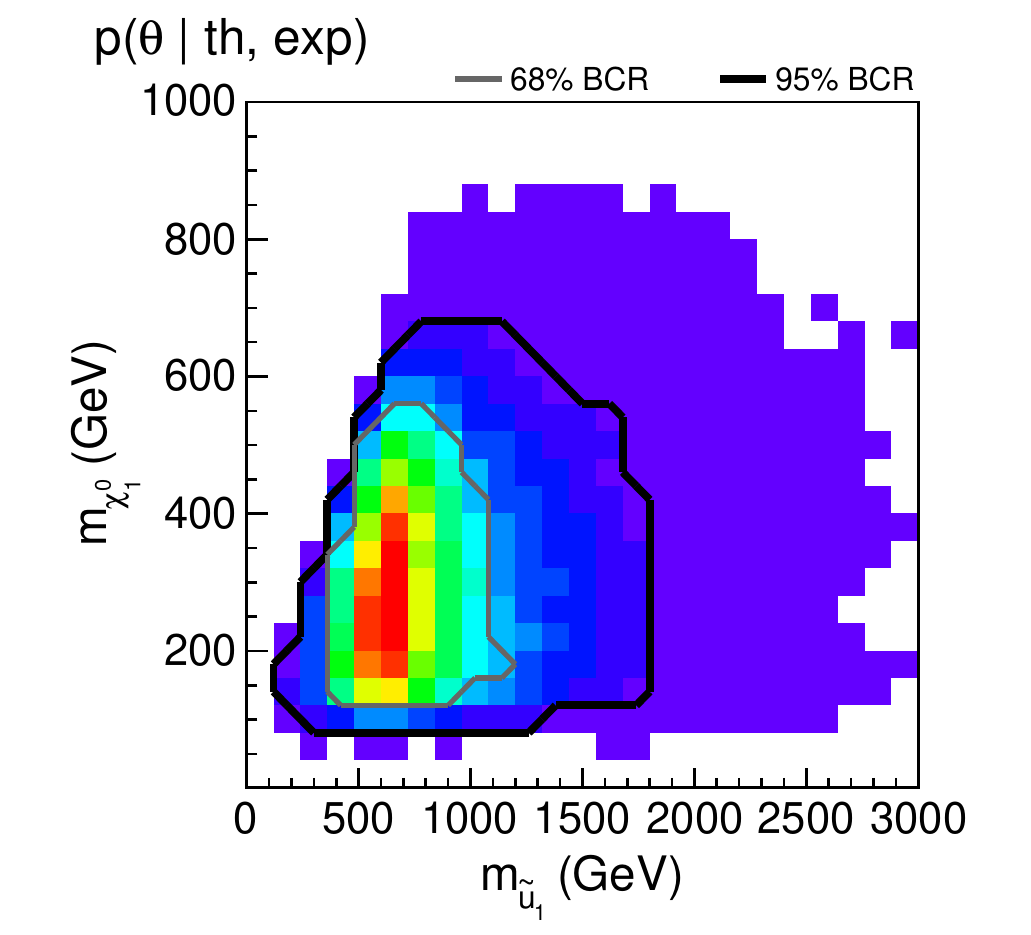}\qquad
		\includegraphics[width=0.4\textwidth,clip=true,trim=270 0 0 0]{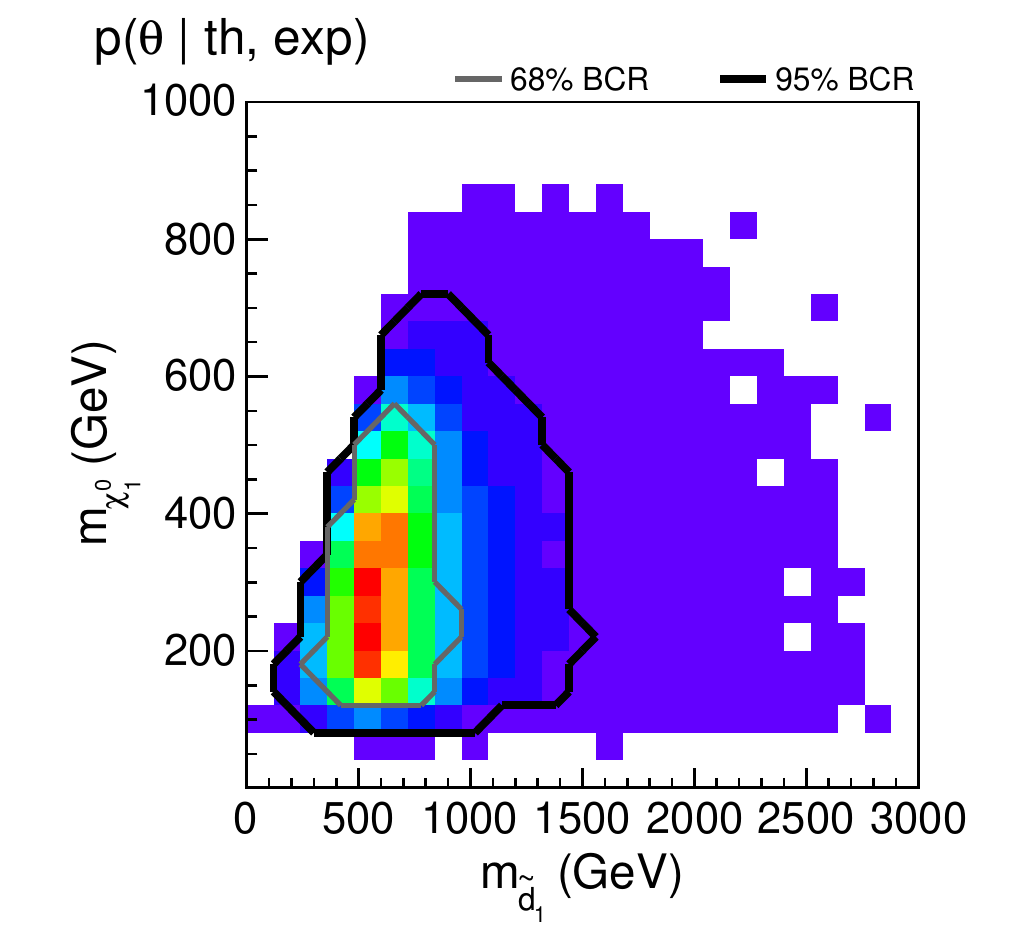}
	\end{center}
	\vspace*{-0.5cm}
	\caption{Correlations between the lightest neutralino mass and the lightest up-and down-type squark masses. Red colour indicates the highest likelihood.}
	\label{fig:massdiff}
\end{figure}

In Figure~\ref{fig:massdiff}, we show the correlations between the masses of the
lightest squark states and the one of the lightest neutralino. For most  ($\sim95\%$)
viable points, the mass difference is well above 50 GeV, which is a favorable
condition for collider searches as the spectrum is not compressed. A
considerable number ($\sim40\%$) of parameter points features $\tilde{u}_1$ masses of about
500 -- 1000~GeV together with neutralino masses of the order of 150 -- 400~GeV.
Such mass configurations are likely to be ruled out by Run I LHC data. This is
accounted for in the next section, where we include collider constraints on the
NMFV MSSM setup and define benchmark scenarios suitable for searches at the LHC
Run II.

\section{Benchmark scenarios \label{sec:Benchmark}}

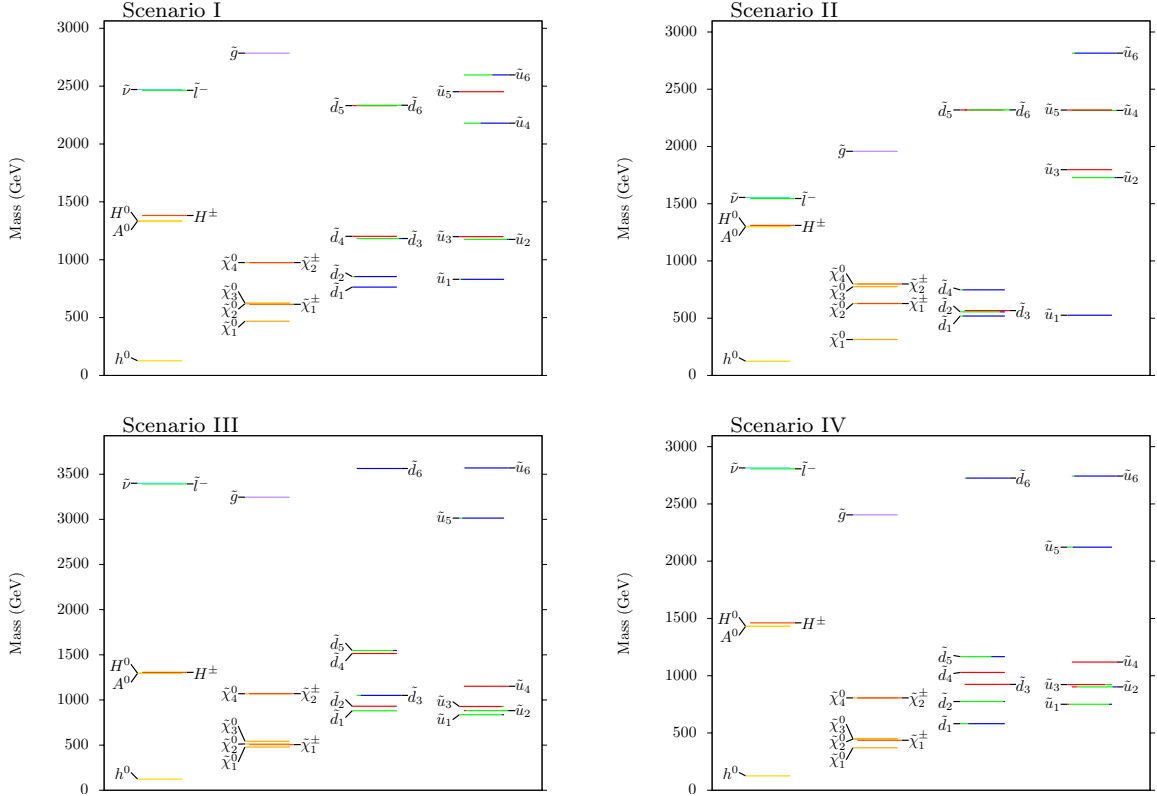
\begin{figure}
\vspace{.5cm}
\setlength{\unitlength}{1mm}
\begin{picture}(250,105)
\put(15,108){\scriptsize Scenario I}
\put(0,60){\includegraphics[width=0.47\textwidth]{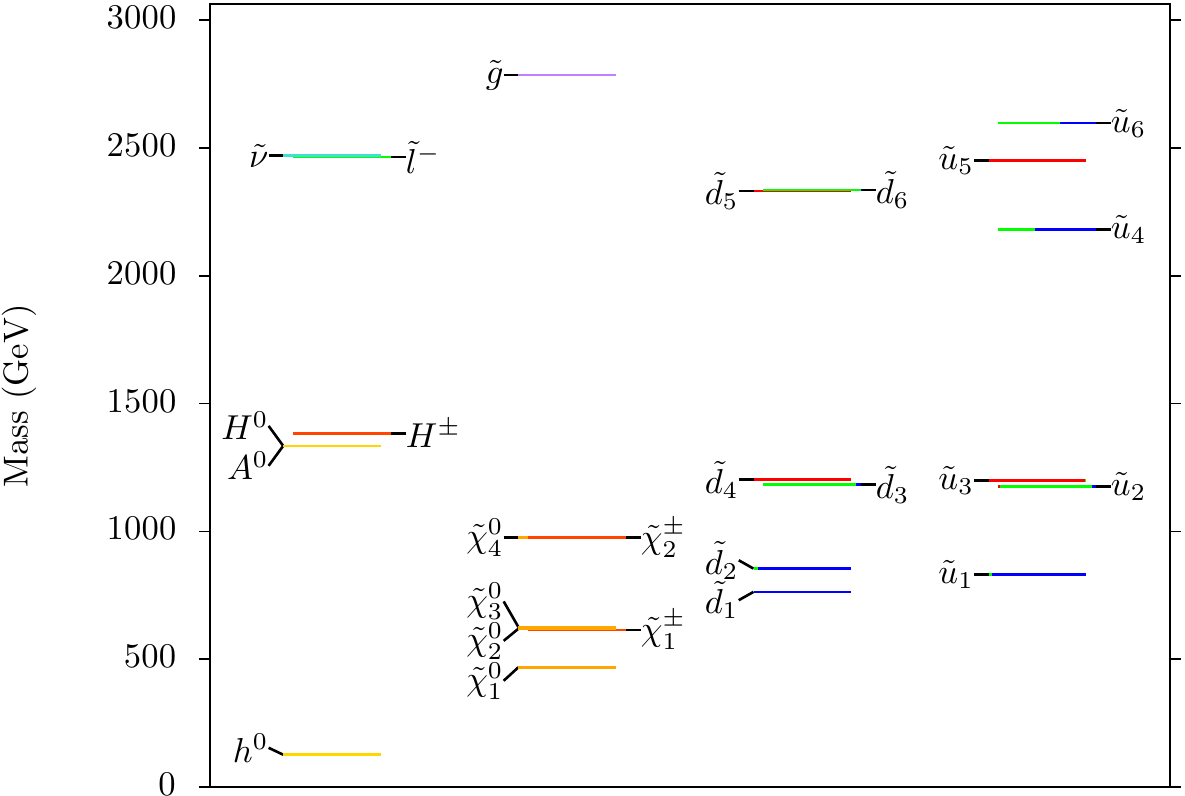}}
\put(95,108){\scriptsize Scenario II}
\put(80,60){\includegraphics[width=0.47\textwidth]{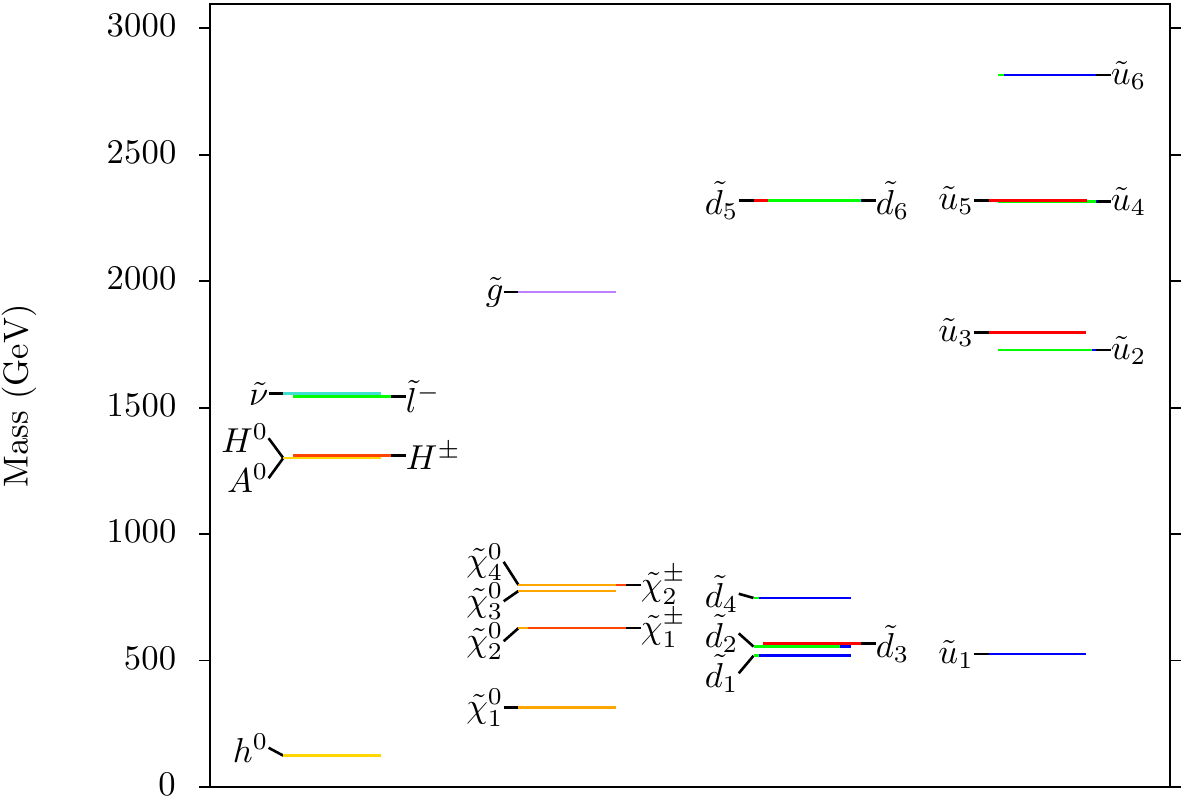}}
\put(15,53){\scriptsize Scenario III}
\put(0,5){\includegraphics[width=0.47\textwidth]{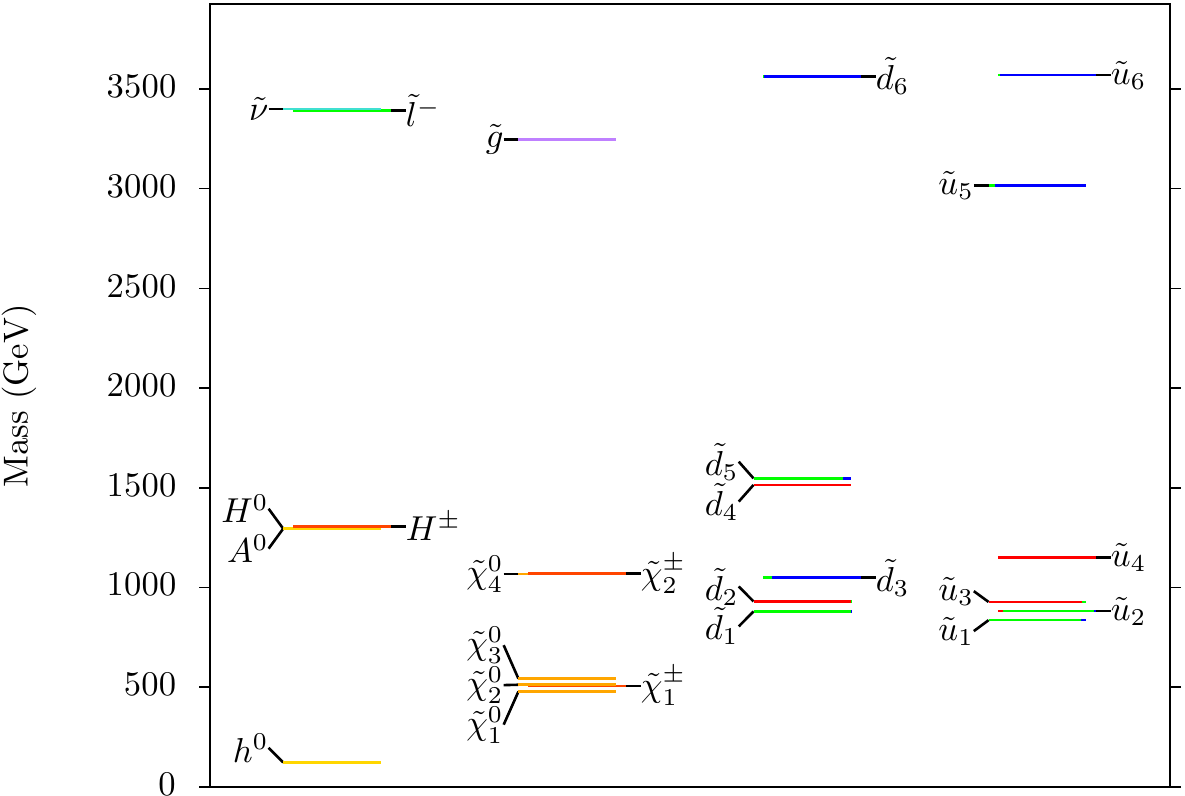}}
\put(95,53){\scriptsize Scenario IV}
\put(80,5){\includegraphics[width=0.47\textwidth]{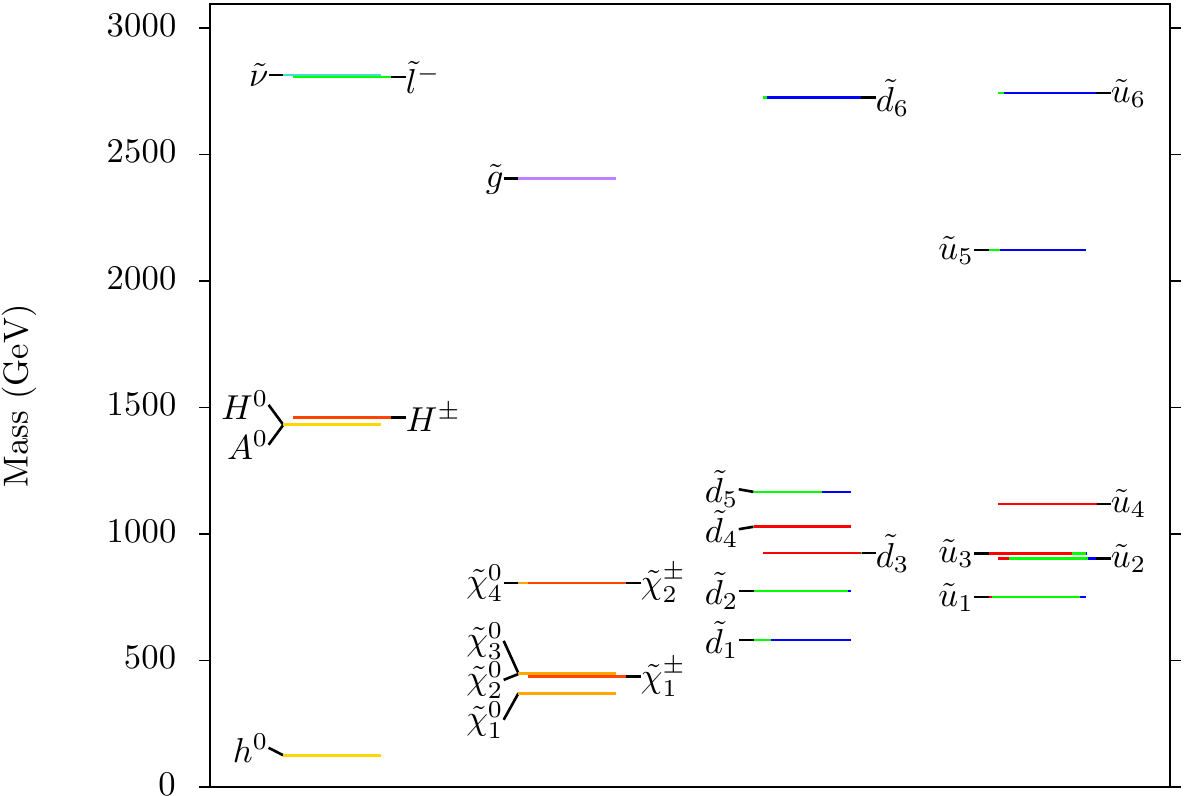}}
\end{picture}
\vspace*{-7mm}
  \caption{Mass spectra of the benchmark scenarios defined in
    Table~\ref{tab:benchmark}. The colour code that has been employed for
    depicting the squark eigenstates indicates their flavour content: the
    \{red, green, blue\} colour corresponds to the \{first, second, third\}
    generation flavours.}
    \label{fig:bmspectra}
\end{figure}

In this section, we identify an ensemble of benchmark scenarios capturing
typical features of the parameter space regions favoured by the constraints
previously investigated. We have ordered all acceptable parameter setups
according to their likelihood and selected four scenarios among the best ones.
Our selection is aimed to cover different phenomenological properties of the
NMFV MSSM and to be relevant for future LHC searches. The input parameters
corresponding to the benchmark scenarios of our choice are indicated in
Table~\ref{tab:benchmark}. In Figure~\ref{fig:bmspectra}, we present the mass
spectra of the four selected scenarios and depict the flavour content of the
different squark eigenstates. We finally show in Table~\ref{tab:benchbranch} the
branching ratios related to the dominant decay modes of the squarks ligther than
about 1~TeV. Additionally, we have verified that the electroweak vacuum is
stable for all selected points by using the programme
{\tt Vevacious}~\cite{Camargo-Molina:2013qva}. We now briefly outline the main
characteristics of the four proposed benchmark scenarios\footnote{The benchmark
scenarios can be provided, under the form of Supersymmetry Les Houches Accord
compliant files~\cite{Allanach:2008qq}, by the authors upon request.}.

\renewcommand{\arraystretch}{1.2}
\begin{table}
	\begin{center}
	\begin{tabular}{c||c|c|c|c}
		Parameter & ~~~I~~~ & ~~II~~ & ~III~ & ~~IV~~ \\
		\hline \hline
		$\alpha_s(m_Z)$  & $1.187\cdot 10^{-1}$ & $1.194\cdot 10^{-1}$ & $1.176\cdot 10^{-1}$ & $1.176\cdot 10^{-1}$ \\
		$m_t^{\rm pole}$ & 176.00~GeV & 175.53~GeV & 173.53~GeV & 174.02~GeV \\
		$m_b(m_b)$       & 4.10~GeV & 4.24~GeV & 4.28~GeV & 4.10~GeV  \\
		\hline \hline
		$M_{\tilde{Q}_{1,2}}$ & 1192.7~GeV & 2288.2~GeV &  637.7~GeV &  753.2~GeV \\
		$M_{\tilde{Q}_{3}}$   &  883.7~GeV &  425.3~GeV & 3483.0~GeV & 2662.7~GeV \\
		$M_{\tilde{U}_{1,2}}$ & 2412.6~GeV & 1757.7~GeV &  934.0~GeV &  984.7~GeV \\
		$M_{\tilde{U}_{3}}$   & 2344.3~GeV & 2753.8~GeV & 2862.2~GeV & 2010.6~GeV \\
		$M_{\tilde{D}_{1,2}}$ & 2295.1~GeV &  551.6~GeV & 1331.1~GeV &  882.7~GeV \\
		$M_{\tilde{D}_{3}}$   &  843.8~GeV &  713.5~GeV &  901.8~GeV &  670.5~GeV \\
		\hline \hline
		$A_f$ & -2424.1~GeV & 1807.3~GeV & 1586.3~GeV & -2833.4~GeV \\
		\hline \hline
		$\tan\beta$ & 17.4 & 21.1 & 29.2 & 34.0 \\
		$\mu$ &  615.7~GeV & 772.8~GeV& 508.1~GeV & 442.7~GeV \\
		$m_A$ & 1334.5~GeV & 1300.3~GeV & 1294.8~GeV & 1431.0~GeV \\
		$M_1$ &  474.5~GeV & 315.3~GeV& 525.2~GeV & 390.0~GeV \\
		$M_{\tilde{\ell}}$ & 2466.5~GeV & 1552.5~GeV & 3396.7~GeV & 2813.4~GeV \\
		\hline \hline
		$\delta_{LL}$   &  $1.4 \cdot 10^{-1}$ & $-4.6 \cdot 10^{-2}$ &  $3.7 \cdot 10^{-1}$ &  $5.9 \cdot 10^{-1}$ \\
		$\delta^u_{RR}$ &  $1.7 \cdot 10^{-1}$ &  $2.2 \cdot 10^{-1}$ &  $7.3 \cdot 10^{-1}$ &  $6.0 \cdot 10^{-1}$ \\
		$\delta^d_{RR}$ &  $1.4 \cdot 10^{-1}$ & $-1.4 \cdot 10^{-1}$ & $-2.9 \cdot 10^{-1}$ & $-7.5 \cdot 10^{-1}$ \\
		$\delta^u_{LR}$ &  $9.2 \cdot 10^{-2}$ &  $3.5 \cdot 10^{-2}$ &  $1.7 \cdot 10^{-1}$ &  $1.0 \cdot 10^{-1}$ \\
		$\delta^d_{LR}$ & $-3.9 \cdot 10^{-2}$ & $-3.6 \cdot 10^{-3}$ & $-6.1 \cdot 10^{-3}$ &  $4.6 \cdot 10^{-3}$ \\
		$\delta^u_{RL}$ & $-9.7 \cdot 10^{-2}$ &  $1.4 \cdot 10^{-2}$ & $-9.9 \cdot 10^{-2}$ & $-7.2 \cdot 10^{-2}$ \\
		$\delta^d_{RL}$ & $-7.6 \cdot 10^{-4}$ & $-1.4 \cdot 10^{-2}$ & $-1.2 \cdot 10^{-3}$ &  $1.6 \cdot 10^{-3}$ \\
	\end{tabular}	
	\end{center}
  \caption{Definition of four benchmark points suitable for phenomenological
    studies of the NMFV MSSM.}
	\label{tab:benchmark}
\end{table}

\begin{table}[!t]
\renewcommand{\arraystretch}{0.94}
	\begin{center}
	\begin{tabular}{c||c|c|c|c}
		Decay & I & II & III & IV \\
		\hline\hline
		$\tilde{u}_1 \to t \tilde{\chi}^0_1$ & 0.14 & 0.99 & 0.06 & 0.09 \\
		$\tilde{u}_1 \to c \tilde{\chi}^0_1$ &      & 0.01 & 0.24 & 0.07 \\
		$\tilde{u}_1 \to t \tilde{\chi}^0_2$ & 0.24 &      & 0.11 &      \\
		$\tilde{u}_1 \to t \tilde{\chi}^0_3$ & 0.43 &      & 0.06 & 0.26 \\
		$\tilde{u}_1 \to t \tilde{\chi}^0_3$ &      &      &      & 0.07 \\
		$\tilde{u}_1 \to c \tilde{\chi}^0_3$ &      &      & 0.27 & 0.10 \\
		$\tilde{u}_1 \to b \tilde{\chi}^+_1$ & 0.19 &      & 0.24 & 0.22 \\
		$\tilde{u}_1 \to s \tilde{\chi}^+_1$ &      &      & 0.02 & 0.17 \\
		$\tilde{u}_1 \to W^+ \tilde{d}_1$    &      &      &      & 0.02 \\
		\hline\hline                                  
		$\tilde{u}_2 \to t \tilde{\chi}^0_1$ &      &      & 0.07 & 0.04 \\
		$\tilde{u}_2 \to c \tilde{\chi}^0_1$ &      &      & 0.22 & 0.32 \\
		$\tilde{u}_2 \to t \tilde{\chi}^0_2$ &      &      & 0.08 & 0.09 \\
		$\tilde{u}_2 \to t \tilde{\chi}^0_3$ &      &      & 0.09 & 0.11 \\
		$\tilde{u}_2 \to c \tilde{\chi}^0_3$ &      &      & 0.35 & 0.06 \\
		$\tilde{u}_2 \to b \tilde{\chi}^+_1$ &      &      & 0.11 & 0.21 \\
		$\tilde{u}_2 \to s \tilde{\chi}^+_1$ &      &      & 0.09 &      \\
		$\tilde{u}_2 \to W^+ \tilde{d}_2$    &      &      &      & 0.06 \\
		$\tilde{u}_2 \to Z^0 \tilde{u}_1$    &      &      &      & 0.05 \\
		$\tilde{u}_2 \to h^0 \tilde{u}_1$    &      &      &      & 0.02 \\
		\hline\hline
		$\tilde{u}_3 \to c \tilde{\chi}^0_1$ &      &      & 0.03 & 0.09 \\         
		$\tilde{u}_3 \to u \tilde{\chi}^0_1$ &      &      & 0.03 & 0.03 \\         
		$\tilde{u}_3 \to t \tilde{\chi}^0_2$ &      &      &      & 0.02 \\         
		$\tilde{u}_3 \to t \tilde{\chi}^0_3$ &      &      &      & 0.02 \\         
		$\tilde{u}_3 \to c \tilde{\chi}^0_3$ &      &      & 0.05 & 0.02 \\         
		$\tilde{u}_3 \to u \tilde{\chi}^0_3$ &      &      & 0.45 & 0.09 \\         
		$\tilde{u}_3 \to c \tilde{\chi}^0_4$ &      &      &      & 0.01 \\         
		$\tilde{u}_3 \to u \tilde{\chi}^0_4$ &      &      &      & 0.15 \\         
		$\tilde{u}_3 \to b \tilde{\chi}^+_1$ &      &      & 0.01 & 0.05 \\         
		$\tilde{u}_3 \to d \tilde{\chi}^+_1$ &      &      & 0.41 & 0.01 \\         
		$\tilde{u}_3 \to d \tilde{\chi}^+_2$ &      &      &      & 0.33 \\         
		$\tilde{u}_3 \to W^+ \tilde{d}_2$    &      &      &      & 0.01 \\
		$\tilde{u}_3 \to Z^0 \tilde{u}_1$    &      &      &      & 0.02 
		\vspace*{3.0cm}
	\end{tabular}
	\qquad
	\begin{tabular}{c||c|c|c|c}
		Decay & I & II & III & IV \\
		\hline\hline
		$\tilde{d}_1 \to b \tilde{\chi}^0_1$ & 0.73 & 0.84 &      & 0.35 \\
		$\tilde{d}_1 \to s \tilde{\chi}^0_1$ &      & 0.16 & 0.40 & 0.09 \\
		$\tilde{d}_1 \to b \tilde{\chi}^0_2$ & 0.15 &      & 0.01 & 0.30 \\
		$\tilde{d}_1 \to s \tilde{\chi}^0_2$ &      &      & 0.02 &      \\
		$\tilde{d}_1 \to b \tilde{\chi}^0_3$ & 0.12 &      &      & 0.25 \\
		$\tilde{d}_1 \to s \tilde{\chi}^0_3$ &      &      & 0.05 &      \\
		$\tilde{d}_1 \to t \tilde{\chi}^-_1$ &      &      & 0.42 &      \\
		$\tilde{d}_1 \to c \tilde{\chi}^-_1$ &      &      & 0.09 &      \\
		\hline\hline
		$\tilde{d}_2 \to b \tilde{\chi}^0_1$ & 0.04 & 0.05 &      & 0.02 \\
		$\tilde{d}_2 \to s \tilde{\chi}^0_1$ &      &      &      & 0.24 \\
		$\tilde{d}_2 \to d \tilde{\chi}^0_1$ &      &      & 0.70 & \\
		$\tilde{d}_2 \to b \tilde{\chi}^0_2$ &      &      &      & 0.04 \\
		$\tilde{d}_2 \to s \tilde{\chi}^0_2$ & 0.04 & 0.95 &      & \\
		$\tilde{d}_2 \to d \tilde{\chi}^0_2$ &      &      & 0.03 & \\
		$\tilde{d}_2 \to d \tilde{\chi}^0_2$ &      &      & 0.08 & \\
		$\tilde{d}_2 \to b \tilde{\chi}^0_3$ &      &      &      & 0.04 \\
		$\tilde{d}_2 \to s \tilde{\chi}^0_3$ & 0.04 &      &      & 0.02 \\
		$\tilde{d}_2 \to t \tilde{\chi}^-_1$ & 0.87 &      &      & 0.51 \\
		$\tilde{d}_2 \to c \tilde{\chi}^-_1$ &      &      &      & 0.09 \\
		$\tilde{d}_2 \to u \tilde{\chi}^-_1$ &      &      & 0.15 & \\
		$\tilde{d}_2 \to W^- \tilde{u}_1$    &      &      & 0.03 & \\
		$\tilde{d}_2 \to Z^0 \tilde{d}_1$    &      &      &      & 0.02 \\
		$\tilde{d}_2 \to h^0 \tilde{d}_1$    &      &      &      & 0.02 \\
		\hline\hline
		$\tilde{d}_3 \to d \tilde{\chi}^0_1$ &      & 1.00 &      & 0.16 \\
		$\tilde{d}_3 \to d \tilde{\chi}^0_3$ &      &      &      & 0.01 \\
		$\tilde{d}_3 \to d \tilde{\chi}^0_4$ &      &      &      & 0.25 \\
		$\tilde{d}_3 \to u \tilde{\chi}^-_1$ &      &      &      & 0.07 \\
		$\tilde{d}_3 \to c \tilde{\chi}^-_2$ &      &      &      & 0.02 \\
		$\tilde{d}_3 \to u \tilde{\chi}^-_2$ &      &      &      & 0.48 \\
		\hline\hline
		$\tilde{d}_4 \to b \tilde{\chi}^0_1$ &      & 0.42 &      & \\
		$\tilde{d}_4 \to s \tilde{\chi}^0_1$ &      & 0.03 &      & \\
		$\tilde{d}_4 \to d \tilde{\chi}^0_1$ &      &      &      & 0.83 \\
		$\tilde{d}_4 \to d \tilde{\chi}^0_3$ &      &      &      & 0.17 \\
		$\tilde{d}_4 \to W^- \tilde{u}_1$    &      & 0.27 &      & \\
		$\tilde{d}_4 \to Z^0 \tilde{d}_1$    &      & 0.13 &      & \\
		$\tilde{d}_4 \to h^0 \tilde{d}_1$    &      & 0.13 &      & \\
		$\tilde{d}_4 \to h^0 \tilde{d}_2$    &      & 0.01 &      & 
	\end{tabular}
	\end{center}
  \caption{Branching ratios associated with the dominant decay modes of the
    up-type (left) and down-type (right) squarks lighter than about 1~TeV for
    the benchmark points defined in Table~\ref{tab:benchmark}. Branching ratios
    below 1\% are not indicated.}
   \label{tab:benchbranch}
\end{table}

\paragraph{Scenario I} This benchmark point presents one up-type and two
down-type squarks with masses below 1~TeV. The lightest up-type squark
is mostly stop-like, although it contains a small scharm component, and has a
mass of 831~GeV which implies a sizeable production cross
section at the LHC. In the sector of the down-type squarks, the two lightest
states are almost purely sbottom-like with masses of 763~GeV and 854~GeV
respectively. Since only the heaviest neutralino and chargino states are heavier
than the three lightest squark states, various decay channels are open so that
the real challenge for future LHC analyses would be to become sensitive to
flavour-violating branching ratios of a few
percents. In this scenario, the electroweak vacuum is long-lived and
has a lifetime larger than the age of the Universe.

\paragraph{Scenario II} In this scenario, only the lightest of the up-type
squarks is expected to lie within the reach of LHC, with a mass of
526~GeV. It is almost a pure stop state with a small charm component. Since
the only lighter superpartner is the lightest neutralino, it will preferably
decay into a $\tilde{\chi}^0_1 t$ system. There are four down-type squarks lying
below 1~TeV, their masses being 519~GeV, 555~GeV, 566~GeV and 747~GeV. These
four states are admixtures of all three flavours and their dominant decay modes
include in particular final states containing the next-to-lightest neutralino
or the lightest chargino. Contrary to the scenario I, the branching ratios
related to flavour-violating decays can reach up to 16~percents, which make them
possibly testable at the LHC. Moreover, the fourth
down-type squark has sizeable branching ratios for decays into the lightest
up-type squark and a $W$-boson as well as into the lightest down-type squark
and either a $Z$-boson or a Higgs boson. Although many squarks are very light,
this scenario evades all LHC Run~I constraints thanks to a heavy
lightest neutralino of 315~GeV. Moreover, the vacuum has been
found to be stable.

\paragraph{Scenario III} This benchmark point features numerous squark mass
eigenstates in the reach of the LHC. It indeed exhibits three up-type squarks
with masses of 836~GeV, 882~GeV and 928~GeV, and three down-type squarks with
masses of 880~GeV, 931~GeV and 1050~GeV. The stop-like states are here the
heaviest ones, and the up-type squark states reachable at LHC only contain up
and charm flavours. Similarly, the heavier $\tilde{d}_3$ and $\tilde{d}_6$
down-type squarks are the only ones containing a sbottom component. This feature
is the direct
consequence of the lower values that are favoured for the $M_{\tilde{Q}_{1,2}}$,
$M_{\tilde{U}_{1,2}}$ and $M_{\tilde{D}_{1,2}}$ parameters, when compared to the
values that are favoured by the third-generation soft parameters $M_{\tilde{Q}_3}$,
$M_{\tilde{U}_3}$ and $M_{\tilde{D}_3}$. Equivalently, this can be seen as an
implication of allowing for flavour-violating entries in the squark mass matrices
(see Section~\ref{sec:ResultsFC}). In addition, all gauginos except the heaviest
neutralino and chargino feature lower masses, so that a variety of decay
channels are open. 
Finally, we found a direction in which, for extremely large field excitations,
the electroweak vacuum is unbounded from below. This
situation is similar to the case of the Standard Model~\cite{Degrassi:2012ry,%
Buttazzo:2013uya}, and higher order corrections to the scalar potential would be
needed to make a conclusive statement.

\paragraph{Scenario IV} Our last scenario features numerous squark states as
well as a complete electroweakino spectrum below 1~TeV. More precisely, the
lighter up-type squarks have masses of 751~GeV, 902~GeV and 923~GeV, while the
lighter down-type squark masses are of 582~GeV, 775~GeV and 924~GeV. In
addition, three other states are not too far above 1~TeV with masses of 1119~GeV
($\tilde{u}_4$), 1029~GeV ($\tilde{d}_4$) and 1167~GeV ($\tilde{d}_5$). The two
lightest up-type squarks consist in this case of a mixture of all three flavours
(with a dominant charm content), which leads to interesting decay patterns, as
shown in Table~\ref{tab:benchbranch}.
Similarly to scenario I, this scenario exhibits a long-lived electroweak
vacuum with a lifetime larger than the age of the Universe.

\section{Conclusion \label{sec:Conclusion}}

We have studied non-minimal flavour-violation in the MSSM by allowing
for flavour mixing between the second and third generation squarks. We have
used a Markov Chain Monte Carlo scanning technique to explore the underlying
parameter space and imposed a set of experimental constraints arising from
$B$-meson and kaon physics. We have additionally enforced the model to
accommodate a light Higgs boson with a mass of 125~GeV.

First and second generation soft squark masses are theoretically restricted
to low values in order to avoid tachyons in the Higgs sector.
As a consequence,
the lighter squarks are often not the stop and sbottom ones, which contrasts
with scenarios of the usual minimally flavour-violating MSSM.
Requiring a theoretically consistent Higgs sector and a light Higgs boson
of about 125~GeV similarly restrict the left-right
and right-left flavour-violating squark mixing parameters
$\delta_{LR/RL}^{u,d}$ to be small.
In contrast, the $\delta_{LL}$ and $\delta_{RR}^d$ NMFV parameters
are mainly constrained by neutral $B$-meson oscillations, and the rare
$B_s \rightarrow \mu \mu$ decay mainly influences $\delta_{RL}^u$.
All other NMFV parameters are left unconstrained by the considered experimental observations.

In view of the recently started second LHC run, we have used our MCMC scan results to propose four benchmark
scenarios allowed by current data that exhibit distinct features and that are
suitable for future analyses of NMFV effects in the MSSM. In most proposed
scenarios, several squarks have masses close to 1~TeV so that they should be
reachable within the next few years.

\acknowledgments
The authors would like to thank the organisers of the `PhysTeV Les Houches'
workshop, where this work has been initiated, for the welcoming and inspiring
atmosphere. We are also grateful to the IIHE IT team for allowing us to use the IIHE
cluster. KDC and NS are supported in part by `FWO-Vlaanderen' aspirant fellowships,
and KDC also thanks the Strategic Research Program `High Energy
Physics' of the Vrije Universiteit Brussel. BF and FM acknowledge partial support
from the Theory-LHC France initiative of the CNRS (IN2P3/INP) and
WP has been supported by the BMBF, project nr.~05H12WWE.

\clearpage

\bibliographystyle{JHEP}
\bibliography{paper}

\providecommand{\href}[2]{#2}\begingroup\raggedright\begin{thebibliography}{10}

\bibitem{Nilles:1983ge}
H.~P. Nilles, {\it {Supersymmetry, Supergravity and Particle Physics}},  {\em
  Phys.Rept.} {\bf 110} (1984) 1--162.

\bibitem{Haber:1984rc}
H.~E. Haber and G.~L. Kane, {\it {The Search for Supersymmetry: Probing Physics
  Beyond the Standard Model}},  {\em Phys.Rept.} {\bf 117} (1985) 75--263.

\bibitem{AtlasSusy}
 {\em {http://twiki.cern.ch/twiki/AtlasPublic/SupersymmetryPublicResults}}.

\bibitem{CmsSusy}
 {\em {http://twiki.cern.ch/twiki/CMSPublic/PhysicsResultsSUS}}.

\bibitem{Hall:1985dx}
L.~J. Hall, V.~A. Kostelecky, and S.~Raby, {\it {New Flavor Violations in
  Supergravity Models}},  {\em Nucl.Phys.} {\bf B267} (1986) 415.

\bibitem{D'Ambrosio:2002ex}
G.~D'Ambrosio, G.~Giudice, G.~Isidori, and A.~Strumia, {\it {Minimal flavor
  violation: An Effective field theory approach}},  {\em Nucl.Phys.} {\bf B645}
  (2002) 155--187, [\href{http://xxx.lanl.gov/abs/hep-ph/0207036}{{\tt
  hep-ph/0207036}}].

\bibitem{Cirigliano:2005ck}
V.~Cirigliano, B.~Grinstein, G.~Isidori, and M.~B. Wise, {\it {Minimal flavor
  violation in the lepton sector}},  {\em Nucl.Phys.} {\bf B728} (2005)
  121--134, [\href{http://xxx.lanl.gov/abs/hep-ph/0507001}{{\tt
  hep-ph/0507001}}].

\bibitem{Gabbiani:1988rb}
F.~Gabbiani and A.~Masiero, {\it {FCNC in Generalized Supersymmetric
  Theories}},  {\em Nucl.Phys.} {\bf B322} (1989) 235.

\bibitem{Buchalla:2008jp}
M.~Artuso, D.~Asner, P.~Ball, E.~Baracchini, G.~Bell, et~al., {\it {$B$, $D$
  and $K$ decays}},  {\em Eur.Phys.J.} {\bf C57} (2008) 309--492,
  [\href{http://xxx.lanl.gov/abs/0801.1833}{{\tt arXiv:0801.1833}}].

\bibitem{Heinemeyer:2004by}
S.~Heinemeyer, W.~Hollik, F.~Merz, and S.~Penaranda, {\it {Electroweak
  precision observables in the MSSM with nonminimal flavor violation}},  {\em
  Eur.Phys.J.} {\bf C37} (2004) 481--493,
  [\href{http://xxx.lanl.gov/abs/hep-ph/0403228}{{\tt hep-ph/0403228}}].

\bibitem{Bozzi:2007me}
G.~Bozzi, B.~Fuks, B.~Herrmann, and M.~Klasen, {\it {Squark and gaugino
  hadroproduction and decays in non-minimal flavour violating supersymmetry}},
  {\em Nucl.Phys.} {\bf B787} (2007) 1--54,
  [\href{http://xxx.lanl.gov/abs/0704.1826}{{\tt arXiv:0704.1826}}].

\bibitem{Dittmaier:2007uw}
S.~Dittmaier, G.~Hiller, T.~Plehn, and M.~Spannowsky, {\it {Charged-Higgs
  Collider Signals with or without Flavor}},  {\em Phys. Rev.} {\bf D77} (2008)
  115001, [\href{http://xxx.lanl.gov/abs/0708.0940}{{\tt arXiv:0708.0940}}].

\bibitem{Fuks:2008ab}
B.~Fuks, B.~Herrmann, and M.~Klasen, {\it {Flavour Violation in Gauge-Mediated
  Supersymmetry Breaking Models: Experimental Constraints and Phenomenology at
  the LHC}},  {\em Nucl.Phys.} {\bf B810} (2009) 266--299,
  [\href{http://xxx.lanl.gov/abs/0808.1104}{{\tt arXiv:0808.1104}}].

\bibitem{Hurth:2009ke}
T.~Hurth and W.~Porod, {\it {Flavour violating squark and gluino decays}},
  {\em JHEP} {\bf 0908} (2009) 087,
  [\href{http://xxx.lanl.gov/abs/0904.4574}{{\tt arXiv:0904.4574}}].

\bibitem{Bartl:2009au}
A.~Bartl et~al., {\it {Impact of squark generation mixing on the search for
  gluinos at LHC}},  {\em Phys.Lett.} {\bf B679} (2009) 260--266,
  [\href{http://xxx.lanl.gov/abs/0905.0132}{{\tt arXiv:0905.0132}}].

\bibitem{Plehn:2009it}
T.~Plehn, M.~Rauch, and M.~Spannowsky, {\it {Understanding Single Tops using
  Jets}},  {\em Phys. Rev.} {\bf D80} (2009) 114027,
  [\href{http://xxx.lanl.gov/abs/0906.1803}{{\tt arXiv:0906.1803}}].

\bibitem{Bartl:2010du}
A.~Bartl et~al., {\it {Impact of squark generation mixing on the search for
  squarks decaying into fermions at LHC}},  {\em Phys.Lett.} {\bf B698} (2011)
  380--388, [\href{http://xxx.lanl.gov/abs/1007.5483}{{\tt arXiv:1007.5483}}].

\bibitem{Bruhnke:2010rh}
M.~Bruhnke, B.~Herrmann, and W.~Porod, {\it {Signatures of bosonic squark
  decays in non-minimally flavour-violating supersymmetry}},  {\em JHEP} {\bf
  1009} (2010) 006, [\href{http://xxx.lanl.gov/abs/1007.2100}{{\tt
  arXiv:1007.2100}}].

\bibitem{Bartl:2011wq}
A.~Bartl et~al., {\it {Flavour violating gluino three-body decays at LHC}},
  {\em Phys.Rev.} {\bf D84} (2011) 115026,
  [\href{http://xxx.lanl.gov/abs/1107.2775}{{\tt arXiv:1107.2775}}].

\bibitem{Fuks:2011dg}
B.~Fuks, B.~Herrmann, and M.~Klasen, {\it {Phenomenology of anomaly-mediated
  supersymmetry breaking scenarios with non-minimal flavour violation}},  {\em
  Phys.Rev.} {\bf D86} (2012) 015002,
  [\href{http://xxx.lanl.gov/abs/1112.4838}{{\tt arXiv:1112.4838}}].

\bibitem{Bartl:2012tx}
A.~Bartl, H.~Eberl, E.~Ginina, B.~Herrmann, K.~Hidaka, et~al., {\it {Flavor
  violating bosonic squark decays at LHC}},  {\em Int.J.Mod.Phys.} {\bf A29}
  (2014) 1450035, [\href{http://xxx.lanl.gov/abs/1212.4688}{{\tt
  arXiv:1212.4688}}].

\bibitem{Blanke:2013uia}
M.~Blanke, G.~F. Giudice, P.~Paradisi, G.~Perez, and J.~Zupan, {\it {Flavoured
  Naturalness}},  {\em JHEP} {\bf 1306} (2013) 022,
  [\href{http://xxx.lanl.gov/abs/1302.7232}{{\tt arXiv:1302.7232}}].

\bibitem{Aebischer:2014lfa}
J.~Aebischer, A.~Crivellin, and C.~Greub, {\it {One-loop SQCD corrections to
  the decay of top squarks to charm and neutralino in the generic MSSM}},  {\em
  Phys. Rev.} {\bf D91} (2015), no.~3 035010,
  [\href{http://xxx.lanl.gov/abs/1410.8459}{{\tt arXiv:1410.8459}}].

\bibitem{Backovic:2015rwa}
M.~Backović, A.~Mariotti, and M.~Spannowsky, {\it {Signs of Tops from Highly
  Mixed Stops}},  {\em JHEP} {\bf 06} (2015) 122,
  [\href{http://xxx.lanl.gov/abs/1504.0092}{{\tt arXiv:1504.0092}}].

\bibitem{Arana-Catania:2014ooa}
M.~Arana-Catania, S.~Heinemeyer, and M.~Herrero, {\it {Updated Constraints on
  General Squark Flavor Mixing}},  {\em Phys.Rev.} {\bf D90} (2014), no.~7
  075003, [\href{http://xxx.lanl.gov/abs/1405.6960}{{\tt arXiv:1405.6960}}].

\bibitem{Kowalska:2014opa}
K.~Kowalska, {\it {Phenomenology of SUSY with General Flavour Violation}},
  {\em JHEP} {\bf 1409} (2014) 139,
  [\href{http://xxx.lanl.gov/abs/1406.0710}{{\tt arXiv:1406.0710}}].

\bibitem{Ciuchini:2007ha}
M.~Ciuchini, A.~Masiero, P.~Paradisi, L.~Silvestrini, S.~Vempati, et~al., {\it
  {Soft SUSY breaking grand unification: Leptons versus quarks on the flavor
  playground}},  {\em Nucl.Phys.} {\bf B783} (2007) 112--142,
  [\href{http://xxx.lanl.gov/abs/hep-ph/0702144}{{\tt hep-ph/0702144}}].

\bibitem{Gabbiani:1996hi}
F.~Gabbiani, E.~Gabrielli, A.~Masiero, and L.~Silvestrini, {\it {A Complete
  analysis of FCNC and CP constraints in general SUSY extensions of the
  standard model}},  {\em Nucl.Phys.} {\bf B477} (1996) 321--352,
  [\href{http://xxx.lanl.gov/abs/hep-ph/9604387}{{\tt hep-ph/9604387}}].

\bibitem{Beringer:1900zz}
{\bf Particle Data Group} Collaboration, J.~Beringer et~al., {\it {Review of
  Particle Physics (RPP)}},  {\em Phys.Rev.} {\bf D86} (2012) 010001.

\bibitem{Porod:2003um}
W.~Porod, {\it {SPheno, a program for calculating supersymmetric spectra, SUSY
  particle decays and SUSY particle production at $e^+ e^-$ colliders}},  {\em
  Comput.Phys.Commun.} {\bf 153} (2003) 275--315,
  [\href{http://xxx.lanl.gov/abs/hep-ph/0301101}{{\tt hep-ph/0301101}}].

\bibitem{Porod:2011nf}
W.~Porod and F.~Staub, {\it {SPheno 3.1: Extensions including flavour,
  CP-phases and models beyond the MSSM}},  {\em Comput.Phys.Commun.} {\bf 183}
  (2012) 2458--2469, [\href{http://xxx.lanl.gov/abs/1104.1573}{{\tt
  arXiv:1104.1573}}].

\bibitem{MCMC1}
A.~A. Markov, {\em {Extension of the limit theorems of probability theory to a
  sum of variables connected in a chain}}.
\newblock reprinted in Appendix B of: R. Howard, {\it Dynamic Probabilistic
  Systems, volume 1: Markov Chains}, John Wiley and Sons, 1971.

\bibitem{Metropolis:1953am}
N.~Metropolis, A.~W. Rosenbluth, M.~N. Rosenbluth, A.~H. Teller, and E.~Teller,
  {\it {Equation of state calculations by fast computing machines}},  {\em J.
  Chem. Phys.} {\bf 21} (1953) 1087--1092.

\bibitem{Hastings:1970aa}
W.~K. Hastings, {\it {Monte Carlo Sampling Methods Using Markov Chains and
  Their Applications}},  {\em Biometrika} {\bf 57} (1970) 97--109.

\bibitem{Amhis:2014hma}
{\bf Heavy Flavor Averaging Group (HFAG)} Collaboration, Y.~Amhis et~al., {\it
  {Averages of $b$-hadron, $c$-hadron, and $\tau$-lepton properties as of
  summer 2014}},  \href{http://xxx.lanl.gov/abs/1412.7515}{{\tt
  arXiv:1412.7515}}.

\bibitem{CMS:2014xfa}
{\bf LHCb, CMS} Collaboration, V.~Khachatryan et~al., {\it {Observation of the
  rare $B^0\_s\to\mu^+\mu^-$ decay from the combined analysis of CMS and LHCb
  data}},  {\em Nature} {\bf 522} (2015) 68--72,
  [\href{http://xxx.lanl.gov/abs/1411.4413}{{\tt arXiv:1411.4413}}].

\bibitem{Aaij:2013iag}
{\bf LHCb} Collaboration, R.~Aaij et~al., {\it {Differential branching fraction
  and angular analysis of the decay $B^{0} \to K^{*0} \mu^{+}\mu^{-}$}},  {\em
  JHEP} {\bf 1308} (2013) 131, [\href{http://xxx.lanl.gov/abs/1304.6325}{{\tt
  arXiv:1304.6325}}].

\bibitem{LHCb:2015dla}
{\bf LHCb} Collaboration, {\it {Angular analysis of the $B^{0} \rightarrow
  K^{*0} \mu^{+} \mu^{-}$ decay}},  LHCb-CONF-2015-002,
  CERN-LHCb-CONF-2015-002.

\bibitem{Lees:2013nxa}
{\bf BaBar} Collaboration, J.~Lees et~al., {\it {Measurement of the $B \to X_s
  \ell^+ \ell^+-$ branching fraction and search for direct CP violation from a
  sum of exclusive final states}},  {\em Phys.Rev.Lett.} {\bf 112} (2014)
  211802, [\href{http://xxx.lanl.gov/abs/1312.5364}{{\tt arXiv:1312.5364}}].

\bibitem{Aad:2012tfa}
{\bf ATLAS} Collaboration, G.~Aad et~al., {\it {Observation of a new particle
  in the search for the Standard Model Higgs boson with the ATLAS detector at
  the LHC}},  {\em Phys.Lett.} {\bf B716} (2012) 1--29,
  [\href{http://xxx.lanl.gov/abs/1207.7214}{{\tt arXiv:1207.7214}}].

\bibitem{Chatrchyan:2012ufa}
{\bf CMS} Collaboration, S.~Chatrchyan et~al., {\it {Observation of a new boson
  at a mass of 125 GeV with the CMS experiment at the LHC}},  {\em Phys.Lett.}
  {\bf B716} (2012) 30--61, [\href{http://xxx.lanl.gov/abs/1207.7235}{{\tt
  arXiv:1207.7235}}].

\bibitem{Mahmoudi:2007vz}
F.~Mahmoudi, {\it {SuperIso: A Program for calculating the isospin asymmetry of
  B $\to$ K* gamma in the MSSM}},  {\em Comput.Phys.Commun.} {\bf 178} (2008)
  745--754, [\href{http://xxx.lanl.gov/abs/0710.2067}{{\tt arXiv:0710.2067}}].

\bibitem{Mahmoudi:2008tp}
F.~Mahmoudi, {\it {SuperIso v2.3: A Program for calculating flavor physics
  observables in Supersymmetry}},  {\em Comput.Phys.Commun.} {\bf 180} (2009)
  1579--1613, [\href{http://xxx.lanl.gov/abs/0808.3144}{{\tt
  arXiv:0808.3144}}].

\bibitem{Mahmoudi:2010iz}
F.~Mahmoudi, S.~Heinemeyer, A.~Arbey, A.~Bharucha, T.~Goto, et~al., {\it
  {Flavour Les Houches Accord: Interfacing Flavour related Codes}},  {\em
  Comput.Phys.Commun.} {\bf 183} (2012) 285--298,
  [\href{http://xxx.lanl.gov/abs/1008.0762}{{\tt arXiv:1008.0762}}].

\bibitem{Misiak:2006zs}
M.~Misiak, H.~Asatrian, K.~Bieri, M.~Czakon, A.~Czarnecki, et~al., {\it
  {Estimate of $B(\bar{B} \to X(s) \gamma)$ at $O(\alpha(s)^2)$}},  {\em
  Phys.Rev.Lett.} {\bf 98} (2007) 022002,
  [\href{http://xxx.lanl.gov/abs/hep-ph/0609232}{{\tt hep-ph/0609232}}].

\bibitem{Misiak:2006ab}
M.~Misiak and M.~Steinhauser, {\it {NNLO QCD corrections to the $\bar{B}\to X_s
  \gamma$ matrix elements using interpolation in m(c)}},  {\em Nucl.Phys.} {\bf
  B764} (2007) 62--82, [\href{http://xxx.lanl.gov/abs/hep-ph/0609241}{{\tt
  hep-ph/0609241}}].

\bibitem{Misiak:2010tk}
M.~Misiak and M.~Poradzinski, {\it {Completing the Calculation of BLM
  corrections to $\bar{B} \to X_s \gamma$}},  {\em Phys.Rev.} {\bf D83} (2011)
  014024, [\href{http://xxx.lanl.gov/abs/1009.5685}{{\tt arXiv:1009.5685}}].

\bibitem{Dai:1996vg}
Y.-B. Dai, C.-S. Huang, and H.-W. Huang, {\it {$B \to X_s \tau^+ \tau^-$ in a
  two Higgs doublet model}},  {\em Phys.Lett.} {\bf B390} (1997) 257--262,
  [\href{http://xxx.lanl.gov/abs/hep-ph/9607389}{{\tt hep-ph/9607389}}].

\bibitem{Ghinculov:2003qd}
A.~Ghinculov, T.~Hurth, G.~Isidori, and Y.~Yao, {\it {The Rare decay $B \to X_s
  l^+ l^-$ to NNLL precision for arbitrary dilepton invariant mass}},  {\em
  Nucl.Phys.} {\bf B685} (2004) 351--392,
  [\href{http://xxx.lanl.gov/abs/hep-ph/0312128}{{\tt hep-ph/0312128}}].

\bibitem{Huber:2005ig}
T.~Huber, E.~Lunghi, M.~Misiak, and D.~Wyler, {\it {Electromagnetic logarithms
  in $\bar{B} \to X_s l^+ l^-$}},  {\em Nucl.Phys.} {\bf B740} (2006) 105--137,
  [\href{http://xxx.lanl.gov/abs/hep-ph/0512066}{{\tt hep-ph/0512066}}].

\bibitem{Huber:2007vv}
T.~Huber, T.~Hurth, and E.~Lunghi, {\it {Logarithmically Enhanced Corrections
  to the Decay Rate and Forward Backward Asymmetry in $\bar{B} \to X_s \ell^+
  \ell^-$}},  {\em Nucl.Phys.} {\bf B802} (2008) 40--62,
  [\href{http://xxx.lanl.gov/abs/0712.3009}{{\tt arXiv:0712.3009}}].

\bibitem{Beneke:2001at}
M.~Beneke, T.~Feldmann, and D.~Seidel, {\it {Systematic approach to exclusive
  $B \to V l^+ l^-$, $V \gamma$ decays}},  {\em Nucl.Phys.} {\bf B612} (2001)
  25--58, [\href{http://xxx.lanl.gov/abs/hep-ph/0106067}{{\tt
  hep-ph/0106067}}].

\bibitem{Beneke:2004dp}
M.~Beneke, T.~Feldmann, and D.~Seidel, {\it {Exclusive radiative and
  electroweak $b \to d$ and $b \to s$ penguin decays at NLO}},  {\em
  Eur.Phys.J.} {\bf C41} (2005) 173--188,
  [\href{http://xxx.lanl.gov/abs/hep-ph/0412400}{{\tt hep-ph/0412400}}].

\bibitem{Kruger:2005ep}
F.~Kruger and J.~Matias, {\it {Probing new physics via the transverse
  amplitudes of $B^0 \to K^{*0} (\to K^- \pi^+) l^+l^-$ at large recoil}},
  {\em Phys.Rev.} {\bf D71} (2005) 094009,
  [\href{http://xxx.lanl.gov/abs/hep-ph/0502060}{{\tt hep-ph/0502060}}].

\bibitem{Egede:2008uy}
U.~Egede, T.~Hurth, J.~Matias, M.~Ramon, and W.~Reece, {\it {New observables in
  the decay mode $\bar{B_d} \to \bar{K^{*0}} l^+ l^-$}},  {\em JHEP} {\bf 0811}
  (2008) 032, [\href{http://xxx.lanl.gov/abs/0807.2589}{{\tt
  arXiv:0807.2589}}].

\bibitem{Egede:2010zc}
U.~Egede, T.~Hurth, J.~Matias, M.~Ramon, and W.~Reece, {\it {New physics reach
  of the decay mode $\bar{B} \to \bar{K}^{*0}\ell^+\ell^-$}},  {\em JHEP} {\bf
  1010} (2010) 056, [\href{http://xxx.lanl.gov/abs/1005.0571}{{\tt
  arXiv:1005.0571}}].

\bibitem{Khodjamirian:2010vf}
A.~Khodjamirian, T.~Mannel, A.~Pivovarov, and Y.-M. Wang, {\it {Charm-loop
  effect in $B \to K^{(*)} \ell^{+} \ell^{-}$ and $B\to K^*\gamma$}},  {\em
  JHEP} {\bf 1009} (2010) 089, [\href{http://xxx.lanl.gov/abs/1006.4945}{{\tt
  arXiv:1006.4945}}].

\bibitem{Dreiner:2012dh}
H.~Dreiner, K.~Nickel, W.~Porod, and F.~Staub, {\it {Full 1-loop calculation of
  BR$(B_{s,d}^0\to \ell \bar \ell)$ in models beyond the MSSM with SARAH and
  SPheno}},  {\em Comput.Phys.Commun.} {\bf 184} (2013) 2604--2617,
  [\href{http://xxx.lanl.gov/abs/1212.5074}{{\tt arXiv:1212.5074}}].

\bibitem{Buras:2002vd}
A.~J. Buras, P.~H. Chankowski, J.~Rosiek, and L.~Slawianowska, {\it {$\Delta
  M_{d,s}, B^0{d,s} \to \mu^{+} \mu^{-}$ and $B \to X_{s} \gamma$ in
  supersymmetry at large $\tan\beta$}},  {\em Nucl.Phys.} {\bf B659} (2003) 3,
  [\href{http://xxx.lanl.gov/abs/hep-ph/0210145}{{\tt hep-ph/0210145}}].

\bibitem{Isidori:2002qe}
G.~Isidori and A.~Retico, {\it {$B_{s,d} \to \ell^{+} \ell^{-}$ and $K_{L} \to
  \ell^{+}\ell^{-}$ in SUSY models with nonminimal sources of flavor mixing}},
  {\em JHEP} {\bf 0209} (2002) 063,
  [\href{http://xxx.lanl.gov/abs/hep-ph/0208159}{{\tt hep-ph/0208159}}].

\bibitem{Baek:2001kh}
S.~Baek, T.~Goto, Y.~Okada, and K.-i. Okumura, {\it {Muon anomalous magnetic
  moment, lepton flavor violation, and flavor changing neutral current
  processes in SUSY GUT with right-handed neutrino}},  {\em Phys.Rev.} {\bf
  D64} (2001) 095001, [\href{http://xxx.lanl.gov/abs/hep-ph/0104146}{{\tt
  hep-ph/0104146}}].

\bibitem{Buras:2001ra}
A.~J. Buras, S.~Jager, and J.~Urban, {\it {Master formulae for Delta F=2 NLO
  QCD factors in the standard model and beyond}},  {\em Nucl.Phys.} {\bf B605}
  (2001) 600--624, [\href{http://xxx.lanl.gov/abs/hep-ph/0102316}{{\tt
  hep-ph/0102316}}].

\bibitem{Herrlich:1996vf}
S.~Herrlich and U.~Nierste, {\it {The Complete $|\Delta S| = 2$ - Hamiltonian
  in the next-to-leading order}},  {\em Nucl.Phys.} {\bf B476} (1996) 27--88,
  [\href{http://xxx.lanl.gov/abs/hep-ph/9604330}{{\tt hep-ph/9604330}}].

\bibitem{Buras:2004qb}
A.~J. Buras, T.~Ewerth, S.~Jager, and J.~Rosiek, {\it {$K^+ \to \pi^+ \nu
  \bar{\nu}$ and $K(L) \to \pi^0 \nu \bar{\nu}$ decays in the general MSSM}},
  {\em Nucl.Phys.} {\bf B714} (2005) 103--136,
  [\href{http://xxx.lanl.gov/abs/hep-ph/0408142}{{\tt hep-ph/0408142}}].

\bibitem{Colangelo:1998pm}
G.~Colangelo and G.~Isidori, {\it {Supersymmetric contributions to rare kaon
  decays: Beyond the single mass insertion approximation}},  {\em JHEP} {\bf
  9809} (1998) 009, [\href{http://xxx.lanl.gov/abs/hep-ph/9808487}{{\tt
  hep-ph/9808487}}].

\bibitem{Crivellin:2011jt}
A.~Crivellin, L.~Hofer, and J.~Rosiek, {\it {Complete resummation of
  chirally-enhanced loop-effects in the MSSM with non-minimal sources of
  flavor-violation}},  {\em JHEP} {\bf 1107} (2011) 017,
  [\href{http://xxx.lanl.gov/abs/1103.4272}{{\tt arXiv:1103.4272}}].

\bibitem{Ibrahim:1999hh}
T.~Ibrahim and P.~Nath, {\it {CP violation and the muon anomaly in N=1
  supergravity}},  {\em Phys.Rev.} {\bf D61} (2000) 095008,
  [\href{http://xxx.lanl.gov/abs/hep-ph/9907555}{{\tt hep-ph/9907555}}].

\bibitem{Pierce:1996zz}
D.~M. Pierce, J.~A. Bagger, K.~T. Matchev, and R.-j. Zhang, {\it {Precision
  corrections in the minimal supersymmetric standard model}},  {\em Nucl.Phys.}
  {\bf B491} (1997) 3--67, [\href{http://xxx.lanl.gov/abs/hep-ph/9606211}{{\tt
  hep-ph/9606211}}].

\bibitem{Degrassi:2001yf}
G.~Degrassi, P.~Slavich, and F.~Zwirner, {\it {On the neutral Higgs boson
  masses in the MSSM for arbitrary stop mixing}},  {\em Nucl.Phys.} {\bf B611}
  (2001) 403--422, [\href{http://xxx.lanl.gov/abs/hep-ph/0105096}{{\tt
  hep-ph/0105096}}].

\bibitem{Brignole:2001jy}
A.~Brignole, G.~Degrassi, P.~Slavich, and F.~Zwirner, {\it {On the
  $O(\alpha(t)^2)$ two loop corrections to the neutral Higgs boson masses in
  the MSSM}},  {\em Nucl.Phys.} {\bf B631} (2002) 195--218,
  [\href{http://xxx.lanl.gov/abs/hep-ph/0112177}{{\tt hep-ph/0112177}}].

\bibitem{Brignole:2002bz}
A.~Brignole, G.~Degrassi, P.~Slavich, and F.~Zwirner, {\it {On the two loop
  sbottom corrections to the neutral Higgs boson masses in the MSSM}},  {\em
  Nucl.Phys.} {\bf B643} (2002) 79--92,
  [\href{http://xxx.lanl.gov/abs/hep-ph/0206101}{{\tt hep-ph/0206101}}].

\bibitem{Dedes:2003km}
A.~Dedes, G.~Degrassi, and P.~Slavich, {\it {On the two loop Yukawa corrections
  to the MSSM Higgs boson masses at large $\tan\beta$}},  {\em Nucl.Phys.} {\bf
  B672} (2003) 144--162, [\href{http://xxx.lanl.gov/abs/hep-ph/0305127}{{\tt
  hep-ph/0305127}}].

\bibitem{Dedes:2002dy}
A.~Dedes and P.~Slavich, {\it {Two loop corrections to radiative electroweak
  symmetry breaking in the MSSM}},  {\em Nucl.Phys.} {\bf B657} (2003)
  333--354, [\href{http://xxx.lanl.gov/abs/hep-ph/0212132}{{\tt
  hep-ph/0212132}}].

\bibitem{Allanach:2004rh}
B.~Allanach, A.~Djouadi, J.~Kneur, W.~Porod, and P.~Slavich, {\it {Precise
  determination of the neutral Higgs boson masses in the MSSM}},  {\em JHEP}
  {\bf 0409} (2004) 044, [\href{http://xxx.lanl.gov/abs/hep-ph/0406166}{{\tt
  hep-ph/0406166}}].

\bibitem{AranaCatania:2011ak}
M.~Arana-Catania, S.~Heinemeyer, M.~Herrero, and S.~Penaranda, {\it {Higgs
  Boson masses and B-Physics Constraints in Non-Minimal Flavor Violating SUSY
  scenarios}},  {\em JHEP} {\bf 1205} (2012) 015,
  [\href{http://xxx.lanl.gov/abs/1109.6232}{{\tt arXiv:1109.6232}}].

\bibitem{gelman1992}
A.~Gelman and D.~B. Rubin, {\it Inference from iterative simulation using
  multiple sequences},  {\em Statist. Sci.} {\bf 7} (11, 1992) 457--472.

\bibitem{Camargo-Molina:2013qva}
J.~Camargo-Molina, B.~O'Leary, W.~Porod, and F.~Staub, {\it
  {$\mathbf{Vevacious}$: A Tool For Finding The Global Minima Of One-Loop
  Effective Potentials With Many Scalars}},  {\em Eur.Phys.J.} {\bf C73} (2013)
  2588, [\href{http://xxx.lanl.gov/abs/1307.1477}{{\tt arXiv:1307.1477}}].

\bibitem{Allanach:2008qq}
B.~C. Allanach et~al., {\it {SUSY Les Houches Accord 2}},  {\em Comput. Phys.
  Commun.} {\bf 180} (2009) 8--25,
  [\href{http://xxx.lanl.gov/abs/0801.0045}{{\tt arXiv:0801.0045}}].

\bibitem{Degrassi:2012ry}
G.~Degrassi, S.~Di~Vita, J.~Elias-Miro, J.~R. Espinosa, G.~F. Giudice, et~al.,
  {\it {Higgs mass and vacuum stability in the Standard Model at NNLO}},  {\em
  JHEP} {\bf 1208} (2012) 098, [\href{http://xxx.lanl.gov/abs/1205.6497}{{\tt
  arXiv:1205.6497}}].

\bibitem{Buttazzo:2013uya}
D.~Buttazzo, G.~Degrassi, P.~P. Giardino, G.~F. Giudice, F.~Sala, et~al., {\it
  {Investigating the near-criticality of the Higgs boson}},  {\em JHEP} {\bf
  1312} (2013) 089, [\href{http://xxx.lanl.gov/abs/1307.3536}{{\tt
  arXiv:1307.3536}}].

\end{thebibliography}\endgroup

\end{document}